\documentclass[twocolumn,superscriptaddress,amsmath,amssymb,aps,longbibliography,prx]{revtex4-2}
\usepackage[T1]{fontenc}
\usepackage{amssymb}
\usepackage{graphicx}
\usepackage{booktabs}
\usepackage{amsmath}
\usepackage{multirow}
\usepackage{blkarray}
\usepackage{nicematrix}
\usepackage{tikz}
\usepackage{mathrsfs}
\usepackage[bookmarks=true,colorlinks,citecolor=blue,urlcolor=blue]{hyperref}
\usepackage{braket}
\usepackage{babel}
\usepackage{blindtext}
\usepackage{soul}
\usepackage[compat=1.1.0]{tikz-feynman}
\usepackage[normalem]{ulem}
\usepackage{physics}

\usepackage{mathtools}
\usepackage{bbm}
\usepackage{tikz}
\usepackage{tabularx}
\usepackage{multirow}
\usepackage{graphicx}

\definecolor{mypurp}{rgb}{0.35, 0, 0.7}

\usepackage{amsthm}
\usepackage{txfonts}

\usepackage[compat=1.1.0]{tikz-feynman}
\theoremstyle{definition}

\newcommand{\myvertex}[4]{%
\raisebox{-0.4\height}{%
\begin{tikzpicture}[scale=0.35]
  \coordinate (c0) at (0,0);
  \coordinate (c1) at (0,2);
  \coordinate (c2) at (2,0);
  \coordinate (c3) at (2,2);
  \coordinate (m01) at  (0.35,1);
   \coordinate (m23) at (1.65,1);
   \coordinate (m02) at (1,0.35);
   \coordinate (m13) at (1,1.65);

  \draw[color=white] (c0) -- (c1) node[midway,color=black] (n01) {$#1$};
  \draw[color=white] (c0) -- (c2) node[midway,color=black] (n02) {$#4$};
  \draw[color=white] (c2) -- (c3) node[midway,color=black] (n23) {$#3$};
  \draw[color=white] (c1) -- (c3) node[midway,color=black] (n13) {$#2$};

  \draw[-,densely dotted,color=gray] (m01) -- (m23);
  \draw[-,densely dotted,color=gray] (m02) -- (m13);
\end{tikzpicture}%
}
}

\newcommand{\myplaqgauge}[4]{%
\raisebox{-0.4\height}{%
\begin{tikzpicture}[scale=0.35]
  \coordinate (c0) at (0,0);
  \coordinate (c1) at (0,2);
  \coordinate (c2) at (2,0);
  \coordinate (c3) at (2,2);

  \draw[color=white] (c0) -- (c1) node[midway,color=black] (n01) {$#1$};
  \draw[color=white] (c0) -- (c2) node[midway,color=black] (n02) {$#4$};
  \draw[color=white] (c2) -- (c3) node[midway,color=black] (n23) {$#3$};
  \draw[color=white] (c1) -- (c3) node[midway,color=black] (n13) {$#2$};
  \draw[densely dotted,gray] (0,0) -- (0,0.65) {};
   \draw[densely dotted,gray] (0,1.35) -- (0,2) {};
\draw[densely dotted,gray] (0,2) -- (0.65,2) {};
\draw[densely dotted,gray] (1.35,2) -- (2,2) {};
 \draw[densely dotted,gray] (2,0.65) -- (2,0) {};
  \draw[densely dotted,gray] (2,2) -- (2,1.35) {};
\draw[densely dotted,gray] (1.35,0) -- (2,0) {};
\draw[densely dotted,gray] (0,0) -- (0.65,0) {};
\end{tikzpicture}%
}
}

\newcommand{\xwtvertex}[4]{%
\raisebox{-0.4\height}{%
\begin{tikzpicture}[scale=0.35]
  \coordinate (c0) at (0,0);
  \coordinate (c1) at (0,2);
  \coordinate (c2) at (2,0);
  \coordinate (c3) at (2,2);
  \coordinate (m01) at  (0.15,1);
   \coordinate (m23) at (1.45,1);
   \coordinate (m02) at (0.8,0.15);
   \coordinate (m13) at (0.8,1.45);

  \draw[color=white] (c0) -- (c1) node[midway,color=black] (n01) {$#1$};
  \draw[color=white] (c0) -- (c2) node[midway,color=black] (n02) {$#4$};
  \draw[color=white] (c2) -- (c3) node[midway,color=black] (n23) {$#3$};
  \draw[color=white] (c1) -- (c3) node[midway,color=black] (n13) {$#2$};

  \draw[-,densely dotted,color=gray] (m01) -- (m23);
  \draw[-,densely dotted,color=gray] (m02) -- (m13);
\end{tikzpicture}%
}
}

\newcommand{\myarraysix}[6]{%
\raisebox{0.2\height}{%
\begin{tikzpicture}[scale=0.6,baseline=(current bounding box.center)]
  \coordinate (A) at (0,1) {};
  \coordinate (D) at (0,0) {};
  \coordinate (E) at (1,0) {};
  \coordinate (F) at (2,0) {};
  \coordinate (B) at (1,1) {};
  \coordinate (C) at (2,1) {};
  \node at (A)  {$#1$};
  \node at (B)  {$#2$};
  \node at (C) {$#3$};
  \node at (D) {$#4$};
  \node at (E) {$#5$};
  \node at (F) {$#6$};
  \draw[densely dotted,gray] (0,0.35) -- (0,0.65) {};
\draw[densely dotted,gray] (0.35,1) -- (0.65,1) {};
\draw[densely dotted,gray] (1,0.65) -- (1,0.35) {};
 \draw[densely dotted,gray] (0.65,0) -- (0.35,0) {};
 \draw[densely dotted,gray] (2,0.65) -- (2,0.35) {};
\draw[densely dotted,gray] (1.35,1) -- (1.65,1) {};
\draw[densely dotted,gray] (1.35,0) -- (1.65,0) {};

\end{tikzpicture}%
}}

\newcommand{\myarraynineur}[3]{%
\raisebox{0.4\height}{%
\begin{tikzpicture}[scale=0.4,baseline=(current bounding box.center)]
  \coordinate (A) at (0,1) {};
  \coordinate (D) at (0,0) {};
  \coordinate (E) at (1,0) {};
  \coordinate (F) at (2,0) {};
  \coordinate (B) at (1,1) {};
  \coordinate (C) at (2,1) {};
  \coordinate (G) at (0,-1) {};
  \coordinate (H) at (1,-1) {};
  \coordinate (I) at (2,-1) {};
  \node at (B)  {$#1$};
  \node at (E) {$#2$};
  \node at (F) {$#3$};
  \draw[densely dotted,gray] (0,0) -- (0,1) {};
\draw[densely dotted,gray] (0,1) -- (0.65,1) {};
\draw[densely dotted,gray] (1,0.65) -- (1,0.35) {};
 \draw[densely dotted,gray] (0.65,0) -- (0,0) {};
 \draw[densely dotted,gray] (2,1) -- (2,0.35) {};
\draw[densely dotted,gray] (1.35,1) -- (2,1) {};
\draw[densely dotted,gray] (1.35,0) -- (1.65,0) {};
 \draw[densely dotted,gray] (2,-1) -- (2,-0.35) {};
 \draw[densely dotted,gray] (1,-1) -- (1,0) {};
\draw[densely dotted,gray] (0,0) -- (0,-1) {};
\draw[densely dotted,gray] (1,-1) -- (2,-1) {};
\draw[densely dotted,gray] (0,-1) -- (1,-1) {};

\end{tikzpicture}%
}}

\newcommand{\myarraynineld}[3]{%
\raisebox{0.0\height}{%
\begin{tikzpicture}[scale=0.4,baseline=(current bounding box.center)]
  \coordinate (A) at (0,1) {};
  \coordinate (D) at (0,0) {};
  \coordinate (E) at (1,0) {};
  \coordinate (F) at (2,0) {};
  \coordinate (B) at (1,1) {};
  \coordinate (C) at (2,1) {};
  \coordinate (G) at (0,-1) {};
  \coordinate (H) at (1,-1) {};
  \coordinate (I) at (2,-1) {};
  \node at (D) {$#1$};
  \node at (E) {$#2$};
  \node at (H) {$#3$};
  \draw[densely dotted,gray] (0,0.35) -- (0,1) {};
\draw[densely dotted,gray] (0,1) -- (2,1) {};
\draw[densely dotted,gray] (1,1) -- (1,0.35) {};
 \draw[densely dotted,gray] (0.65,0) -- (0.35,0) {};
 \draw[densely dotted,gray] (2,-1) -- (2,1) {};
\draw[densely dotted,gray] (1.35,0) -- (2,0) {};
 \draw[densely dotted,gray] (1,-0.65) -- (1,-0.35) {};
\draw[densely dotted,gray] (0,-0.35) -- (0,-1) {};
\draw[densely dotted,gray] (1.35,-1) -- (2,-1) {};
\draw[densely dotted,gray] (0,-1) -- (0.65,-1) {};
\end{tikzpicture}%
}}

\newcommand{\myarrayninerd}[3]{%
\raisebox{0.0\height}{%
\begin{tikzpicture}[scale=0.4,baseline=(current bounding box.center)]
  \coordinate (A) at (0,1) {};
  \coordinate (D) at (0,0) {};
  \coordinate (E) at (1,0) {};
  \coordinate (F) at (2,0) {};
  \coordinate (B) at (1,1) {};
  \coordinate (C) at (2,1) {};
  \coordinate (G) at (0,-1) {};
  \coordinate (H) at (1,-1) {};
  \coordinate (I) at (2,-1) {};
  \node at (E)  {$#1$};
  \node at (F)  {$#2$};
  \node at (H) {$#3$};
\draw[densely dotted,gray] (0,1) -- (2,1) {};
\draw[densely dotted,gray] (1,1) -- (1,0.35) {};
 \draw[densely dotted,gray] (0.65,0) -- (0,0) {};
 \draw[densely dotted,gray] (2,1) -- (2,0.35) {};
\draw[densely dotted,gray] (1.35,0) -- (1.65,0) {};
 \draw[densely dotted,gray] (2,-1) -- (2,-0.35) {};
 \draw[densely dotted,gray] (1,-0.65) -- (1,-0.35) {};
\draw[densely dotted,gray] (0,-1) -- (0,1) {};
\draw[densely dotted,gray] (1.35,-1) -- (2,-1) {};
\draw[densely dotted,gray] (0,-1) -- (0.65,-1) {};

\end{tikzpicture}%
}}

\newcommand{\myarraynineul}[3]{%
\raisebox{0.4\height}{%
\begin{tikzpicture}[scale=0.4,baseline=(current bounding box.center)]
  \coordinate (A) at (0,1) {};
  \coordinate (D) at (0,0) {};
  \coordinate (E) at (1,0) {};
  \coordinate (F) at (2,0) {};
  \coordinate (B) at (1,1) {};
  \coordinate (C) at (2,1) {};
  \coordinate (G) at (0,-1) {};
  \coordinate (H) at (1,-1) {};
  \coordinate (I) at (2,-1) {};
  \node at (B)  {$#1$};
  \node at (D)  {$#2$};
  \node at (E) {$#3$};
  \draw[densely dotted,gray] (0,0.35) -- (0,1) {};
\draw[densely dotted,gray] (0,1) -- (0.65,1) {};
\draw[densely dotted,gray] (1,0.65) -- (1,0.35) {};
 \draw[densely dotted,gray] (0.65,0) -- (0.35,0) {};
 \draw[densely dotted,gray] (2,-1) -- (2,1) {};
\draw[densely dotted,gray] (1.35,1) -- (2,1) {};
\draw[densely dotted,gray] (1.35,0) -- (2,0) {};
 \draw[densely dotted,gray] (1,-1) -- (1,-0.35) {};
\draw[densely dotted,gray] (0,-0.35) -- (0,-1) {};
\draw[densely dotted,gray] (0,-1) -- (2,-1) {};
\end{tikzpicture}%
}}

\newcommand{\myarrayninem}[1]{%
\raisebox{0.4\height}{%
\begin{tikzpicture}[scale=0.4,baseline=(current bounding box.center)]
  \coordinate (A) at (0,1) {};
  \coordinate (D) at (0,0) {};
  \coordinate (E) at (1,0) {};
  \coordinate (F) at (2,0) {};
  \coordinate (B) at (1,1) {};
  \coordinate (C) at (2,1) {};
  \coordinate (G) at (0,-1) {};
  \coordinate (H) at (1,-1) {};
  \coordinate (I) at (2,-1) {};
  \node at (E)  {$#1$};
  \draw[densely dotted,gray] (0,-1) -- (0,1) {};
\draw[densely dotted,gray] (0,1) -- (2,1) {};
\draw[densely dotted,gray] (1,1) -- (1,0.35) {};
 \draw[densely dotted,gray] (0.65,0) -- (0,0) {};
 \draw[densely dotted,gray] (2,-1) -- (2,1) {};
\draw[densely dotted,gray] (1.35,1) -- (2,1) {};
\draw[densely dotted,gray] (1.35,0) -- (2,0) {};
 \draw[densely dotted,gray] (1,-1) -- (1,-0.35) {};
\draw[densely dotted,gray] (0,-0.35) -- (0,-1) {};
\draw[densely dotted,gray] (0,-1) -- (2,-1) {};
\end{tikzpicture}%
}}

\newcommand{\myarraynineulnom}[2]{%
\raisebox{0.4\height}{%
\begin{tikzpicture}[scale=0.4,baseline=(current bounding box.center)]
  \coordinate (A) at (0,1) {};
  \coordinate (D) at (0,0) {};
  \coordinate (E) at (1,0) {};
  \coordinate (F) at (2,0) {};
  \coordinate (B) at (1,1) {};
  \coordinate (C) at (2,1) {};
  \coordinate (G) at (0,-1) {};
  \coordinate (H) at (1,-1) {};
  \coordinate (I) at (2,-1) {};
  \node at (B)  {$#1$};
  \node at (D)  {$#2$};
  \draw[densely dotted,gray] (0,0.35) -- (0,1) {};
\draw[densely dotted,gray] (0,1) -- (0.65,1) {};
 \draw[densely dotted,gray] (2,0) -- (0.35,0) {};
 \draw[densely dotted,gray] (2,-1) -- (2,1) {};
\draw[densely dotted,gray] (1.35,1) -- (2,1) {};
 \draw[densely dotted,gray] (1,-1) -- (1,0.65) {};
\draw[densely dotted,gray] (0,-0.35) -- (0,-1) {};
\draw[densely dotted,gray] (0,-1) -- (2,-1) {};
\end{tikzpicture}%
}}

\newcommand{\myarrayplaq}[4]{%
\raisebox{0.2\height}{%
\begin{tikzpicture}[scale=0.4,baseline=(current bounding box.center)]
  \coordinate (A) at (0,1) {};
  \coordinate (D) at (1,1) {};
  \coordinate (C) at (1,0) {};
  \coordinate (B) at (0,0) {};
  \node at (A)  {$#1$};
  \node at (B)  {$#2$};
  \node at (C) {$#3$};
  \node at (D) {$#4$};
  \draw[densely dotted,gray] (0,0.35) -- (0,0.65) {};
\draw[densely dotted,gray] (0.35,1) -- (0.65,1) {};
\draw[densely dotted,gray] (1,0.6) -- (1,0.4) {};
 \draw[densely dotted,gray] (0.6,0) -- (0.4,0) {};

\end{tikzpicture}%
}}

\newcommand{\myarray}[4]{#1 #2 #3 #4}
\newcommand{\nameO}[0]{conserved operator }
\newcommand{\nameOs}[0]{conserved operators }
\newcommand{\nameOsend}[0]{conserved operators}

\begin{document}

\title{Extracting \nameOs from a projected entangled pair state} 

\newcommand{\mcqst}[0]{Munich Center for Quantum Science and Technology (MCQST), Schellingstraße 4, 80799 München, Germany}

\author{Wen-Tao Xu}
\affiliation{Technical University of Munich, TUM School of Natural Sciences, Physics Department, 85748 Garching, Germany}
\affiliation{\mcqst}

\author{Miguel Fr\'ias P\'erez}
\affiliation{ICFO-Institut  de  Ciencies  Fotoniques,  The  Barcelona  Institute  of  Science  and  Technology, 08860 Castelldefels (Barcelona), Spain}
\affiliation{Max-Planck-Institut f{\"u}r Quantenoptik, Hans-Kopfermann-Straße 1, D-85748 Garching, Germany}
\affiliation{\mcqst}

\author{Mingru Yang}
\email{mingruy@uci.edu}
\affiliation{\mcqst}
\affiliation{Max-Planck-Institut f{\"u}r Quantenoptik, Hans-Kopfermann-Straße 1, D-85748 Garching, Germany}

\begin{abstract}
Given a tensor network state, how can we determine conserved operators (including Hamiltonians) for which the state is an eigenstate? We answer this question by presenting a method to extract geometrically $k$-local \nameOs that have the given infinite projected entangled pair state (iPEPS) in 2D as an (approximate) eigenstate. The key ingredient is the evaluation of the static structure factors of multi-site operators through differentiating the generating function. These generating functions define a manifold of the given tensor network state deformed by some parameters, endowed with a quantum geometry, where conserved operators correspond to vanishing fidelity susceptibility. Despite the approximation errors, we show that our method is still able to extract from exact or variational iPEPS to good precision both frustration-free and non-frustration-free parent Hamiltonians that are beyond the standard construction and obtain better locality. In particular, we find a 4-site-plaquette local Hamiltonian that approximately has the short-range RVB state as the ground state. Moreover, we find a Hamiltonian for which the deformed toric code state at arbitrary string tension is an excited eigenstate with the same energy, thereby potentially realizing quantum many-body scars.

\end{abstract}
\date{\today}

\maketitle
\textbf{Introduction.} 
Given access only to a Gibbs state~\cite{Anshu2021,learncommuting,Haah2024,10.1145/3618260.3649619,PhysRevX.8.031029,Qi2019determininglocal,PhysRevLett.122.020504} or to the time evolution~\cite{PhysRevLett.130.200403,PRXQuantum.5.010350} generated by an unknown Hamiltonian, how can we reconstruct the underlying Hamiltonian from measurement data? This inverse problem, often termed as Hamiltonian learning~\cite{Anshu2024}, has been brought to the forefront due to the advances of quantum simulators~\cite{Feynman1982,RevModPhys.86.153} and noisy intermediate-scale quantum (NISQ) devices~\cite{Preskill_NISQ_2018} in recent years, where noises or decoherence from interacting with the environment are unavoidable, making it crucial to certify~\cite{PhysRevLett.112.190501,Wang2017} that the analog or digital platforms implement the intended models.

A complementary motivation arises in the study of strongly correlated quantum matter. Understanding phenomena such as high-temperature superconductivity relies on deriving low-energy effective Hamiltonian $H_{\mathrm{eff}}$ that is simplified but still captures the essential physics of complicated ab initio descriptions. For weak perturbations $H_{\mathrm{eff}}$ can be derived by the Schrieffer–Wolff transformation~\cite{SW_1966,S_W_transformation_2011}, whereas for strongly correlated systems this downfolding is highly nontrivial: competing interactions can appear at comparable scales, and it is often unclear which terms are indispensable in low-energy physics~\cite{Jiang_white_2023}. 
This raises a natural question: can one infer $H_{\mathrm{eff}}$ directly from experimental observables at low temperatures or from classical simulations of ground state of more microscopic models? In addition, symmetries play an essential role in organizing low-energy physics, though often hidden or emergent~\cite{Yang_2023}, and therefore it would be equally desirable to have schemes that uncover conserved operators (including, but not limited to, the Hamiltonian itself) in an automated, data-driven manner. On the other hand, ansatz wave functions such as the Laughlin state~\cite{Laughlin_1983} or the RVB state~\cite{Anderson_RVB_1987} has been proposed to describe interesting physics, and it would also be useful to have a way to systematically determine Hamiltonians for which those ansatz is an eigenstate, rather than postulating models simply by physical intuition. Relatedly, interests on quantum many-body scars~\cite{scar_2018} also require discoveries of new models that can embed certain low-entanglement states as excited states~\cite{Sanjay_2020,PhysRevB.107.224312,PhysRevE.92.012128,PhysRevB.94.144208,PhysRevResearch.2.023348,Sanjay_2025}.

\begin{figure}
    \centering
    \includegraphics[width=\linewidth]{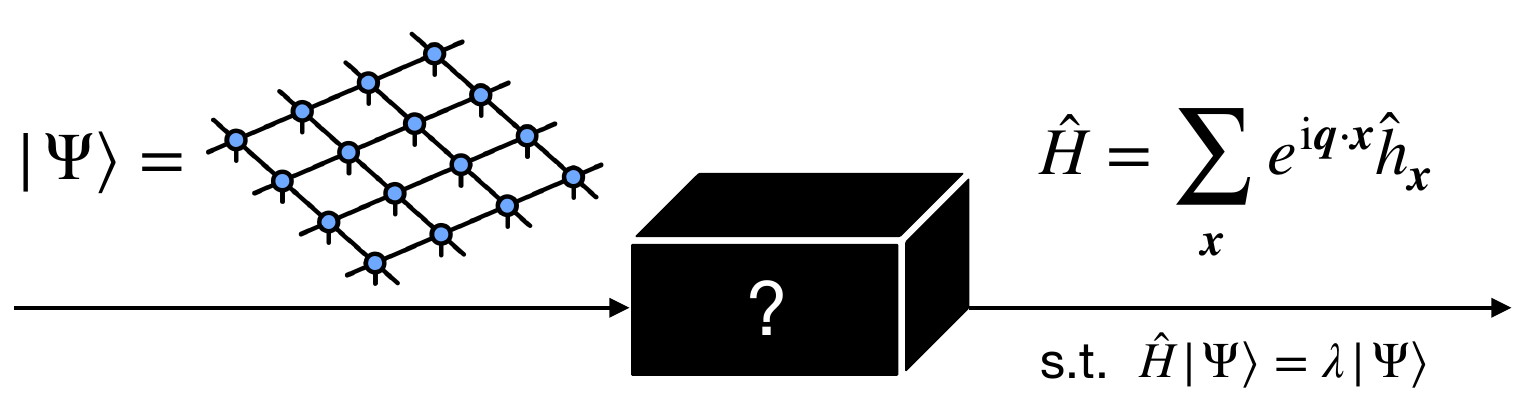}
    \caption{Given a uniform PEPS $|\Psi\rangle$ as input, our method outputs a geometrically $k$-local operator $\hat{H}$ which has the PEPS as an eigenstate.}
    \label{fig:HL}
\end{figure}

Tensor network states provide a particularly suitable playground for these questions. Matrix product states (MPS)~\cite{Fannes1992,PhysRevLett.69.2863,PhysRevB.48.10345} and projected entangled pair states (PEPS)~\cite{maeshima:2001,verstraete2004} efficiently capture ground states of 1D~\cite{Hastings_2007,arad2013arealawsubexponentialalgorithm,PhysRevB.73.094423} and 2D~\cite{Andras_2015} gapped local Hamiltonians and furnish a structured, physically motivated manifold of states for many-body systems~\cite{RMP_MPS_PEPS_2021}. Since ground states are Gibbs states of the same Hamiltonian at the zero temperature limit, a natural question would be: given a tensor network state, how can we identify \nameOs for which this state is an eigenstate? Previous Hamiltonian learning theories established bounds on sample and time complexity that are exponential in inverse temperature in the low temperature regime~\cite{Anshu2021,10.1145/3618260.3649619}, which indicates reconstructing Hamiltonian from its ground state is extremely hard in the worst case, because ground states are believed to lack sufficient information about the full Hamiltonian~\cite{Haah2024}. Nevertheless, in 1D, efficient algorithms have been developed to learn geometrically $k$-local Hamiltonians (or other conserved operators) from MPS~\cite{Yang_2023}. Extending such approaches to 2D PEPS is substantially more challenging: exact contraction of PEPS is \#P-complete~\cite{Schuch_2007}, and practical calculations of observables in physically relevant systems rely on approximate schemes~\cite{1D_TEBD_2008,CTMRG_1,CTMRG_corboz,TRG_2007,window_MPS_2022}, which introduce controlled but non-negligible errors in correlation functions and static structure factors. It is therefore unclear whether we can still extract information about the underlying Hamiltonian or other conserved operators under such approximation errors.

In this work, we clarify this issue by providing a classical algorithm to extract \nameOs from a given PEPS whose static two-point correlations are computable within the approximate contraction schemes (Fig.~\ref{fig:HL}). Our only input is the static structure factors of the state, which is not only accessible numerically by tensor network methods or Monte Carlo simulations, but also directly measurable in scattering experiments~\cite{RevModPhys.79.175,RevModPhys.96.015003}. From this input, our method outputs geometrically $k$-local \nameOs (including Hamiltonians) $\hat{H}$'s for which the PEPS $\ket{\Psi}$ is an eigenstate, i.e. $\hat{H}\ket{\Psi}=\lambda\ket{\Psi}$. Without loss of generality, we assume that the system is translationally invariant and the target \nameO can be expressed as $\hat{H}=\sum_{\pmb{x}}e^{i\pmb{q}\cdot\pmb{x}}\hat{h}_{\pmb{x}}$, where $\pmb{q}$ denotes momentum and $\hat{h}_{\pmb{x}}$ is the same geometrically $k$-local Hermitian operator for each $\pmb{x}$. The computational cost of the procedure is comparable to that of evaluating the static structure factor itself and thus scales polynomially in the PEPS bond dimension. In contrast to conventional parent-Hamiltonian constructions~\cite{10.5555/2016976.2016982,SCHUCH_2010,RMP_MPS_PEPS_2021}, the operators we obtain need not be frustration free, need not yield the given state as its ground state, have much better locality, and can capture both exact or emergent \nameOs associated with the given state. 

\textbf{Conserved operators as kernel of the static structure factor matrix.}
Ref.~\cite{Yang_2023} pointed out that \nameOs of a quantum state can be solved via variational optimization of the static structure factors, which is equivalent to an eigenvalue problem~\cite{PhysRevX.8.031029,Qi2019determininglocal,Yang_2023}. Here we consider the static structure factors for $\{\hat{o}^{\alpha}_{\pmb{x}}\}$ that forms a complete orthonormal ($\mathrm{Tr}(\hat{o}^{\alpha}_{\pmb{x}}\hat{o}^{\beta}_{\pmb{x}})=\delta_{\alpha\beta}$) basis set of geometrically $k$-local Hermitian operators at $\pmb{x}$
\begin{align}\label{eq:def_struct_factor}
     S_{\alpha,\beta}(\pmb{q})=&\sum_{\pmb{x}}e^{-i\pmb{q}\cdot\pmb{x}}\left(\frac{\bra{\Psi}\hat{o}^{\alpha}_{\pmb{x}}\hat{o}^{\beta}_{\pmb{0}}\ket{\Psi}}{\langle\Psi |\Psi\rangle} - \frac{\bra{\Psi}\hat{o}^{\alpha}_{\pmb{x}}\ket{\Psi}}{\langle\Psi |\Psi\rangle}\frac{\bra{\Psi}\hat{o}^{\beta}_{\pmb{0}}\ket{\Psi}}{\langle\Psi |\Psi\rangle}\right),
\end{align}
which are simply the Fourier transformed connected correlation functions. The \nameOs $\{\hat{H}^{[i]}\}$ can be determined by solving the kernel of the static structure factor matrix $\mathcal{S}\equiv (S+S^\mathsf{T})/2$. Because $\mathcal{S}$ is real symmetric and semi-positive definite, the eigenvalues are real and non-negative, and it is guaranteed that the eigenvectors will be real up to an overall phase factor~\cite{Yang_2023}. From the eigenequation $\mathcal{S}\pmb{h}=s\pmb{h}$, the $i$-th exact (approximate) solution $\hat{H}^{[i]}=\sum_{\pmb{x}}e^{i\pmb{q}\cdot\pmb{x}}\hat{h}_{\pmb{x}}^{[i]}$, where $\hat{h}_{\pmb{x}}^{[i]}=\sum_\alpha h_\alpha^{[i]} \hat{o}^{\alpha}_{\pmb{x}}$, can be obtained from the eigenvector $\pmb{h}^{[i]}=(h^{[i]}_1,h^{[i]}_2,\cdots)^\mathsf{T}$ associated with the $i$-th exact (approximate) zero eigenvalue $s_i$ (kernel space) of $\mathcal{S}$. The eigenvalues $s$ are equal to the quantum fluctuation per site of $\hat{H}$ in the state $\ket{\Psi}$ after normalizing $\pmb{h}$. When $\mathcal{S}$ is a small matrix we can fully diagonalize it; otherwise we can use the Lanczos method to find the lowest several eigenvalues and eigenvectors~\cite{Yang_2023}.

Let us remark some additional properties of the solutions. First, it was found that not all eigenvectors in the null space of $\mathcal{S}$ are physical~\cite{Yang_2023}. In fact, for each $\pmb{q}$, the null space of $\mathcal{S}$ has a trivial subspace $\{\pmb{h}^{\mathrm{tri}}\}$ such that $\hat{h}^{\text{tri}}_{\pmb{x}}\neq 0$ but $\hat{H}^{\text{tri}}=\sum_{\pmb{x}} e^{i\pmb{q}\cdot\pmb{x}}\hat{h}^{\text{tri}}_{\pmb{x}}=0$, or $\hat{h}^{\text{tri}}_{\pmb{x}}\propto \mathbbm{1}$. We summarize the 1-site, 2-site, and 4-site-plaquette trivial solutions in the supplementary material (SM)~\cite{appendix}. Second, the $n$-site solutions contain all $m$-site solutions if the $m$-site support is within the $n$-site support.

\textbf{Static structure factor of multi-site operator in iPEPS.}
We focus on the input states being 2D pure quantum states that can be represented by infinite PEPS (iPEPS)~\cite{maeshima:2001,verstraete2004}, which comprise descriptions of rich physical phenomena~\cite{RMP_MPS_PEPS_2021} such as gapped non-chiral topologically ordered states~\cite{SCHUCH_2010,Schuch_2011_classify},
(2+0)D critical states with area-law entanglement entropy but power-law decaying correlations~\cite{Verstraete_2006}, long-range ordered cat states and so on. With finite correlation length scaling~\cite{PhysRevLett.129.200601,Xu_huang_2025}, iPEPS can also be used to study (2+1)D criticality~\cite{Corboz_Finite_IPEPS_2018,Lauchli_Finite_IPEPS_2018}. 

Various approaches on evaluating the static structure factors of iPEPS have been proposed, including (i) summing a geometric
series of channel operators built from the boundary-MPS
approach~\cite{PEPS_tangent_space_2015,iPEPS_Laurens_2016}, (ii) summing diagrams in the corner transfer matrix renormalization
group (CTMRG) method~\cite{CTMRG_1,corboz_2020,Corboz_2022}, (iii) window MPS method~\cite{window_MPS_2022}, (iv) automatic differentiating iPEPS tensors~\cite{Juraj_structure_factor_2023}, (v) the generating function method~\cite{Gen_func_2021,Generating_iPEPS_2024}, and (vi) large unit cell CTMRG~\cite{Corboz_spec_func_2024}. Notice that previously people have only used these methods to evaluate structure factors of on-site operators. In this work we adopt the generating function method to calculate $\mathcal{S}$, since it is easier to generalize to the multi-site operator case compared to other approaches, and avoids dealing with many tensor-network diagrams by hands.

Now we reformulate the generating function method to include the case of $\hat{o}^{\alpha}_{\pmb{x}}$ being multi-site. To deal with the infinite sums in Eq.~\eqref{eq:def_struct_factor},  we use the relation
\begin{equation}
    \sum_{\pmb{x}} e^{-i\pmb{q}\cdot\pmb{x}} \hat{o}^{\alpha}_{\pmb{x}}=\frac{\mathrm{d}}{\mathrm{d}\mu}\left(\prod_{\pmb{x}}(\mathbbm{1}+\mu e^{-i\pmb{q}\cdot\pmb{x}} \hat{o}^{\alpha}_{\pmb{x}})\right)\Bigg|_{\mu=0}.
\end{equation}
When $\hat{o}^{\alpha}_{\pmb{x}}$ is a multi-site operator, $\prod_{\pmb{x}}(\mathbbm{1}+\mu e^{-i\pmb{q}\cdot \pmb{x}} \hat{o}^{\alpha}_{\pmb{x}})$ will be an infinite projected entangled pair operator (iPEPO) whose bond dimension is larger than 1. It can be verified that the iPEPO bound dimension will be the smallest if $\hat{o}^{\alpha}$ is chosen to be a tensor product of on-site operators. The generating function is then defined as~\cite{Gen_func_2021,Generating_iPEPS_2024}
\begin{equation}
\bra{G^{\alpha}(\mu,\pmb{q})}=\bra{\Psi}\prod_{\pmb{x}}(\mathbbm{1}+\mu e^{-i\pmb{q}\cdot\pmb{x}} \hat{o}^{\alpha}_{\pmb{x}}),
\end{equation}
where $\ket{\Psi}$ is an \emph{unnormalized} 1-site unit cell iPEPS. 
When the momentum $\pmb{q}$ is commensurate, i.e. $\pmb{q}=(\frac{2\pi n}{L_x}, \frac{2\pi m}{L_y})$ with $n,m\in\mathbbm{N}$, $L_x,L_y\in\mathbbm{Z}^+$, $n<L_x$, $m<L_y$, $\mathrm{gcd}(n,L_x)=1$, and $\mathrm{gcd}(m,L_y)=1$, the generating function $\bra{G^{\alpha}(\mu,\pmb{q})}$ has a unit cell size $L_x\times L_y$~\cite{Juraj_structure_factor_2023}. Actually, the generating function is not unique and can be alternatively chosen to be a unitary circuit, $\prod_{\pmb{x}}\exp(\mu e^{-i\pmb{q}\cdot\pmb{x}} \hat{o}^{\alpha}_{\pmb{x}})$, applied to $|\Psi\rangle$, if $\mu e^{-i\pmb{q}}$ is restricted to be purely imaginary~\cite{appendix}. The static structure factor can be expressed as~\cite{appendix}
\begin{align}\label{Eq:iPEPS_struct_factor}
   S_{\alpha,\beta}(\pmb{q}) &=\frac{\mathrm{d}}{\mathrm{d}\mu}\frac{\bra{G^{\alpha}(\mu,\pmb{q})}\hat{o}^{\beta}_{\pmb{0}}\ket{\Psi}}{\braket{G^{\alpha}(\mu,\pmb{q})}{\Psi}}\Bigg|_{\mu=0},
\end{align} 
which can be calculated by differentiation of (multi-site unit cell) CTMRG~\cite{CTMRG_1,CTMRG_corboz}. Since $\mu$ is a scalar, the derivative can be implemented by either automatic differentiation~\cite{AD_2019} or finite difference. We find that automatic differentiation is generally unstable for multi-site $\{\hat{o}^{\alpha}\}$~\cite{appendix}, so we use the central 5-point finite difference formula instead to achieve higher precision. It is also straightforward to generalize to the case that $\ket{\Psi}$ has multi-site unit cells~\cite{appendix}. In addition, the iPEPS representation of topologically ordered states and cat states is non-injective~\cite{SCHUCH_2010}, CTMRG of such iPEPS may converge to different environment fixed points, which needs to be taken care of~\cite{appendix} when evaluating $\mathcal{S}$. Similar to the complexity of the approximate contraction of iPEPS~\cite{Orus_TNS_review_2019}, the complexity of evaluating $\mathcal{S}$ is  $O(D'^3D^6\chi^3)$, where $D$ and $D'$ are the iPEPS and iPEPO bond dimensions, respectively. In fact, the generating functions form a manifold of the tensor network state $|\Psi\rangle$ deformed by the parameter $\mu$ and induce a \emph{quantum geometry}. Interestingly, the static structure factor matrix $\mathcal{S}$ is proportional to the quantum metric tensor~\cite{appendix} and \nameOs correspond to zero fidelity susceptibility~\cite{Fidelity_2020}.

In addition to the error induced by the finite difference step, it also comes from the approximate contraction of the iPEPS overlap in Eq.~(\ref{Eq:iPEPS_struct_factor}) using CTMRG, where the error is controlled by the environment bond dimension $\chi$ upon full convergence of CTMRG iterations. For iPEPS with finite correlation length, a finite $\chi$ is already enough to suppress the error exponentially, while for iPEPS with divergent correlation length one can perform scaling analysis with $\chi$~\cite{PhysRevLett.129.200601}. Heuristically, the sufficient number of CTMRG iterations should be several times the correlation length $\xi(\chi)$ induced by $\chi$. For critical iPEPS, it converges much slower because $\xi(\chi)\sim \chi^{\kappa}$, where $\kappa$ is the finite entanglement scaling exponent~\cite{Pollmann_2011}. Moreover, when the iPEPS is obtained from variational optimization~\cite{iPEPS_Laurens_2016,iPEPS_corboz_2016}, the full convergence of correlation functions is generally very hard to reach, and thus this suboptimality can cause additional errors~\cite{AD_2019}. Regardless of those errors, we will show that our approach still works through multiple representative examples.

\textbf{Benchmark on exact PEPS: 2D AKLT state.}
We first benchmark our method on the 2D spin-2 Affleck-Kennedy-Lieb-Tasaki (AKLT) state~\cite{systematic_SU2_PEPS_2016} on the square lattice, which is a natural generalization of the 1D spin-1 AKLT state~\cite{AKLT_1987}. It can be exactly expressed as a PEPS with virtual (physical) bond dimension $D=2$ ($d=5$)~\cite{appendix}, and it has an exact SO(3) symmetry generated by $\sum_iS_i^{\alpha}$, where $\alpha=x,y,z$ and $S^{\alpha}$ are the spin-$2$ operators.

\begin{figure}
    \centering
    \includegraphics[width=\linewidth]{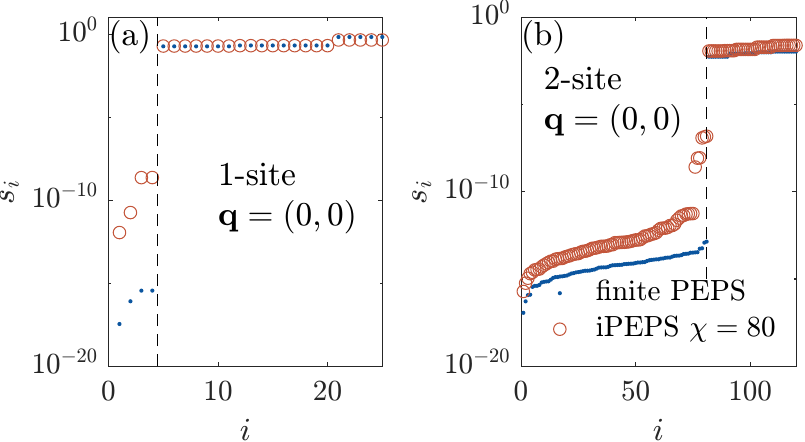}
    \caption{\textbf{Spectrum of the structure factor matrix $\mathcal{S}$ from the 2D AKLT state.} (a) Spectrum of the 1-site $\mathcal{S}$. (b) Spectrum of the 2-site $\mathcal{S}$. We compare the spectra from exactly contracting a finite periodic PEPS with a system size $4\times 4$ (blue dots) and approximately contracting the iPEPS using CTMRG with environment bond dimension $\chi=80$ and finite difference step size $\delta=10^{-4}$ (red circle), which gives the same solutions because they are exact and frustration free.
  }
    \label{fig:AKLT_site}
\end{figure}

Fig.~\ref{fig:AKLT_site}a shows the spectrum of the 1-site $\mathcal{S}$. There are 4 zero eigenvalues (up to precision $10^{-9}$), and the associated eigenvectors are the identity and the three spin-2 operators $S^x$, $S^y$, and $S^z$. Fig.~\ref{fig:AKLT_site}b shows the 2-site result after projecting out the 1-site and the 2-site trivial solutions. There remain 81 zero eigenvalues (up to precision $10^{-7}$), which correspond to the projectors onto the channel $4$ due to the fusion of two neighboring spin-2's, i.e. $2\otimes 2=0\oplus1\oplus2\oplus3\oplus4$~\cite{appendix}, and they enumerate all non-trivial frustration-free Hamiltonians whose common eigenstate is the 2D AKLT state~\cite{Sanjay_2020}. No non-frustration-free solutions are found up to 2-site.

\textbf{Benchmark on variational iPEPS: 2D quantum XX model.}
To demonstrate that our method works well not only for finding frustration-free parent Hamiltonians of a PEPS with exact form, we apply our method to the variationally optimized ground states of a spin-1/2 XX model in a staggered $Z$ field on the 2D square lattice, and verify if it can find the non-frustration-free ($J\neq 0$)
model Hamiltonian
\begin{equation}\label{eq:XY_H_mod}
    H(J)=-J\sum_{\langle i,j \rangle}(X_iX_j +Y_iY_j)-(1-J)\sum_{i}(-1)^i Z_i,
\end{equation}
where $\{X,Y,Z\}$ are the Pauli matrices and $(-1)^i$ denotes that the sign of the $Z$ field is opposite on the two sublattices of the bipartite square lattice. The model has a U(1) symmetry generated by $\sum_iZ_i$. Its ground state has a U(1) symmetric paramagnetic phase and an easy-plane (XY) ferromagnetic phase where the U(1) symmetry is broken spontaneously. We first use the gradient-based variational optimization~\cite{iPEPS_Laurens_2016,iPEPS_corboz_2016,AD_2019} to obtain a 1-site unit cell $C_{4v}$ symmetric iPEPS approximating the ground state, and find two phases which are separated by a quantum critical point at $J=J_c\approx0.336$~\cite{appendix}, which is known to be in the XY universality class~\cite{XX_quantum_simulator_2024}. Although in the ferromagnetic phase all spins can spontaneously point towards a random direction in the $x-y$ plane, in practice we use real iPEPS tensors to force the spin to point to the $x$ direction.

\begin{figure}
     \centering
     \includegraphics[width=\linewidth]{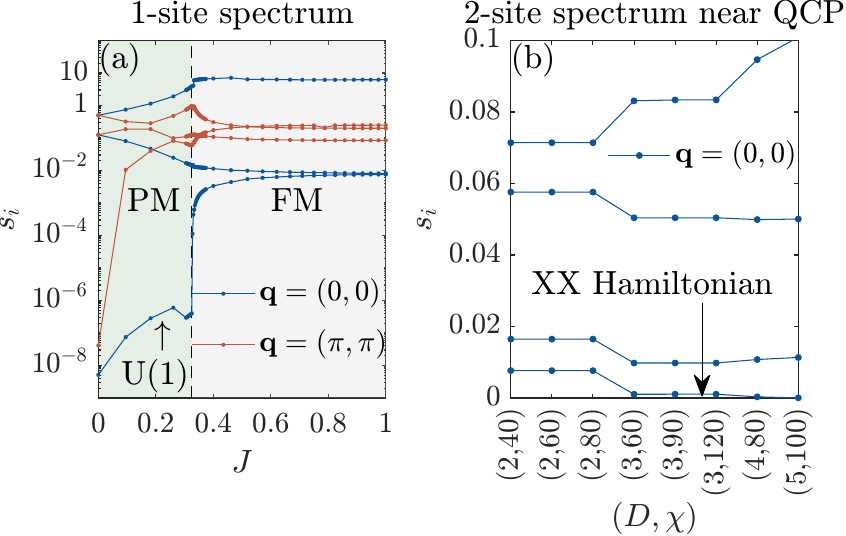}
     \caption{\textbf{Spectrum of the structure factor matrix $\mathcal{S}$ from the variational iPEPS for the ground state of the 2D XX model in a staggered $Z$ field.} (a) Spectra of the 1-site $\mathcal{S}$ evaluated with $(D,\chi)=(3,60)$ and the finite difference step $\delta=10^{-4}$, where $D$ is the iPEPS bond dimension and $\chi$ is the CTMRG environment bond dimension. `U(1)' denotes the corresponding eigenvector is the generator $\sum_i Z_i$.  When $J=0$, two eigenvalues are $0$ up to precision $10^{-7}$, and the corresponding eigenvectors are $\sum_iZ_i$ (blue) and $\sum_{i}(-1)^i Z_i$ (red), which are consistent with the fact that the ground state is a simple product state.
     (b) Low lying spectra of the 2-site $\mathcal{S}$ at $J=1/3$ near the critical point, where $\delta=5\times 10^{-3}$. 
     }
     \label{fig:XX}
\end{figure}

From this iPEPS we then diagonalize $\mathcal{S}$ at momentum $\pmb{q}=(0,0)$ and $\pmb{q}=(\pi,\pi)$. The spectrum of the 1-site $\mathcal{S}$ from iPEPS with $D=3$ and CTMRG environment bond dimension $\chi=60$ is shown in Fig.~\ref{fig:XX}a, where the trivial solution (identity operator) has been excluded. When $J<J_c$, an eigenvalue corresponding to the U(1) symmetry generator $\sum_iZ_i$ is found to be $0$ up to high precision ($10^{-6}$) until $J=J_c$. When $J>J_c$, the two lowest eigenvalues (blue) correspond to $a X-b Z$ and $b X+ aZ$, where $a,b$ depend on $J$. Their linear combinations give two approximate solutions $\sum_i X_i$ and $\sum_i Z_i$, the former is the magnetization order parameter and the latter is the U(1) symmetry generator. Therefore, even when the U(1) symmetry is broken spontaneously, signatures of the U(1) symmetry can still be found by our method. The underlying reason is that the gapless Goldstone modes arising from the spontaneous U(1) symmetry breaking indicate substantial transverse fluctuations of the order parameter. Since the ferromagnetic ground state is not far from a coherent spin state which saturates the uncertainty relation, large variance of $\sum_i Y_i$ will imply small variance of $\sum_i Z_i$.

The spectrum of the 2-site $\mathcal{S}$ from iPEPS with different bond dimension $(D,\chi)$ at $J=1/3$ near the quantum criticality is shown in Fig.~\ref{fig:XX}(b) (excluding all 1-site and trivial 2-site solutions). We find one eigenvalue that shrinks to zero with increasing bond dimensions. The corresponding eigenvector for $(D,\chi)=(5,100)$ gives the Hamiltonian $H(1/3)-0.0001\sum_{\langle ij\rangle}Z_iZ_j-0.0006\sum_{i}(-1)^i Z_i$, which approximates the original Hamiltonian in Eq.~\eqref{eq:XY_H_mod} to the 4th digit. Therefore, our method can extract non-frustration-free \nameOs of variational iPEPS, which is beyond the standard frustration-free construction~\cite{RMP_MPS_PEPS_2021}. 

\textbf{Approximate local parent Hamiltonian of the RVB state.} 
Now let us apply our method to an iPEPS with finite $D$ but a divergent correlation length.  We consider the short-range resonating valence bond (RVB) state on the square lattice as an illustration~\cite{Anderson_RVB_1987}. This wavefunction has the form $\sum_{C}\prod_{\langle i,j\rangle\in C}\left(\ket{\uparrow_i\downarrow_j}-\ket{\downarrow_i\uparrow_j}\right)$, where $C$ denotes all possible nearest-neighbor dimer coverings on the square lattice, and can be exactly expressed as an iPEPS with $d=2$ (spin-$1/2$) and $D=3$ (spin $0\oplus 1/2$)~\cite{RVB_PEPS_2012,systematic_SU2_PEPS_2016,appendix}. It has an exact SU(2) symmetry generated by $\sum_iS_i^{\alpha}$, where $\alpha=x,y,z$ and $S^{\alpha}$ are the spin-$1/2$ operators. The short-range RVB state on the square lattice has been confirmed to be critical from previous numerics~\cite{RVB_fabian_2010,RVB_sandvik_2011}, and its equal-time correlators are known to be described by the $(2+0)$D free boson conformal field theory~\cite{ardonne_2004,Robustness_U_1_2024}.

\begin{table}
\centering
\caption{\textbf{Results from the RVB iPEPS.} $\chi$ is the environment bond dimension of the CTMRG contracting the RVB PEPS. $s_{\mathrm{min}}$ is the smallest eigenvalue of the 4-site $\mathcal{S}$ at $\pmb{q}=(0,0)$ after excluding the smaller-size and trivial solutions. $J_1,J_2,Q_1$ and $Q_2$ are coefficients defined in Eq.~\eqref{eq:H_JJQQ} and obtained from the eigenvector corresponding to $s_{\mathrm{min}}$. $E_{\text{RVB}}$ is the energy expectation value per site of $\hat{H}$ on the RVB iPEPS. $E_{\text{GS}}$ is the ground state energy per site of $\hat{H}$ obtained from variational iPEPS initialized by the RVB iPEPS. We use a finite difference step $\delta=0.01$.}
\label{tab:Results_RVB}
\begin{tabular}{cccccccc}
\toprule
$\chi$ & 80 & 100 & 120 & 140  & 150& 160  \\
\midrule
$s_{\mathrm{min}}/ 10^{-3}$    & $4.018$ & $3.135$ & $1.613$  & $4.831$ & $2.776$     & $3.108$     \\
$J_{1}$         & 1       & 1       &  1       & 1       &   1   & 1 \\
$J_{2}$         & 0.4825  &  0.4624 &  0.4625  & 0.5252  &   0.5000   & 0.5002   \\
$Q_{1}$         & -0.5610 & -0.5860 & -0.6255  & -0.5435 &   -0.4596   & -0.4741   \\
$Q_{2}$         & 0.4020   & 0.4607  & 0.4712   & 0.0719  & 0.2422     & 0.2260    \\
$E_{\text{GS}}$         & -0.6085   & -0.6234  & -0.6342& -0.5768  & -0.5809   &  -0.5819  \\
$E_{\text{RVB}}$       & -0.6004   & -0.6149   & -0.6259   & -0.5651  & -0.5714     & -0.5722    \\
\bottomrule
\end{tabular}
\end{table}

We focus on looking for 4-site-plaquette conserved operators at $\pmb{q}=(0,0)$. Utilizing the SU(2) symmetry (which could be found from the 1-site solutions), we can reduce the size of $\mathcal{S}$ from $256^2$ to $39^2$~\cite{appendix}. The 4-site solutions with both the SU(2) symmetry and the $C_{4v}$ symmetry must take the following form
\begin{align}\label{eq:H_JJQQ}
    \hat{H}=J_{1}\sum_{\langle ij\rangle}\pmb{S}_i\cdot \pmb{S}_j+J_{2}\sum_{\langle \langle ij\rangle\rangle}\pmb{S}_i\cdot \pmb{S}_j\notag\\
    +Q_{1}\sum_p\sum_{\langle ij\rangle,\langle kl\rangle\in p }(\pmb{S}_i\cdot \pmb{S}_j)(\pmb{S}_k\cdot \pmb{S}_l)\notag\\
    +Q_{2}\sum_p\sum_{\langle\langle ij\rangle\rangle,\langle\langle kl\rangle\rangle\in p}(\pmb{S}_i\cdot \pmb{S}_j)(\pmb{S}_k\cdot \pmb{S}_l),
\end{align}
where in the latter two terms, $i,j,k,l$ are the four different sites in the plaquette $p$, and $J_1$, $J_2$, $Q_1$, and $Q_2$ are coefficients to be determined by the approximate kernel of the 4-site $\mathcal{S}$.

Since the RVB state has divergent correlation length, it requires thousands of CTMRG iterations to evaluate $\mathcal{S}$. The lowest eigenvalue $s_{\mathrm{min}}$ excluding all smaller-size and trivial solutions, as well as the coefficients determined by the corresponding eigenvectors, are shown in Tab.~\ref{tab:Results_RVB} for various $\chi$. Since $s_{\mathrm{min}}$ which characterizes the quantum fluctuation of the solution $\hat{H}$ per site is small, the RVB state can be regarded as a good eigenstate of $\hat{H}$. Moreover, we compare the energy expectation value $E_{\text{RVB}}=\langle\mathrm{RVB}|\hat{H}|\mathrm{RVB}\rangle/\langle\mathrm{RVB}|\mathrm{RVB}\rangle$ and the ground state energy $E_{\text{GS}}$ of $\hat{H}$ obtained by variational optimization, and find they are very close to each other (see the last two rows of Tab.~\ref{tab:Results_RVB}), so the non-frustration-free $\hat{H}$ we obtained is indeed a good parent Hamiltonian of the RVB state, and it will be much easier to be engineered for state preparation, because it has better locality compared to the standard frustration-free construction~\cite{Satoshi_Fujimoto_2005,Fendly_2010,twelve_site_H_kagome_RVB_2014,Locality_optim_2022}, which should be at least 8-site.

\textbf{Hamiltonian having the deformed Ising or toric code wavefunctions as excited states.} 
In previous examples, we have used our method to find frustration-free or non-frustration-free parent Hamiltonian of exact iPEPS and variational iPEPS, gapped or gapless. In this section, we show the Hamiltonian we obtain can also have the iPEPS as an excited state. We consider a family of 2D deformed Ising wavefunctions on the square lattice, $\ket{\Psi_{\text{Ising}}(\beta)}=\exp\left(\frac{\beta}{2}\sum_{\langle i,j\rangle}Z_iZ_j\right)\ket{++\cdots+}$, where $\beta$ is a turning parameter. They can be exactly expressed as PEPS with $D=2$~\cite{computation_power_2006}. Solving the kernel of the 4-site $\mathcal{S}$ of the PEPS for a $4\times 4$ periodic system already gives interesting result. The lowest eigenvalue of the 4-site $\mathcal{S}$ at $\pmb{q}=(0,0)$ excluding all trivial solutions is zero up to machine precision for all $\beta$, and the corresponding eigenvector gives a Hamiltonian $\hat{H}=\sum_{\pmb{x}} \hat{h}_{\pmb{x}}$ with 
\begin{align}\label{eq:regroup}
    \hat{h}_{\pmb{x}}&=\myarraynineur{Z}{P_{-}}{Z}+\myarraynineld{Z}{P_{-}}{Z}-\myarrayninerd{P_{-}}{Z}{Z}-
   \myarraynineul{Z}{Z}{P_{-}}
\end{align}
after regrouping, where $P_{-}=I-X$. It obviously satisfies $\hat{h}_{\pmb{x}}\ket{+++\cdots+}=0$ and can be proved to commute with the deformation~\cite{appendix}, i.e. $\left[\hat{h}_{\pmb{x}},\sum_{\langle ij\rangle}Z_iZ_j\right]=0$, for any system size. Therefore, $\hat{h}_{\pmb{x}}\ket{\Psi_{\text{Ising}}(\beta)}=0$ and automatically $\hat{H}\ket{\Psi_{\text{Ising}}(\beta)}=0$, $\forall\beta$. Interestingly, $\ket{\Psi_{\text{Ising}}(\beta)}$'s whose entanglement entropies satisfy the area law are not the ground states of $\hat{H}$, but actually lie in the middle of its spectrum (the spectrum of $\hat{H}$ is symmetric about $0$). $\ket{\Psi_{\text{Ising}}(\beta)}$ can be regarded as quantum many-body scars~\cite{scar_2018,Sanjay_2020} of $\hat{H}$, which we discuss in SM~\cite{appendix} and leave for thorough study in the future.

Furthermore, since the deformed Ising wavefunctions are dual to the deformed toric code states $\ket{\Psi_{\text{TC}}(\beta)}=\prod_e \exp(\frac{\beta}{2}Z_e)\ket{\text{TC}}$~\cite{Deforemd_TC_2008,Trebst_2009}, where $\ket{\text{TC}}$ is a ground state of the toric code model~\cite{kitaev_2002}, we can get a dual Hamiltonian $\hat{H}_{\text{dual}}$ from $\hat{H}$, with its local term being 
\begin{align}
    \hat{h}_{\pmb{x}}^{\text{dual}}=&2\myvertex{I}{Z}{Z}{I}-2\myvertex{I}{I}{Z}{Z}+4\left(\xwtvertex{\sigma^{-}}{\sigma^{-}}{\sigma^{+}}{\sigma^{+}}-\xwtvertex{\sigma^{-}}{\sigma^{+}}{\sigma^{+}}{\sigma^{-}}+\mathrm{h.c.}\right),  \notag   
\end{align}    
where $\sigma^{\pm}=(X\pm iY)/2$, such that the family of the deformed toric code states are $E=0$ excited eigenstates of $\hat{H}_{\text{dual}}$ (see SM~\cite{appendix}).

\textbf{Conclusion and outlook.}
In this work, we show that given a uniform PEPS, geometrically $k$-local conserved operators (including Hamiltonians) which have the PEPS as an (approximate) eigenstate can be obtained by solving the (approximate) kernel of the static structure factor matrices, where the static structure factors are evaluated by differentiating the generating functions. We benchmark our method by extracting the frustration-free parent Hamiltonians of the 2D AKLT state and the non-frustration-free parent Hamiltonian and symmetries of the variationally optimized ground state of a 2D XX model. Applying our method to the critical short-range RVB state on the square lattice, we show that our method can be used to obtain an approximate 4-site-plaquette local parent Hamiltonian. Moreover, we find a Hamiltonian that has all the deformed Ising wavefunctions (or deformed toric code state after gauging) as excited eigenstates in the middle of the spectrum, which could potentially be quantum many-body scars. Our results demonstrate that learning Hamiltonians from iPEPS is feasible despite the errors in the approximate contraction schemes and the state itself being only an approximate ground state.  

Some open questions remain. First, for a larger support size $k$, one can parameterize the local term $h_{\pmb{x}}$ in terms of a tensor network operator with a small bond dimension to reduce the computational complexity. Second, the generating function method to evaluate the 4-site $\mathcal{S}$ of the critical RVB iPEPS has limited precision. One may try evaluating $\mathcal{S}$ using the Monte Carlo method~\cite{Sandvik_2006,RVB_fabian_2010} on finite systems and extrapolating the lowest non-trivial eigenvalue and eigenvector to the thermodynamic limit. It would be interesting to apply this approach to the long-range RVB states that serve as ground states of gapped $\mathbb{Z}_2$ spin liquids~\cite{LR_RVB_Z2_2018}. Third, it would be desirable if a more accurate and stable method could be developed to calculate the static structure factor for critical iPEPS, so as to enable a scaling analysis similar to the MPS~\cite{Yang_2023} and pave the way for applying the method to extract emergent continuous symmetries in 2D, e.g. at the deconfined quantum critical points~\cite{Norther_DQCP_2019}.

In many physically relevant cases, the ground state actually contains enough information--encoded in its static correlations~\cite{Yang_2023}, entanglement Hamiltonian~\cite{PhysRevB.96.054425,cocchiarella2025excitedstateslocaleffective,hu2020emergentuniversalitycriticalquantum}, and transfer operator~\cite{Zauner_2015}--to reconstruct the Hamiltonian and other conserved operators, which correctly describe the physics up to certain energy cutoff or temperature. Since our results demonstrate that learning Hamiltonians from various physically relevant tensor network states can be done in polynomial time, it would be useful to tighten the learning complexity bound~\cite{10.1145/3618260.3649619} at zero temperature under these restrictions. It would also be interesting to give a precise analysis of how the errors scale with the cutoff distance in the correlation and explore if there is an error threshold such that above which one cannot reconstruct the Hamiltonian anymore. For PEPS whose norm can be mapped to 2D classical partition function~\cite{Verstraete_2006}, it would be intriguing to check the connection between the classical Hamiltonian learning~\cite{learncommuting,Haah2024} and the quantum state tomography~\cite{Anshu2024} for this class of PEPS. Generalizing our method to extract Hamiltonians from a PEPO that represents a Gibbs state at finite temperature~\cite{Hastings_2006,PhysRevLett.100.070502,Andras_2015,PhysRevX.11.011047,x2gx-wsp1} will be a direction worth exploring in the future.

\textbf{Acknowledgements.}
We thank Juraj Hasik, Bram Vanhecke, Anna Francuz, Sanjay Moudgalya, and Nathanan Tantivasadakarn for helpful comments. The CTMRG calculations in this work is performed using PEPS-torch~\cite{peps-torch}.  W.-T.X. acknowledges support from the Munich Quantum Valley, which is supported by the Bavarian
state government with funds from the Hightech Agenda
Bayern Plus. M.F.P. acknowledges support from the Government of Spain (Quantum in Spain, Severo Ochoa CEX2019-000910-S and FUNQIP), Fundació Cellex, Fundació Mir-Puig and Generalitat de Catalunya (CERCA program). M.Y. is supported by the Distinguished Postdoc Fellowship of the Munich Center for Quantum Science and Technology (MCQST) funded by DFG under the Excellence Strategy EXC2111-390814868.

\textbf{Data availability.}
The code and data are available on Zenodo~\cite{Zenodo}.

\bibliographystyle{apsrev4-1}
\bibliography{ref.bib}

\begin{thebibliography}{109}%
\makeatletter
\providecommand \@ifxundefined [1]{%
 \@ifx{#1\undefined}
}%
\providecommand \@ifnum [1]{%
 \ifnum #1\expandafter \@firstoftwo
 \else \expandafter \@secondoftwo
 \fi
}%
\providecommand \@ifx [1]{%
 \ifx #1\expandafter \@firstoftwo
 \else \expandafter \@secondoftwo
 \fi
}%
\providecommand \natexlab [1]{#1}%
\providecommand \enquote  [1]{``#1''}%
\providecommand \bibnamefont  [1]{#1}%
\providecommand \bibfnamefont [1]{#1}%
\providecommand \citenamefont [1]{#1}%
\providecommand \href@noop [0]{\@secondoftwo}%
\providecommand \href [0]{\begingroup \@sanitize@url \@href}%
\providecommand \@href[1]{\@@startlink{#1}\@@href}%
\providecommand \@@href[1]{\endgroup#1\@@endlink}%
\providecommand \@sanitize@url [0]{\catcode `\\12\catcode `\$12\catcode `\&12\catcode `\#12\catcode `\^12\catcode `\_12\catcode `\%12\relax}%
\providecommand \@@startlink[1]{}%
\providecommand \@@endlink[0]{}%
\providecommand \url  [0]{\begingroup\@sanitize@url \@url }%
\providecommand \@url [1]{\endgroup\@href {#1}{\urlprefix }}%
\providecommand \urlprefix  [0]{URL }%
\providecommand \Eprint [0]{\href }%
\providecommand \doibase [0]{http://dx.doi.org/}%
\providecommand \selectlanguage [0]{\@gobble}%
\providecommand \bibinfo  [0]{\@secondoftwo}%
\providecommand \bibfield  [0]{\@secondoftwo}%
\providecommand \translation [1]{[#1]}%
\providecommand \BibitemOpen [0]{}%
\providecommand \bibitemStop [0]{}%
\providecommand \bibitemNoStop [0]{.\EOS\space}%
\providecommand \EOS [0]{\spacefactor3000\relax}%
\providecommand \BibitemShut  [1]{\csname bibitem#1\endcsname}%
\let\auto@bib@innerbib\@empty
\bibitem [{\citenamefont {Anshu}\ \emph {et~al.}(2021{\natexlab{a}})\citenamefont {Anshu}, \citenamefont {Arunachalam}, \citenamefont {Kuwahara},\ and\ \citenamefont {Soleimanifar}}]{Anshu2021}%
  \BibitemOpen
  \bibfield  {author} {\bibinfo {author} {\bibfnamefont {A.}~\bibnamefont {Anshu}}, \bibinfo {author} {\bibfnamefont {S.}~\bibnamefont {Arunachalam}}, \bibinfo {author} {\bibfnamefont {T.}~\bibnamefont {Kuwahara}}, \ and\ \bibinfo {author} {\bibfnamefont {M.}~\bibnamefont {Soleimanifar}},\ }\href {\doibase 10.1038/s41567-021-01232-0} {\bibfield  {journal} {\bibinfo  {journal} {Nature Physics}\ }\textbf {\bibinfo {volume} {17}},\ \bibinfo {pages} {931} (\bibinfo {year} {2021}{\natexlab{a}})}\BibitemShut {NoStop}%
\bibitem [{\citenamefont {Anshu}\ \emph {et~al.}(2021{\natexlab{b}})\citenamefont {Anshu}, \citenamefont {Arunachalam}, \citenamefont {Kuwahara},\ and\ \citenamefont {Soleimanifar}}]{learncommuting}%
  \BibitemOpen
  \bibfield  {author} {\bibinfo {author} {\bibfnamefont {A.}~\bibnamefont {Anshu}}, \bibinfo {author} {\bibfnamefont {S.}~\bibnamefont {Arunachalam}}, \bibinfo {author} {\bibfnamefont {T.}~\bibnamefont {Kuwahara}}, \ and\ \bibinfo {author} {\bibfnamefont {M.}~\bibnamefont {Soleimanifar}},\ }\href@noop {} {\enquote {\bibinfo {title} {Efficient learning of commuting hamiltonians on lattices},}\ }\bibinfo {howpublished} {\url{https://anuraganshu.seas.harvard.edu/links}} (\bibinfo {year} {2021}{\natexlab{b}}),\ \bibinfo {note} {accessed: 2025-11-18}\BibitemShut {NoStop}%
\bibitem [{\citenamefont {Haah}\ \emph {et~al.}(2024)\citenamefont {Haah}, \citenamefont {Kothari},\ and\ \citenamefont {Tang}}]{Haah2024}%
  \BibitemOpen
  \bibfield  {author} {\bibinfo {author} {\bibfnamefont {J.}~\bibnamefont {Haah}}, \bibinfo {author} {\bibfnamefont {R.}~\bibnamefont {Kothari}}, \ and\ \bibinfo {author} {\bibfnamefont {E.}~\bibnamefont {Tang}},\ }\href {\doibase 10.1038/s41567-023-02376-x} {\bibfield  {journal} {\bibinfo  {journal} {Nature Physics}\ }\textbf {\bibinfo {volume} {20}},\ \bibinfo {pages} {1027} (\bibinfo {year} {2024})}\BibitemShut {NoStop}%
\bibitem [{\citenamefont {Bakshi}\ \emph {et~al.}(2024)\citenamefont {Bakshi}, \citenamefont {Liu}, \citenamefont {Moitra},\ and\ \citenamefont {Tang}}]{10.1145/3618260.3649619}%
  \BibitemOpen
  \bibfield  {author} {\bibinfo {author} {\bibfnamefont {A.}~\bibnamefont {Bakshi}}, \bibinfo {author} {\bibfnamefont {A.}~\bibnamefont {Liu}}, \bibinfo {author} {\bibfnamefont {A.}~\bibnamefont {Moitra}}, \ and\ \bibinfo {author} {\bibfnamefont {E.}~\bibnamefont {Tang}},\ }in\ \href {\doibase 10.1145/3618260.3649619} {\emph {\bibinfo {booktitle} {Proceedings of the 56th Annual ACM Symposium on Theory of Computing}}},\ \bibinfo {series and number} {STOC 2024}\ (\bibinfo  {publisher} {Association for Computing Machinery},\ \bibinfo {address} {New York, NY, USA},\ \bibinfo {year} {2024})\ p.\ \bibinfo {pages} {1470–1477}\BibitemShut {NoStop}%
\bibitem [{\citenamefont {Chertkov}\ and\ \citenamefont {Clark}(2018)}]{PhysRevX.8.031029}%
  \BibitemOpen
  \bibfield  {author} {\bibinfo {author} {\bibfnamefont {E.}~\bibnamefont {Chertkov}}\ and\ \bibinfo {author} {\bibfnamefont {B.~K.}\ \bibnamefont {Clark}},\ }\href {\doibase 10.1103/PhysRevX.8.031029} {\bibfield  {journal} {\bibinfo  {journal} {Phys. Rev. X}\ }\textbf {\bibinfo {volume} {8}},\ \bibinfo {pages} {031029} (\bibinfo {year} {2018})}\BibitemShut {NoStop}%
\bibitem [{\citenamefont {Qi}\ and\ \citenamefont {Ranard}(2019)}]{Qi2019determininglocal}%
  \BibitemOpen
  \bibfield  {author} {\bibinfo {author} {\bibfnamefont {X.-L.}\ \bibnamefont {Qi}}\ and\ \bibinfo {author} {\bibfnamefont {D.}~\bibnamefont {Ranard}},\ }\href {\doibase 10.22331/q-2019-07-08-159} {\bibfield  {journal} {\bibinfo  {journal} {{Quantum}}\ }\textbf {\bibinfo {volume} {3}},\ \bibinfo {pages} {159} (\bibinfo {year} {2019})}\BibitemShut {NoStop}%
\bibitem [{\citenamefont {Bairey}\ \emph {et~al.}(2019)\citenamefont {Bairey}, \citenamefont {Arad},\ and\ \citenamefont {Lindner}}]{PhysRevLett.122.020504}%
  \BibitemOpen
  \bibfield  {author} {\bibinfo {author} {\bibfnamefont {E.}~\bibnamefont {Bairey}}, \bibinfo {author} {\bibfnamefont {I.}~\bibnamefont {Arad}}, \ and\ \bibinfo {author} {\bibfnamefont {N.~H.}\ \bibnamefont {Lindner}},\ }\href {\doibase 10.1103/PhysRevLett.122.020504} {\bibfield  {journal} {\bibinfo  {journal} {Phys. Rev. Lett.}\ }\textbf {\bibinfo {volume} {122}},\ \bibinfo {pages} {020504} (\bibinfo {year} {2019})}\BibitemShut {NoStop}%
\bibitem [{\citenamefont {Huang}\ \emph {et~al.}(2023)\citenamefont {Huang}, \citenamefont {Tong}, \citenamefont {Fang},\ and\ \citenamefont {Su}}]{PhysRevLett.130.200403}%
  \BibitemOpen
  \bibfield  {author} {\bibinfo {author} {\bibfnamefont {H.-Y.}\ \bibnamefont {Huang}}, \bibinfo {author} {\bibfnamefont {Y.}~\bibnamefont {Tong}}, \bibinfo {author} {\bibfnamefont {D.}~\bibnamefont {Fang}}, \ and\ \bibinfo {author} {\bibfnamefont {Y.}~\bibnamefont {Su}},\ }\href {\doibase 10.1103/PhysRevLett.130.200403} {\bibfield  {journal} {\bibinfo  {journal} {Phys. Rev. Lett.}\ }\textbf {\bibinfo {volume} {130}},\ \bibinfo {pages} {200403} (\bibinfo {year} {2023})}\BibitemShut {NoStop}%
\bibitem [{\citenamefont {Zhan}\ \emph {et~al.}(2024)\citenamefont {Zhan}, \citenamefont {Elben}, \citenamefont {Huang},\ and\ \citenamefont {Tong}}]{PRXQuantum.5.010350}%
  \BibitemOpen
  \bibfield  {author} {\bibinfo {author} {\bibfnamefont {Y.}~\bibnamefont {Zhan}}, \bibinfo {author} {\bibfnamefont {A.}~\bibnamefont {Elben}}, \bibinfo {author} {\bibfnamefont {H.-Y.}\ \bibnamefont {Huang}}, \ and\ \bibinfo {author} {\bibfnamefont {Y.}~\bibnamefont {Tong}},\ }\href {\doibase 10.1103/PRXQuantum.5.010350} {\bibfield  {journal} {\bibinfo  {journal} {PRX Quantum}\ }\textbf {\bibinfo {volume} {5}},\ \bibinfo {pages} {010350} (\bibinfo {year} {2024})}\BibitemShut {NoStop}%
\bibitem [{\citenamefont {Anshu}\ and\ \citenamefont {Arunachalam}(2024)}]{Anshu2024}%
  \BibitemOpen
  \bibfield  {author} {\bibinfo {author} {\bibfnamefont {A.}~\bibnamefont {Anshu}}\ and\ \bibinfo {author} {\bibfnamefont {S.}~\bibnamefont {Arunachalam}},\ }\href {\doibase 10.1038/s42254-023-00662-4} {\bibfield  {journal} {\bibinfo  {journal} {Nature Reviews Physics}\ }\textbf {\bibinfo {volume} {6}},\ \bibinfo {pages} {59} (\bibinfo {year} {2024})}\BibitemShut {NoStop}%
\bibitem [{\citenamefont {Feynman}(1982)}]{Feynman1982}%
  \BibitemOpen
  \bibfield  {author} {\bibinfo {author} {\bibfnamefont {R.~P.}\ \bibnamefont {Feynman}},\ }\href {\doibase 10.1007/BF02650179} {\bibfield  {journal} {\bibinfo  {journal} {International Journal of Theoretical Physics}\ }\textbf {\bibinfo {volume} {21}},\ \bibinfo {pages} {467} (\bibinfo {year} {1982})}\BibitemShut {NoStop}%
\bibitem [{\citenamefont {Georgescu}\ \emph {et~al.}(2014)\citenamefont {Georgescu}, \citenamefont {Ashhab},\ and\ \citenamefont {Nori}}]{RevModPhys.86.153}%
  \BibitemOpen
  \bibfield  {author} {\bibinfo {author} {\bibfnamefont {I.~M.}\ \bibnamefont {Georgescu}}, \bibinfo {author} {\bibfnamefont {S.}~\bibnamefont {Ashhab}}, \ and\ \bibinfo {author} {\bibfnamefont {F.}~\bibnamefont {Nori}},\ }\href {\doibase 10.1103/RevModPhys.86.153} {\bibfield  {journal} {\bibinfo  {journal} {Rev. Mod. Phys.}\ }\textbf {\bibinfo {volume} {86}},\ \bibinfo {pages} {153} (\bibinfo {year} {2014})}\BibitemShut {NoStop}%
\bibitem [{\citenamefont {Preskill}(2018)}]{Preskill_NISQ_2018}%
  \BibitemOpen
  \bibfield  {author} {\bibinfo {author} {\bibfnamefont {J.}~\bibnamefont {Preskill}},\ }\href {\doibase 10.22331/q-2018-08-06-79} {\bibfield  {journal} {\bibinfo  {journal} {{Quantum}}\ }\textbf {\bibinfo {volume} {2}},\ \bibinfo {pages} {79} (\bibinfo {year} {2018})}\BibitemShut {NoStop}%
\bibitem [{\citenamefont {Wiebe}\ \emph {et~al.}(2014)\citenamefont {Wiebe}, \citenamefont {Granade}, \citenamefont {Ferrie},\ and\ \citenamefont {Cory}}]{PhysRevLett.112.190501}%
  \BibitemOpen
  \bibfield  {author} {\bibinfo {author} {\bibfnamefont {N.}~\bibnamefont {Wiebe}}, \bibinfo {author} {\bibfnamefont {C.}~\bibnamefont {Granade}}, \bibinfo {author} {\bibfnamefont {C.}~\bibnamefont {Ferrie}}, \ and\ \bibinfo {author} {\bibfnamefont {D.~G.}\ \bibnamefont {Cory}},\ }\href {\doibase 10.1103/PhysRevLett.112.190501} {\bibfield  {journal} {\bibinfo  {journal} {Phys. Rev. Lett.}\ }\textbf {\bibinfo {volume} {112}},\ \bibinfo {pages} {190501} (\bibinfo {year} {2014})}\BibitemShut {NoStop}%
\bibitem [{\citenamefont {Wang}\ \emph {et~al.}(2017)\citenamefont {Wang}, \citenamefont {Paesani}, \citenamefont {Santagati}, \citenamefont {Knauer}, \citenamefont {Gentile}, \citenamefont {Wiebe}, \citenamefont {Petruzzella}, \citenamefont {O'Brien}, \citenamefont {Rarity}, \citenamefont {Laing},\ and\ \citenamefont {Thompson}}]{Wang2017}%
  \BibitemOpen
  \bibfield  {author} {\bibinfo {author} {\bibfnamefont {J.}~\bibnamefont {Wang}}, \bibinfo {author} {\bibfnamefont {S.}~\bibnamefont {Paesani}}, \bibinfo {author} {\bibfnamefont {R.}~\bibnamefont {Santagati}}, \bibinfo {author} {\bibfnamefont {S.}~\bibnamefont {Knauer}}, \bibinfo {author} {\bibfnamefont {A.~A.}\ \bibnamefont {Gentile}}, \bibinfo {author} {\bibfnamefont {N.}~\bibnamefont {Wiebe}}, \bibinfo {author} {\bibfnamefont {M.}~\bibnamefont {Petruzzella}}, \bibinfo {author} {\bibfnamefont {J.~L.}\ \bibnamefont {O'Brien}}, \bibinfo {author} {\bibfnamefont {J.~G.}\ \bibnamefont {Rarity}}, \bibinfo {author} {\bibfnamefont {A.}~\bibnamefont {Laing}}, \ and\ \bibinfo {author} {\bibfnamefont {M.~G.}\ \bibnamefont {Thompson}},\ }\href {\doibase 10.1038/nphys4074} {\bibfield  {journal} {\bibinfo  {journal} {Nature Physics}\ }\textbf {\bibinfo {volume} {13}},\ \bibinfo {pages} {551} (\bibinfo {year} {2017})}\BibitemShut {NoStop}%
\bibitem [{\citenamefont {Schrieffer}\ and\ \citenamefont {Wolff}(1966)}]{SW_1966}%
  \BibitemOpen
  \bibfield  {author} {\bibinfo {author} {\bibfnamefont {J.~R.}\ \bibnamefont {Schrieffer}}\ and\ \bibinfo {author} {\bibfnamefont {P.~A.}\ \bibnamefont {Wolff}},\ }\href {\doibase 10.1103/PhysRev.149.491} {\bibfield  {journal} {\bibinfo  {journal} {Phys. Rev.}\ }\textbf {\bibinfo {volume} {149}},\ \bibinfo {pages} {491} (\bibinfo {year} {1966})}\BibitemShut {NoStop}%
\bibitem [{\citenamefont {Bravyi}\ \emph {et~al.}(2011)\citenamefont {Bravyi}, \citenamefont {DiVincenzo},\ and\ \citenamefont {Loss}}]{S_W_transformation_2011}%
  \BibitemOpen
  \bibfield  {author} {\bibinfo {author} {\bibfnamefont {S.}~\bibnamefont {Bravyi}}, \bibinfo {author} {\bibfnamefont {D.~P.}\ \bibnamefont {DiVincenzo}}, \ and\ \bibinfo {author} {\bibfnamefont {D.}~\bibnamefont {Loss}},\ }\href {\doibase https://doi.org/10.1016/j.aop.2011.06.004} {\bibfield  {journal} {\bibinfo  {journal} {Annals of Physics}\ }\textbf {\bibinfo {volume} {326}},\ \bibinfo {pages} {2793} (\bibinfo {year} {2011})}\BibitemShut {NoStop}%
\bibitem [{\citenamefont {Jiang}\ \emph {et~al.}(2023)\citenamefont {Jiang}, \citenamefont {Scalapino},\ and\ \citenamefont {White}}]{Jiang_white_2023}%
  \BibitemOpen
  \bibfield  {author} {\bibinfo {author} {\bibfnamefont {S.}~\bibnamefont {Jiang}}, \bibinfo {author} {\bibfnamefont {D.~J.}\ \bibnamefont {Scalapino}}, \ and\ \bibinfo {author} {\bibfnamefont {S.~R.}\ \bibnamefont {White}},\ }\href {\doibase 10.1103/PhysRevB.108.L161111} {\bibfield  {journal} {\bibinfo  {journal} {Phys. Rev. B}\ }\textbf {\bibinfo {volume} {108}},\ \bibinfo {pages} {L161111} (\bibinfo {year} {2023})}\BibitemShut {NoStop}%
\bibitem [{\citenamefont {Yang}\ \emph {et~al.}(2023)\citenamefont {Yang}, \citenamefont {Vanhecke},\ and\ \citenamefont {Schuch}}]{Yang_2023}%
  \BibitemOpen
  \bibfield  {author} {\bibinfo {author} {\bibfnamefont {M.}~\bibnamefont {Yang}}, \bibinfo {author} {\bibfnamefont {B.}~\bibnamefont {Vanhecke}}, \ and\ \bibinfo {author} {\bibfnamefont {N.}~\bibnamefont {Schuch}},\ }\href {\doibase 10.1103/PhysRevLett.131.036505} {\bibfield  {journal} {\bibinfo  {journal} {Phys. Rev. Lett.}\ }\textbf {\bibinfo {volume} {131}},\ \bibinfo {pages} {036505} (\bibinfo {year} {2023})}\BibitemShut {NoStop}%
\bibitem [{\citenamefont {Laughlin}(1983)}]{Laughlin_1983}%
  \BibitemOpen
  \bibfield  {author} {\bibinfo {author} {\bibfnamefont {R.~B.}\ \bibnamefont {Laughlin}},\ }\href {\doibase 10.1103/PhysRevLett.50.1395} {\bibfield  {journal} {\bibinfo  {journal} {Phys. Rev. Lett.}\ }\textbf {\bibinfo {volume} {50}},\ \bibinfo {pages} {1395} (\bibinfo {year} {1983})}\BibitemShut {NoStop}%
\bibitem [{\citenamefont {Anderson}(1987)}]{Anderson_RVB_1987}%
  \BibitemOpen
  \bibfield  {author} {\bibinfo {author} {\bibfnamefont {P.~W.}\ \bibnamefont {Anderson}},\ }\href {\doibase 10.1126/science.235.4793.1196} {\bibfield  {journal} {\bibinfo  {journal} {Science}\ }\textbf {\bibinfo {volume} {235}},\ \bibinfo {pages} {1196} (\bibinfo {year} {1987})}\BibitemShut {NoStop}%
\bibitem [{\citenamefont {Turner}\ \emph {et~al.}(2018)\citenamefont {Turner}, \citenamefont {Michailidis}, \citenamefont {Abanin}, \citenamefont {Serbyn},\ and\ \citenamefont {Papi{\'c}}}]{scar_2018}%
  \BibitemOpen
  \bibfield  {author} {\bibinfo {author} {\bibfnamefont {C.~J.}\ \bibnamefont {Turner}}, \bibinfo {author} {\bibfnamefont {A.~A.}\ \bibnamefont {Michailidis}}, \bibinfo {author} {\bibfnamefont {D.~A.}\ \bibnamefont {Abanin}}, \bibinfo {author} {\bibfnamefont {M.}~\bibnamefont {Serbyn}}, \ and\ \bibinfo {author} {\bibfnamefont {Z.}~\bibnamefont {Papi{\'c}}},\ }\href {\doibase 10.1038/s41567-018-0137-5} {\bibfield  {journal} {\bibinfo  {journal} {Nature Physics}\ }\textbf {\bibinfo {volume} {14}},\ \bibinfo {pages} {745} (\bibinfo {year} {2018})}\BibitemShut {NoStop}%
\bibitem [{\citenamefont {Moudgalya}\ \emph {et~al.}(2020)\citenamefont {Moudgalya}, \citenamefont {O'Brien}, \citenamefont {Bernevig}, \citenamefont {Fendley},\ and\ \citenamefont {Regnault}}]{Sanjay_2020}%
  \BibitemOpen
  \bibfield  {author} {\bibinfo {author} {\bibfnamefont {S.}~\bibnamefont {Moudgalya}}, \bibinfo {author} {\bibfnamefont {E.}~\bibnamefont {O'Brien}}, \bibinfo {author} {\bibfnamefont {B.~A.}\ \bibnamefont {Bernevig}}, \bibinfo {author} {\bibfnamefont {P.}~\bibnamefont {Fendley}}, \ and\ \bibinfo {author} {\bibfnamefont {N.}~\bibnamefont {Regnault}},\ }\href {\doibase 10.1103/PhysRevB.102.085120} {\bibfield  {journal} {\bibinfo  {journal} {Phys. Rev. B}\ }\textbf {\bibinfo {volume} {102}},\ \bibinfo {pages} {085120} (\bibinfo {year} {2020})}\BibitemShut {NoStop}%
\bibitem [{\citenamefont {Moudgalya}\ and\ \citenamefont {Motrunich}(2023)}]{PhysRevB.107.224312}%
  \BibitemOpen
  \bibfield  {author} {\bibinfo {author} {\bibfnamefont {S.}~\bibnamefont {Moudgalya}}\ and\ \bibinfo {author} {\bibfnamefont {O.~I.}\ \bibnamefont {Motrunich}},\ }\href {\doibase 10.1103/PhysRevB.107.224312} {\bibfield  {journal} {\bibinfo  {journal} {Phys. Rev. B}\ }\textbf {\bibinfo {volume} {107}},\ \bibinfo {pages} {224312} (\bibinfo {year} {2023})}\BibitemShut {NoStop}%
\bibitem [{\citenamefont {Kim}\ \emph {et~al.}(2015)\citenamefont {Kim}, \citenamefont {Ba\~nuls}, \citenamefont {Cirac}, \citenamefont {Hastings},\ and\ \citenamefont {Huse}}]{PhysRevE.92.012128}%
  \BibitemOpen
  \bibfield  {author} {\bibinfo {author} {\bibfnamefont {H.}~\bibnamefont {Kim}}, \bibinfo {author} {\bibfnamefont {M.~C.}\ \bibnamefont {Ba\~nuls}}, \bibinfo {author} {\bibfnamefont {J.~I.}\ \bibnamefont {Cirac}}, \bibinfo {author} {\bibfnamefont {M.~B.}\ \bibnamefont {Hastings}}, \ and\ \bibinfo {author} {\bibfnamefont {D.~A.}\ \bibnamefont {Huse}},\ }\href {\doibase 10.1103/PhysRevE.92.012128} {\bibfield  {journal} {\bibinfo  {journal} {Phys. Rev. E}\ }\textbf {\bibinfo {volume} {92}},\ \bibinfo {pages} {012128} (\bibinfo {year} {2015})}\BibitemShut {NoStop}%
\bibitem [{\citenamefont {O'Brien}\ \emph {et~al.}(2016)\citenamefont {O'Brien}, \citenamefont {Abanin}, \citenamefont {Vidal},\ and\ \citenamefont {Papi\ifmmode~\acute{c}\else \'{c}\fi{}}}]{PhysRevB.94.144208}%
  \BibitemOpen
  \bibfield  {author} {\bibinfo {author} {\bibfnamefont {T.~E.}\ \bibnamefont {O'Brien}}, \bibinfo {author} {\bibfnamefont {D.~A.}\ \bibnamefont {Abanin}}, \bibinfo {author} {\bibfnamefont {G.}~\bibnamefont {Vidal}}, \ and\ \bibinfo {author} {\bibfnamefont {Z.}~\bibnamefont {Papi\ifmmode~\acute{c}\else \'{c}\fi{}}},\ }\href {\doibase 10.1103/PhysRevB.94.144208} {\bibfield  {journal} {\bibinfo  {journal} {Phys. Rev. B}\ }\textbf {\bibinfo {volume} {94}},\ \bibinfo {pages} {144208} (\bibinfo {year} {2016})}\BibitemShut {NoStop}%
\bibitem [{\citenamefont {Chertkov}\ \emph {et~al.}(2020)\citenamefont {Chertkov}, \citenamefont {Villalonga},\ and\ \citenamefont {Clark}}]{PhysRevResearch.2.023348}%
  \BibitemOpen
  \bibfield  {author} {\bibinfo {author} {\bibfnamefont {E.}~\bibnamefont {Chertkov}}, \bibinfo {author} {\bibfnamefont {B.}~\bibnamefont {Villalonga}}, \ and\ \bibinfo {author} {\bibfnamefont {B.~K.}\ \bibnamefont {Clark}},\ }\href {\doibase 10.1103/PhysRevResearch.2.023348} {\bibfield  {journal} {\bibinfo  {journal} {Phys. Rev. Res.}\ }\textbf {\bibinfo {volume} {2}},\ \bibinfo {pages} {023348} (\bibinfo {year} {2020})}\BibitemShut {NoStop}%
\bibitem [{\citenamefont {Gioia}\ \emph {et~al.}(2025)\citenamefont {Gioia}, \citenamefont {Moudgalya},\ and\ \citenamefont {Motrunich}}]{Sanjay_2025}%
  \BibitemOpen
  \bibfield  {author} {\bibinfo {author} {\bibfnamefont {L.}~\bibnamefont {Gioia}}, \bibinfo {author} {\bibfnamefont {S.}~\bibnamefont {Moudgalya}}, \ and\ \bibinfo {author} {\bibfnamefont {O.~I.}\ \bibnamefont {Motrunich}},\ }\href {https://arxiv.org/abs/2510.24713} {\enquote {\bibinfo {title} {Distinct types of parent hamiltonians for quantum states: Insights from the $w$ state as a quantum many-body scar},}\ } (\bibinfo {year} {2025}),\ \Eprint {http://arxiv.org/abs/2510.24713} {arXiv:2510.24713 [quant-ph]} \BibitemShut {NoStop}%
\bibitem [{\citenamefont {Fannes}\ \emph {et~al.}(1992)\citenamefont {Fannes}, \citenamefont {Nachtergaele},\ and\ \citenamefont {Werner}}]{Fannes1992}%
  \BibitemOpen
  \bibfield  {author} {\bibinfo {author} {\bibfnamefont {M.}~\bibnamefont {Fannes}}, \bibinfo {author} {\bibfnamefont {B.}~\bibnamefont {Nachtergaele}}, \ and\ \bibinfo {author} {\bibfnamefont {R.~F.}\ \bibnamefont {Werner}},\ }\href {\doibase 10.1007/BF02099178} {\bibfield  {journal} {\bibinfo  {journal} {Communications in Mathematical Physics}\ }\textbf {\bibinfo {volume} {144}},\ \bibinfo {pages} {443} (\bibinfo {year} {1992})}\BibitemShut {NoStop}%
\bibitem [{\citenamefont {White}(1992)}]{PhysRevLett.69.2863}%
  \BibitemOpen
  \bibfield  {author} {\bibinfo {author} {\bibfnamefont {S.~R.}\ \bibnamefont {White}},\ }\href {\doibase 10.1103/PhysRevLett.69.2863} {\bibfield  {journal} {\bibinfo  {journal} {Phys. Rev. Lett.}\ }\textbf {\bibinfo {volume} {69}},\ \bibinfo {pages} {2863} (\bibinfo {year} {1992})}\BibitemShut {NoStop}%
\bibitem [{\citenamefont {White}(1993)}]{PhysRevB.48.10345}%
  \BibitemOpen
  \bibfield  {author} {\bibinfo {author} {\bibfnamefont {S.~R.}\ \bibnamefont {White}},\ }\href {\doibase 10.1103/PhysRevB.48.10345} {\bibfield  {journal} {\bibinfo  {journal} {Phys. Rev. B}\ }\textbf {\bibinfo {volume} {48}},\ \bibinfo {pages} {10345} (\bibinfo {year} {1993})}\BibitemShut {NoStop}%
\bibitem [{\citenamefont {Maeshima}\ \emph {et~al.}(2001)\citenamefont {Maeshima}, \citenamefont {Hieida}, \citenamefont {Akutsu}, \citenamefont {Nishino},\ and\ \citenamefont {Okunishi}}]{maeshima:2001}%
  \BibitemOpen
  \bibfield  {author} {\bibinfo {author} {\bibfnamefont {N.}~\bibnamefont {Maeshima}}, \bibinfo {author} {\bibfnamefont {Y.}~\bibnamefont {Hieida}}, \bibinfo {author} {\bibfnamefont {Y.}~\bibnamefont {Akutsu}}, \bibinfo {author} {\bibfnamefont {T.}~\bibnamefont {Nishino}}, \ and\ \bibinfo {author} {\bibfnamefont {K.}~\bibnamefont {Okunishi}},\ }\href {\doibase 10.1103/PhysRevE.64.016705} {\bibfield  {journal} {\bibinfo  {journal} {Phys. Rev. E}\ }\textbf {\bibinfo {volume} {64}},\ \bibinfo {pages} {016705} (\bibinfo {year} {2001})}\BibitemShut {NoStop}%
\bibitem [{\citenamefont {Verstraete}\ and\ \citenamefont {Cirac}(2004)}]{verstraete2004}%
  \BibitemOpen
  \bibfield  {author} {\bibinfo {author} {\bibfnamefont {F.}~\bibnamefont {Verstraete}}\ and\ \bibinfo {author} {\bibfnamefont {J.~I.}\ \bibnamefont {Cirac}},\ }\href {https://arxiv.org/abs/cond-mat/0407066} {\enquote {\bibinfo {title} {Renormalization algorithms for quantum-many body systems in two and higher dimensions},}\ } (\bibinfo {year} {2004}),\ \Eprint {http://arxiv.org/abs/cond-mat/0407066} {arXiv:cond-mat/0407066 [cond-mat.str-el]} \BibitemShut {NoStop}%
\bibitem [{\citenamefont {Hastings}(2007)}]{Hastings_2007}%
  \BibitemOpen
  \bibfield  {author} {\bibinfo {author} {\bibfnamefont {M.~B.}\ \bibnamefont {Hastings}},\ }\href {\doibase 10.1088/1742-5468/2007/08/P08024} {\bibfield  {journal} {\bibinfo  {journal} {Journal of Statistical Mechanics: Theory and Experiment}\ }\textbf {\bibinfo {volume} {2007}},\ \bibinfo {pages} {P08024} (\bibinfo {year} {2007})}\BibitemShut {NoStop}%
\bibitem [{\citenamefont {Arad}\ \emph {et~al.}(2013)\citenamefont {Arad}, \citenamefont {Kitaev}, \citenamefont {Landau},\ and\ \citenamefont {Vazirani}}]{arad2013arealawsubexponentialalgorithm}%
  \BibitemOpen
  \bibfield  {author} {\bibinfo {author} {\bibfnamefont {I.}~\bibnamefont {Arad}}, \bibinfo {author} {\bibfnamefont {A.}~\bibnamefont {Kitaev}}, \bibinfo {author} {\bibfnamefont {Z.}~\bibnamefont {Landau}}, \ and\ \bibinfo {author} {\bibfnamefont {U.}~\bibnamefont {Vazirani}},\ }\href {https://arxiv.org/abs/1301.1162} {\enquote {\bibinfo {title} {An area law and sub-exponential algorithm for 1d systems},}\ } (\bibinfo {year} {2013}),\ \Eprint {http://arxiv.org/abs/1301.1162} {arXiv:1301.1162 [quant-ph]} \BibitemShut {NoStop}%
\bibitem [{\citenamefont {Verstraete}\ and\ \citenamefont {Cirac}(2006)}]{PhysRevB.73.094423}%
  \BibitemOpen
  \bibfield  {author} {\bibinfo {author} {\bibfnamefont {F.}~\bibnamefont {Verstraete}}\ and\ \bibinfo {author} {\bibfnamefont {J.~I.}\ \bibnamefont {Cirac}},\ }\href {\doibase 10.1103/PhysRevB.73.094423} {\bibfield  {journal} {\bibinfo  {journal} {Phys. Rev. B}\ }\textbf {\bibinfo {volume} {73}},\ \bibinfo {pages} {094423} (\bibinfo {year} {2006})}\BibitemShut {NoStop}%
\bibitem [{\citenamefont {Molnar}\ \emph {et~al.}(2015)\citenamefont {Molnar}, \citenamefont {Schuch}, \citenamefont {Verstraete},\ and\ \citenamefont {Cirac}}]{Andras_2015}%
  \BibitemOpen
  \bibfield  {author} {\bibinfo {author} {\bibfnamefont {A.}~\bibnamefont {Molnar}}, \bibinfo {author} {\bibfnamefont {N.}~\bibnamefont {Schuch}}, \bibinfo {author} {\bibfnamefont {F.}~\bibnamefont {Verstraete}}, \ and\ \bibinfo {author} {\bibfnamefont {J.~I.}\ \bibnamefont {Cirac}},\ }\href {\doibase 10.1103/PhysRevB.91.045138} {\bibfield  {journal} {\bibinfo  {journal} {Phys. Rev. B}\ }\textbf {\bibinfo {volume} {91}},\ \bibinfo {pages} {045138} (\bibinfo {year} {2015})}\BibitemShut {NoStop}%
\bibitem [{\citenamefont {Cirac}\ \emph {et~al.}(2021)\citenamefont {Cirac}, \citenamefont {P\'erez-Garc\'{\i}a}, \citenamefont {Schuch},\ and\ \citenamefont {Verstraete}}]{RMP_MPS_PEPS_2021}%
  \BibitemOpen
  \bibfield  {author} {\bibinfo {author} {\bibfnamefont {J.~I.}\ \bibnamefont {Cirac}}, \bibinfo {author} {\bibfnamefont {D.}~\bibnamefont {P\'erez-Garc\'{\i}a}}, \bibinfo {author} {\bibfnamefont {N.}~\bibnamefont {Schuch}}, \ and\ \bibinfo {author} {\bibfnamefont {F.}~\bibnamefont {Verstraete}},\ }\href {\doibase 10.1103/RevModPhys.93.045003} {\bibfield  {journal} {\bibinfo  {journal} {Rev. Mod. Phys.}\ }\textbf {\bibinfo {volume} {93}},\ \bibinfo {pages} {045003} (\bibinfo {year} {2021})}\BibitemShut {NoStop}%
\bibitem [{\citenamefont {Schuch}\ \emph {et~al.}(2007)\citenamefont {Schuch}, \citenamefont {Wolf}, \citenamefont {Verstraete},\ and\ \citenamefont {Cirac}}]{Schuch_2007}%
  \BibitemOpen
  \bibfield  {author} {\bibinfo {author} {\bibfnamefont {N.}~\bibnamefont {Schuch}}, \bibinfo {author} {\bibfnamefont {M.~M.}\ \bibnamefont {Wolf}}, \bibinfo {author} {\bibfnamefont {F.}~\bibnamefont {Verstraete}}, \ and\ \bibinfo {author} {\bibfnamefont {J.~I.}\ \bibnamefont {Cirac}},\ }\href {\doibase 10.1103/PhysRevLett.98.140506} {\bibfield  {journal} {\bibinfo  {journal} {Phys. Rev. Lett.}\ }\textbf {\bibinfo {volume} {98}},\ \bibinfo {pages} {140506} (\bibinfo {year} {2007})}\BibitemShut {NoStop}%
\bibitem [{\citenamefont {Or\'us}\ and\ \citenamefont {Vidal}(2008)}]{1D_TEBD_2008}%
  \BibitemOpen
  \bibfield  {author} {\bibinfo {author} {\bibfnamefont {R.}~\bibnamefont {Or\'us}}\ and\ \bibinfo {author} {\bibfnamefont {G.}~\bibnamefont {Vidal}},\ }\href {\doibase 10.1103/PhysRevB.78.155117} {\bibfield  {journal} {\bibinfo  {journal} {Phys. Rev. B}\ }\textbf {\bibinfo {volume} {78}},\ \bibinfo {pages} {155117} (\bibinfo {year} {2008})}\BibitemShut {NoStop}%
\bibitem [{\citenamefont {Nishino}\ and\ \citenamefont {Okunishi}(1996)}]{CTMRG_1}%
  \BibitemOpen
  \bibfield  {author} {\bibinfo {author} {\bibfnamefont {T.}~\bibnamefont {Nishino}}\ and\ \bibinfo {author} {\bibfnamefont {K.}~\bibnamefont {Okunishi}},\ }\href {\doibase 10.1143/JPSJ.65.891} {\bibfield  {journal} {\bibinfo  {journal} {Journal of the Physical Society of Japan}\ }\textbf {\bibinfo {volume} {65}},\ \bibinfo {pages} {891} (\bibinfo {year} {1996})}\BibitemShut {NoStop}%
\bibitem [{\citenamefont {Corboz}\ \emph {et~al.}(2011)\citenamefont {Corboz}, \citenamefont {White}, \citenamefont {Vidal},\ and\ \citenamefont {Troyer}}]{CTMRG_corboz}%
  \BibitemOpen
  \bibfield  {author} {\bibinfo {author} {\bibfnamefont {P.}~\bibnamefont {Corboz}}, \bibinfo {author} {\bibfnamefont {S.~R.}\ \bibnamefont {White}}, \bibinfo {author} {\bibfnamefont {G.}~\bibnamefont {Vidal}}, \ and\ \bibinfo {author} {\bibfnamefont {M.}~\bibnamefont {Troyer}},\ }\href {\doibase 10.1103/PhysRevB.84.041108} {\bibfield  {journal} {\bibinfo  {journal} {Phys. Rev. B}\ }\textbf {\bibinfo {volume} {84}},\ \bibinfo {pages} {041108} (\bibinfo {year} {2011})}\BibitemShut {NoStop}%
\bibitem [{\citenamefont {Levin}\ and\ \citenamefont {Nave}(2007)}]{TRG_2007}%
  \BibitemOpen
  \bibfield  {author} {\bibinfo {author} {\bibfnamefont {M.}~\bibnamefont {Levin}}\ and\ \bibinfo {author} {\bibfnamefont {C.~P.}\ \bibnamefont {Nave}},\ }\href {\doibase 10.1103/PhysRevLett.99.120601} {\bibfield  {journal} {\bibinfo  {journal} {Phys. Rev. Lett.}\ }\textbf {\bibinfo {volume} {99}},\ \bibinfo {pages} {120601} (\bibinfo {year} {2007})}\BibitemShut {NoStop}%
\bibitem [{\citenamefont {Vanderstraeten}\ \emph {et~al.}(2022)\citenamefont {Vanderstraeten}, \citenamefont {Burgelman}, \citenamefont {Ponsioen}, \citenamefont {Van~Damme}, \citenamefont {Vanhecke}, \citenamefont {Corboz}, \citenamefont {Haegeman},\ and\ \citenamefont {Verstraete}}]{window_MPS_2022}%
  \BibitemOpen
  \bibfield  {author} {\bibinfo {author} {\bibfnamefont {L.}~\bibnamefont {Vanderstraeten}}, \bibinfo {author} {\bibfnamefont {L.}~\bibnamefont {Burgelman}}, \bibinfo {author} {\bibfnamefont {B.}~\bibnamefont {Ponsioen}}, \bibinfo {author} {\bibfnamefont {M.}~\bibnamefont {Van~Damme}}, \bibinfo {author} {\bibfnamefont {B.}~\bibnamefont {Vanhecke}}, \bibinfo {author} {\bibfnamefont {P.}~\bibnamefont {Corboz}}, \bibinfo {author} {\bibfnamefont {J.}~\bibnamefont {Haegeman}}, \ and\ \bibinfo {author} {\bibfnamefont {F.}~\bibnamefont {Verstraete}},\ }\href {\doibase 10.1103/PhysRevB.105.195140} {\bibfield  {journal} {\bibinfo  {journal} {Phys. Rev. B}\ }\textbf {\bibinfo {volume} {105}},\ \bibinfo {pages} {195140} (\bibinfo {year} {2022})}\BibitemShut {NoStop}%
\bibitem [{\citenamefont {Devereaux}\ and\ \citenamefont {Hackl}(2007)}]{RevModPhys.79.175}%
  \BibitemOpen
  \bibfield  {author} {\bibinfo {author} {\bibfnamefont {T.~P.}\ \bibnamefont {Devereaux}}\ and\ \bibinfo {author} {\bibfnamefont {R.}~\bibnamefont {Hackl}},\ }\href {\doibase 10.1103/RevModPhys.79.175} {\bibfield  {journal} {\bibinfo  {journal} {Rev. Mod. Phys.}\ }\textbf {\bibinfo {volume} {79}},\ \bibinfo {pages} {175} (\bibinfo {year} {2007})}\BibitemShut {NoStop}%
\bibitem [{\citenamefont {Boschini}\ \emph {et~al.}(2024)\citenamefont {Boschini}, \citenamefont {Zonno},\ and\ \citenamefont {Damascelli}}]{RevModPhys.96.015003}%
  \BibitemOpen
  \bibfield  {author} {\bibinfo {author} {\bibfnamefont {F.}~\bibnamefont {Boschini}}, \bibinfo {author} {\bibfnamefont {M.}~\bibnamefont {Zonno}}, \ and\ \bibinfo {author} {\bibfnamefont {A.}~\bibnamefont {Damascelli}},\ }\href {\doibase 10.1103/RevModPhys.96.015003} {\bibfield  {journal} {\bibinfo  {journal} {Rev. Mod. Phys.}\ }\textbf {\bibinfo {volume} {96}},\ \bibinfo {pages} {015003} (\bibinfo {year} {2024})}\BibitemShut {NoStop}%
\bibitem [{\citenamefont {Perez-Garcia}\ \emph {et~al.}(2008)\citenamefont {Perez-Garcia}, \citenamefont {Verstraete}, \citenamefont {Wolf},\ and\ \citenamefont {Cirac}}]{10.5555/2016976.2016982}%
  \BibitemOpen
  \bibfield  {author} {\bibinfo {author} {\bibfnamefont {D.}~\bibnamefont {Perez-Garcia}}, \bibinfo {author} {\bibfnamefont {F.}~\bibnamefont {Verstraete}}, \bibinfo {author} {\bibfnamefont {M.~M.}\ \bibnamefont {Wolf}}, \ and\ \bibinfo {author} {\bibfnamefont {J.~I.}\ \bibnamefont {Cirac}},\ }\href {https://dl.acm.org/doi/abs/10.5555/2016976.2016982} {\bibfield  {journal} {\bibinfo  {journal} {Quantum Info. Comput.}\ }\textbf {\bibinfo {volume} {8}},\ \bibinfo {pages} {650–663} (\bibinfo {year} {2008})}\BibitemShut {NoStop}%
\bibitem [{\citenamefont {Schuch}\ \emph {et~al.}(2010)\citenamefont {Schuch}, \citenamefont {Cirac},\ and\ \citenamefont {Pérez-García}}]{SCHUCH_2010}%
  \BibitemOpen
  \bibfield  {author} {\bibinfo {author} {\bibfnamefont {N.}~\bibnamefont {Schuch}}, \bibinfo {author} {\bibfnamefont {I.}~\bibnamefont {Cirac}}, \ and\ \bibinfo {author} {\bibfnamefont {D.}~\bibnamefont {Pérez-García}},\ }\href {\doibase https://doi.org/10.1016/j.aop.2010.05.008} {\bibfield  {journal} {\bibinfo  {journal} {Annals of Physics}\ }\textbf {\bibinfo {volume} {325}},\ \bibinfo {pages} {2153} (\bibinfo {year} {2010})}\BibitemShut {NoStop}%
\bibitem [{app()}]{appendix}%
  \BibitemOpen
  \href@noop {} {}\bibinfo {note} {See Supplemental Material for the trivial solutions for the static structure factor matrix $\mathcal{S}$, derivation of the relation between the static structure factor and the generating function, the relation between $\mathcal{S}$ and the fidelity suspectibility, evaluation of $\mathcal{S}$ for multi-site operators with iPEPS, parent Hamiltonians of the AKLT state on the square lattice, iPEPS study of the phase transition of the XX model in the staggered $Z$ field, supplementary results for the RVB state, and supplementary results for the deformed Ising wavefunctions as well as evaluation of $\mathcal{S}$ for non-injective iPEPS. Supplemental Material also includes Refs.~\cite{Anna_2025,Sandvik_heisenberg_1997,Fidelity_toric_code,Scar_null_space_2021,Wegner_duality_1971,Trebst_2007,U_1_SET_TC_2023,CIRAC_MPDO_2017,haegeman2015shadows,Corboz_TC_2020,Xu_2025}}\BibitemShut {NoStop}%
\bibitem [{\citenamefont {Schuch}\ \emph {et~al.}(2011)\citenamefont {Schuch}, \citenamefont {P\'erez-Garc\'{\i}a},\ and\ \citenamefont {Cirac}}]{Schuch_2011_classify}%
  \BibitemOpen
  \bibfield  {author} {\bibinfo {author} {\bibfnamefont {N.}~\bibnamefont {Schuch}}, \bibinfo {author} {\bibfnamefont {D.}~\bibnamefont {P\'erez-Garc\'{\i}a}}, \ and\ \bibinfo {author} {\bibfnamefont {I.}~\bibnamefont {Cirac}},\ }\href {\doibase 10.1103/PhysRevB.84.165139} {\bibfield  {journal} {\bibinfo  {journal} {Phys. Rev. B}\ }\textbf {\bibinfo {volume} {84}},\ \bibinfo {pages} {165139} (\bibinfo {year} {2011})}\BibitemShut {NoStop}%
\bibitem [{\citenamefont {Verstraete}\ \emph {et~al.}(2006{\natexlab{a}})\citenamefont {Verstraete}, \citenamefont {Wolf}, \citenamefont {Perez-Garcia},\ and\ \citenamefont {Cirac}}]{Verstraete_2006}%
  \BibitemOpen
  \bibfield  {author} {\bibinfo {author} {\bibfnamefont {F.}~\bibnamefont {Verstraete}}, \bibinfo {author} {\bibfnamefont {M.~M.}\ \bibnamefont {Wolf}}, \bibinfo {author} {\bibfnamefont {D.}~\bibnamefont {Perez-Garcia}}, \ and\ \bibinfo {author} {\bibfnamefont {J.~I.}\ \bibnamefont {Cirac}},\ }\href {\doibase 10.1103/PhysRevLett.96.220601} {\bibfield  {journal} {\bibinfo  {journal} {Phys. Rev. Lett.}\ }\textbf {\bibinfo {volume} {96}},\ \bibinfo {pages} {220601} (\bibinfo {year} {2006}{\natexlab{a}})}\BibitemShut {NoStop}%
\bibitem [{\citenamefont {Vanhecke}\ \emph {et~al.}(2022)\citenamefont {Vanhecke}, \citenamefont {Hasik}, \citenamefont {Verstraete},\ and\ \citenamefont {Vanderstraeten}}]{PhysRevLett.129.200601}%
  \BibitemOpen
  \bibfield  {author} {\bibinfo {author} {\bibfnamefont {B.}~\bibnamefont {Vanhecke}}, \bibinfo {author} {\bibfnamefont {J.}~\bibnamefont {Hasik}}, \bibinfo {author} {\bibfnamefont {F.}~\bibnamefont {Verstraete}}, \ and\ \bibinfo {author} {\bibfnamefont {L.}~\bibnamefont {Vanderstraeten}},\ }\href {\doibase 10.1103/PhysRevLett.129.200601} {\bibfield  {journal} {\bibinfo  {journal} {Phys. Rev. Lett.}\ }\textbf {\bibinfo {volume} {129}},\ \bibinfo {pages} {200601} (\bibinfo {year} {2022})}\BibitemShut {NoStop}%
\bibitem [{\citenamefont {Xu}\ and\ \citenamefont {Huang}(2025)}]{Xu_huang_2025}%
  \BibitemOpen
  \bibfield  {author} {\bibinfo {author} {\bibfnamefont {W.-T.}\ \bibnamefont {Xu}}\ and\ \bibinfo {author} {\bibfnamefont {R.-Z.}\ \bibnamefont {Huang}},\ }\href {\doibase 10.1103/PhysRevLett.134.146503} {\bibfield  {journal} {\bibinfo  {journal} {Phys. Rev. Lett.}\ }\textbf {\bibinfo {volume} {134}},\ \bibinfo {pages} {146503} (\bibinfo {year} {2025})}\BibitemShut {NoStop}%
\bibitem [{\citenamefont {Corboz}\ \emph {et~al.}(2018)\citenamefont {Corboz}, \citenamefont {Czarnik}, \citenamefont {Kapteijns},\ and\ \citenamefont {Tagliacozzo}}]{Corboz_Finite_IPEPS_2018}%
  \BibitemOpen
  \bibfield  {author} {\bibinfo {author} {\bibfnamefont {P.}~\bibnamefont {Corboz}}, \bibinfo {author} {\bibfnamefont {P.}~\bibnamefont {Czarnik}}, \bibinfo {author} {\bibfnamefont {G.}~\bibnamefont {Kapteijns}}, \ and\ \bibinfo {author} {\bibfnamefont {L.}~\bibnamefont {Tagliacozzo}},\ }\href {\doibase 10.1103/PhysRevX.8.031031} {\bibfield  {journal} {\bibinfo  {journal} {Phys. Rev. X}\ }\textbf {\bibinfo {volume} {8}},\ \bibinfo {pages} {031031} (\bibinfo {year} {2018})}\BibitemShut {NoStop}%
\bibitem [{\citenamefont {Rader}\ and\ \citenamefont {L\"auchli}(2018)}]{Lauchli_Finite_IPEPS_2018}%
  \BibitemOpen
  \bibfield  {author} {\bibinfo {author} {\bibfnamefont {M.}~\bibnamefont {Rader}}\ and\ \bibinfo {author} {\bibfnamefont {A.~M.}\ \bibnamefont {L\"auchli}},\ }\href {\doibase 10.1103/PhysRevX.8.031030} {\bibfield  {journal} {\bibinfo  {journal} {Phys. Rev. X}\ }\textbf {\bibinfo {volume} {8}},\ \bibinfo {pages} {031030} (\bibinfo {year} {2018})}\BibitemShut {NoStop}%
\bibitem [{\citenamefont {Vanderstraeten}\ \emph {et~al.}(2015)\citenamefont {Vanderstraeten}, \citenamefont {Mari\"en}, \citenamefont {Verstraete},\ and\ \citenamefont {Haegeman}}]{PEPS_tangent_space_2015}%
  \BibitemOpen
  \bibfield  {author} {\bibinfo {author} {\bibfnamefont {L.}~\bibnamefont {Vanderstraeten}}, \bibinfo {author} {\bibfnamefont {M.}~\bibnamefont {Mari\"en}}, \bibinfo {author} {\bibfnamefont {F.}~\bibnamefont {Verstraete}}, \ and\ \bibinfo {author} {\bibfnamefont {J.}~\bibnamefont {Haegeman}},\ }\href {\doibase 10.1103/PhysRevB.92.201111} {\bibfield  {journal} {\bibinfo  {journal} {Phys. Rev. B}\ }\textbf {\bibinfo {volume} {92}},\ \bibinfo {pages} {201111} (\bibinfo {year} {2015})}\BibitemShut {NoStop}%
\bibitem [{\citenamefont {Vanderstraeten}\ \emph {et~al.}(2016)\citenamefont {Vanderstraeten}, \citenamefont {Haegeman}, \citenamefont {Corboz},\ and\ \citenamefont {Verstraete}}]{iPEPS_Laurens_2016}%
  \BibitemOpen
  \bibfield  {author} {\bibinfo {author} {\bibfnamefont {L.}~\bibnamefont {Vanderstraeten}}, \bibinfo {author} {\bibfnamefont {J.}~\bibnamefont {Haegeman}}, \bibinfo {author} {\bibfnamefont {P.}~\bibnamefont {Corboz}}, \ and\ \bibinfo {author} {\bibfnamefont {F.}~\bibnamefont {Verstraete}},\ }\href {\doibase 10.1103/PhysRevB.94.155123} {\bibfield  {journal} {\bibinfo  {journal} {Phys. Rev. B}\ }\textbf {\bibinfo {volume} {94}},\ \bibinfo {pages} {155123} (\bibinfo {year} {2016})}\BibitemShut {NoStop}%
\bibitem [{\citenamefont {Ponsioen}\ and\ \citenamefont {Corboz}(2020)}]{corboz_2020}%
  \BibitemOpen
  \bibfield  {author} {\bibinfo {author} {\bibfnamefont {B.}~\bibnamefont {Ponsioen}}\ and\ \bibinfo {author} {\bibfnamefont {P.}~\bibnamefont {Corboz}},\ }\href {\doibase 10.1103/PhysRevB.101.195109} {\bibfield  {journal} {\bibinfo  {journal} {Phys. Rev. B}\ }\textbf {\bibinfo {volume} {101}},\ \bibinfo {pages} {195109} (\bibinfo {year} {2020})}\BibitemShut {NoStop}%
\bibitem [{\citenamefont {Ponsioen}\ \emph {et~al.}(2022)\citenamefont {Ponsioen}, \citenamefont {Assaad},\ and\ \citenamefont {Corboz}}]{Corboz_2022}%
  \BibitemOpen
  \bibfield  {author} {\bibinfo {author} {\bibfnamefont {B.}~\bibnamefont {Ponsioen}}, \bibinfo {author} {\bibfnamefont {F.~F.}\ \bibnamefont {Assaad}}, \ and\ \bibinfo {author} {\bibfnamefont {P.}~\bibnamefont {Corboz}},\ }\href {\doibase 10.21468/SciPostPhys.12.1.006} {\bibfield  {journal} {\bibinfo  {journal} {SciPost Phys.}\ }\textbf {\bibinfo {volume} {12}},\ \bibinfo {pages} {006} (\bibinfo {year} {2022})}\BibitemShut {NoStop}%
\bibitem [{\citenamefont {Ponsioen}\ \emph {et~al.}(2023)\citenamefont {Ponsioen}, \citenamefont {Hasik},\ and\ \citenamefont {Corboz}}]{Juraj_structure_factor_2023}%
  \BibitemOpen
  \bibfield  {author} {\bibinfo {author} {\bibfnamefont {B.}~\bibnamefont {Ponsioen}}, \bibinfo {author} {\bibfnamefont {J.}~\bibnamefont {Hasik}}, \ and\ \bibinfo {author} {\bibfnamefont {P.}~\bibnamefont {Corboz}},\ }\href {\doibase 10.1103/PhysRevB.108.195111} {\bibfield  {journal} {\bibinfo  {journal} {Phys. Rev. B}\ }\textbf {\bibinfo {volume} {108}},\ \bibinfo {pages} {195111} (\bibinfo {year} {2023})}\BibitemShut {NoStop}%
\bibitem [{\citenamefont {Tu}\ \emph {et~al.}(2021)\citenamefont {Tu}, \citenamefont {Wu}, \citenamefont {Schuch}, \citenamefont {Kawashima},\ and\ \citenamefont {Chen}}]{Gen_func_2021}%
  \BibitemOpen
  \bibfield  {author} {\bibinfo {author} {\bibfnamefont {W.-L.}\ \bibnamefont {Tu}}, \bibinfo {author} {\bibfnamefont {H.-K.}\ \bibnamefont {Wu}}, \bibinfo {author} {\bibfnamefont {N.}~\bibnamefont {Schuch}}, \bibinfo {author} {\bibfnamefont {N.}~\bibnamefont {Kawashima}}, \ and\ \bibinfo {author} {\bibfnamefont {J.-Y.}\ \bibnamefont {Chen}},\ }\href {\doibase 10.1103/PhysRevB.103.205155} {\bibfield  {journal} {\bibinfo  {journal} {Phys. Rev. B}\ }\textbf {\bibinfo {volume} {103}},\ \bibinfo {pages} {205155} (\bibinfo {year} {2021})}\BibitemShut {NoStop}%
\bibitem [{\citenamefont {Tu}\ \emph {et~al.}(2024)\citenamefont {Tu}, \citenamefont {Vanderstraeten}, \citenamefont {Schuch}, \citenamefont {Lee}, \citenamefont {Kawashima},\ and\ \citenamefont {Chen}}]{Generating_iPEPS_2024}%
  \BibitemOpen
  \bibfield  {author} {\bibinfo {author} {\bibfnamefont {W.-L.}\ \bibnamefont {Tu}}, \bibinfo {author} {\bibfnamefont {L.}~\bibnamefont {Vanderstraeten}}, \bibinfo {author} {\bibfnamefont {N.}~\bibnamefont {Schuch}}, \bibinfo {author} {\bibfnamefont {H.-Y.}\ \bibnamefont {Lee}}, \bibinfo {author} {\bibfnamefont {N.}~\bibnamefont {Kawashima}}, \ and\ \bibinfo {author} {\bibfnamefont {J.-Y.}\ \bibnamefont {Chen}},\ }\href {\doibase 10.1103/PRXQuantum.5.010335} {\bibfield  {journal} {\bibinfo  {journal} {PRX Quantum}\ }\textbf {\bibinfo {volume} {5}},\ \bibinfo {pages} {010335} (\bibinfo {year} {2024})}\BibitemShut {NoStop}%
\bibitem [{\citenamefont {Arias~Espinoza}\ and\ \citenamefont {Corboz}(2024)}]{Corboz_spec_func_2024}%
  \BibitemOpen
  \bibfield  {author} {\bibinfo {author} {\bibfnamefont {J.~D.}\ \bibnamefont {Arias~Espinoza}}\ and\ \bibinfo {author} {\bibfnamefont {P.}~\bibnamefont {Corboz}},\ }\href {\doibase 10.1103/PhysRevB.110.094314} {\bibfield  {journal} {\bibinfo  {journal} {Phys. Rev. B}\ }\textbf {\bibinfo {volume} {110}},\ \bibinfo {pages} {094314} (\bibinfo {year} {2024})}\BibitemShut {NoStop}%
\bibitem [{\citenamefont {Liao}\ \emph {et~al.}(2019)\citenamefont {Liao}, \citenamefont {Liu}, \citenamefont {Wang},\ and\ \citenamefont {Xiang}}]{AD_2019}%
  \BibitemOpen
  \bibfield  {author} {\bibinfo {author} {\bibfnamefont {H.-J.}\ \bibnamefont {Liao}}, \bibinfo {author} {\bibfnamefont {J.-G.}\ \bibnamefont {Liu}}, \bibinfo {author} {\bibfnamefont {L.}~\bibnamefont {Wang}}, \ and\ \bibinfo {author} {\bibfnamefont {T.}~\bibnamefont {Xiang}},\ }\href {\doibase 10.1103/PhysRevX.9.031041} {\bibfield  {journal} {\bibinfo  {journal} {Phys. Rev. X}\ }\textbf {\bibinfo {volume} {9}},\ \bibinfo {pages} {031041} (\bibinfo {year} {2019})}\BibitemShut {NoStop}%
\bibitem [{\citenamefont {Orús}(2019)}]{Orus_TNS_review_2019}%
  \BibitemOpen
  \bibfield  {author} {\bibinfo {author} {\bibfnamefont {R.}~\bibnamefont {Orús}},\ }\href {\doibase 10.1038/s42254-019-0086-7} {\bibfield  {journal} {\bibinfo  {journal} {Nature Reviews Physics}\ }\textbf {\bibinfo {volume} {1}},\ \bibinfo {pages} {538} (\bibinfo {year} {2019})}\BibitemShut {NoStop}%
\bibitem [{\citenamefont {GU}(2010)}]{Fidelity_2020}%
  \BibitemOpen
  \bibfield  {author} {\bibinfo {author} {\bibfnamefont {S.-J.}\ \bibnamefont {GU}},\ }\href {\doibase 10.1142/S0217979210056335} {\bibfield  {journal} {\bibinfo  {journal} {International Journal of Modern Physics B}\ }\textbf {\bibinfo {volume} {24}},\ \bibinfo {pages} {4371} (\bibinfo {year} {2010})}\BibitemShut {NoStop}%
\bibitem [{\citenamefont {Pollmann}\ \emph {et~al.}(2009)\citenamefont {Pollmann}, \citenamefont {Mukerjee}, \citenamefont {Turner},\ and\ \citenamefont {Moore}}]{Pollmann_2011}%
  \BibitemOpen
  \bibfield  {author} {\bibinfo {author} {\bibfnamefont {F.}~\bibnamefont {Pollmann}}, \bibinfo {author} {\bibfnamefont {S.}~\bibnamefont {Mukerjee}}, \bibinfo {author} {\bibfnamefont {A.~M.}\ \bibnamefont {Turner}}, \ and\ \bibinfo {author} {\bibfnamefont {J.~E.}\ \bibnamefont {Moore}},\ }\href {\doibase 10.1103/PhysRevLett.102.255701} {\bibfield  {journal} {\bibinfo  {journal} {Phys. Rev. Lett.}\ }\textbf {\bibinfo {volume} {102}},\ \bibinfo {pages} {255701} (\bibinfo {year} {2009})}\BibitemShut {NoStop}%
\bibitem [{\citenamefont {Corboz}(2016)}]{iPEPS_corboz_2016}%
  \BibitemOpen
  \bibfield  {author} {\bibinfo {author} {\bibfnamefont {P.}~\bibnamefont {Corboz}},\ }\href {\doibase 10.1103/PhysRevB.94.035133} {\bibfield  {journal} {\bibinfo  {journal} {Phys. Rev. B}\ }\textbf {\bibinfo {volume} {94}},\ \bibinfo {pages} {035133} (\bibinfo {year} {2016})}\BibitemShut {NoStop}%
\bibitem [{\citenamefont {Mambrini}\ \emph {et~al.}(2016)\citenamefont {Mambrini}, \citenamefont {Or\'us},\ and\ \citenamefont {Poilblanc}}]{systematic_SU2_PEPS_2016}%
  \BibitemOpen
  \bibfield  {author} {\bibinfo {author} {\bibfnamefont {M.}~\bibnamefont {Mambrini}}, \bibinfo {author} {\bibfnamefont {R.}~\bibnamefont {Or\'us}}, \ and\ \bibinfo {author} {\bibfnamefont {D.}~\bibnamefont {Poilblanc}},\ }\href {\doibase 10.1103/PhysRevB.94.205124} {\bibfield  {journal} {\bibinfo  {journal} {Phys. Rev. B}\ }\textbf {\bibinfo {volume} {94}},\ \bibinfo {pages} {205124} (\bibinfo {year} {2016})}\BibitemShut {NoStop}%
\bibitem [{\citenamefont {Affleck}\ \emph {et~al.}(1987)\citenamefont {Affleck}, \citenamefont {Kennedy}, \citenamefont {Lieb},\ and\ \citenamefont {Tasaki}}]{AKLT_1987}%
  \BibitemOpen
  \bibfield  {author} {\bibinfo {author} {\bibfnamefont {I.}~\bibnamefont {Affleck}}, \bibinfo {author} {\bibfnamefont {T.}~\bibnamefont {Kennedy}}, \bibinfo {author} {\bibfnamefont {E.~H.}\ \bibnamefont {Lieb}}, \ and\ \bibinfo {author} {\bibfnamefont {H.}~\bibnamefont {Tasaki}},\ }\href {\doibase 10.1103/PhysRevLett.59.799} {\bibfield  {journal} {\bibinfo  {journal} {Phys. Rev. Lett.}\ }\textbf {\bibinfo {volume} {59}},\ \bibinfo {pages} {799} (\bibinfo {year} {1987})}\BibitemShut {NoStop}%
\bibitem [{\citenamefont {Andersen}\ \emph {et~al.}(2024)\citenamefont {Andersen}, \citenamefont {Astrakhantsev}, \citenamefont {Karamlou}, \citenamefont {Berndtsson}, \citenamefont {Motruk},\ and\ \citenamefont {et~al.}}]{XX_quantum_simulator_2024}%
  \BibitemOpen
  \bibfield  {author} {\bibinfo {author} {\bibfnamefont {T.~I.}\ \bibnamefont {Andersen}}, \bibinfo {author} {\bibfnamefont {N.}~\bibnamefont {Astrakhantsev}}, \bibinfo {author} {\bibfnamefont {A.~H.}\ \bibnamefont {Karamlou}}, \bibinfo {author} {\bibfnamefont {J.}~\bibnamefont {Berndtsson}}, \bibinfo {author} {\bibfnamefont {J.}~\bibnamefont {Motruk}}, \ and\ \bibinfo {author} {\bibnamefont {et~al.}},\ }\href@noop {} {\enquote {\bibinfo {title} {Thermalization and criticality on an analog-digital quantum simulator},}\ } (\bibinfo {year} {2024}),\ \Eprint {http://arxiv.org/abs/2405.17385} {arXiv:2405.17385 [quant-ph]} \BibitemShut {NoStop}%
\bibitem [{\citenamefont {Schuch}\ \emph {et~al.}(2012)\citenamefont {Schuch}, \citenamefont {Poilblanc}, \citenamefont {Cirac},\ and\ \citenamefont {P\'erez-Garc\'{\i}a}}]{RVB_PEPS_2012}%
  \BibitemOpen
  \bibfield  {author} {\bibinfo {author} {\bibfnamefont {N.}~\bibnamefont {Schuch}}, \bibinfo {author} {\bibfnamefont {D.}~\bibnamefont {Poilblanc}}, \bibinfo {author} {\bibfnamefont {J.~I.}\ \bibnamefont {Cirac}}, \ and\ \bibinfo {author} {\bibfnamefont {D.}~\bibnamefont {P\'erez-Garc\'{\i}a}},\ }\href {\doibase 10.1103/PhysRevB.86.115108} {\bibfield  {journal} {\bibinfo  {journal} {Phys. Rev. B}\ }\textbf {\bibinfo {volume} {86}},\ \bibinfo {pages} {115108} (\bibinfo {year} {2012})}\BibitemShut {NoStop}%
\bibitem [{\citenamefont {Albuquerque}\ and\ \citenamefont {Alet}(2010)}]{RVB_fabian_2010}%
  \BibitemOpen
  \bibfield  {author} {\bibinfo {author} {\bibfnamefont {A.~F.}\ \bibnamefont {Albuquerque}}\ and\ \bibinfo {author} {\bibfnamefont {F.}~\bibnamefont {Alet}},\ }\href {\doibase 10.1103/PhysRevB.82.180408} {\bibfield  {journal} {\bibinfo  {journal} {Phys. Rev. B}\ }\textbf {\bibinfo {volume} {82}},\ \bibinfo {pages} {180408} (\bibinfo {year} {2010})}\BibitemShut {NoStop}%
\bibitem [{\citenamefont {Tang}\ \emph {et~al.}(2011)\citenamefont {Tang}, \citenamefont {Sandvik},\ and\ \citenamefont {Henley}}]{RVB_sandvik_2011}%
  \BibitemOpen
  \bibfield  {author} {\bibinfo {author} {\bibfnamefont {Y.}~\bibnamefont {Tang}}, \bibinfo {author} {\bibfnamefont {A.~W.}\ \bibnamefont {Sandvik}}, \ and\ \bibinfo {author} {\bibfnamefont {C.~L.}\ \bibnamefont {Henley}},\ }\href {\doibase 10.1103/PhysRevB.84.174427} {\bibfield  {journal} {\bibinfo  {journal} {Phys. Rev. B}\ }\textbf {\bibinfo {volume} {84}},\ \bibinfo {pages} {174427} (\bibinfo {year} {2011})}\BibitemShut {NoStop}%
\bibitem [{\citenamefont {Ardonne}\ \emph {et~al.}(2004)\citenamefont {Ardonne}, \citenamefont {Fendley},\ and\ \citenamefont {Fradkin}}]{ardonne_2004}%
  \BibitemOpen
  \bibfield  {author} {\bibinfo {author} {\bibfnamefont {E.}~\bibnamefont {Ardonne}}, \bibinfo {author} {\bibfnamefont {P.}~\bibnamefont {Fendley}}, \ and\ \bibinfo {author} {\bibfnamefont {E.}~\bibnamefont {Fradkin}},\ }\href {\doibase 10.1016/j.aop.2004.01.004} {\bibfield  {journal} {\bibinfo  {journal} {Annals of Physics}\ }\textbf {\bibinfo {volume} {310}},\ \bibinfo {pages} {493} (\bibinfo {year} {2004})}\BibitemShut {NoStop}%
\bibitem [{\citenamefont {Dreyer}\ \emph {et~al.}(2024)\citenamefont {Dreyer}, \citenamefont {Vanderstraeten}, \citenamefont {Chen}, \citenamefont {Verresen},\ and\ \citenamefont {Schuch}}]{Robustness_U_1_2024}%
  \BibitemOpen
  \bibfield  {author} {\bibinfo {author} {\bibfnamefont {H.}~\bibnamefont {Dreyer}}, \bibinfo {author} {\bibfnamefont {L.}~\bibnamefont {Vanderstraeten}}, \bibinfo {author} {\bibfnamefont {J.-Y.}\ \bibnamefont {Chen}}, \bibinfo {author} {\bibfnamefont {R.}~\bibnamefont {Verresen}}, \ and\ \bibinfo {author} {\bibfnamefont {N.}~\bibnamefont {Schuch}},\ }\href {\doibase 10.1103/PhysRevB.109.195161} {\bibfield  {journal} {\bibinfo  {journal} {Phys. Rev. B}\ }\textbf {\bibinfo {volume} {109}},\ \bibinfo {pages} {195161} (\bibinfo {year} {2024})}\BibitemShut {NoStop}%
\bibitem [{\citenamefont {Fujimoto}(2005)}]{Satoshi_Fujimoto_2005}%
  \BibitemOpen
  \bibfield  {author} {\bibinfo {author} {\bibfnamefont {S.}~\bibnamefont {Fujimoto}},\ }\href {\doibase 10.1103/PhysRevB.72.024429} {\bibfield  {journal} {\bibinfo  {journal} {Phys. Rev. B}\ }\textbf {\bibinfo {volume} {72}},\ \bibinfo {pages} {024429} (\bibinfo {year} {2005})}\BibitemShut {NoStop}%
\bibitem [{\citenamefont {Cano}\ and\ \citenamefont {Fendley}(2010)}]{Fendly_2010}%
  \BibitemOpen
  \bibfield  {author} {\bibinfo {author} {\bibfnamefont {J.}~\bibnamefont {Cano}}\ and\ \bibinfo {author} {\bibfnamefont {P.}~\bibnamefont {Fendley}},\ }\href {\doibase 10.1103/PhysRevLett.105.067205} {\bibfield  {journal} {\bibinfo  {journal} {Phys. Rev. Lett.}\ }\textbf {\bibinfo {volume} {105}},\ \bibinfo {pages} {067205} (\bibinfo {year} {2010})}\BibitemShut {NoStop}%
\bibitem [{\citenamefont {Zhou}\ \emph {et~al.}(2014)\citenamefont {Zhou}, \citenamefont {Wildeboer},\ and\ \citenamefont {Seidel}}]{twelve_site_H_kagome_RVB_2014}%
  \BibitemOpen
  \bibfield  {author} {\bibinfo {author} {\bibfnamefont {Z.}~\bibnamefont {Zhou}}, \bibinfo {author} {\bibfnamefont {J.}~\bibnamefont {Wildeboer}}, \ and\ \bibinfo {author} {\bibfnamefont {A.}~\bibnamefont {Seidel}},\ }\href {\doibase 10.1103/PhysRevB.89.035123} {\bibfield  {journal} {\bibinfo  {journal} {Phys. Rev. B}\ }\textbf {\bibinfo {volume} {89}},\ \bibinfo {pages} {035123} (\bibinfo {year} {2014})}\BibitemShut {NoStop}%
\bibitem [{\citenamefont {Giudici}\ \emph {et~al.}(2022)\citenamefont {Giudici}, \citenamefont {Cirac},\ and\ \citenamefont {Schuch}}]{Locality_optim_2022}%
  \BibitemOpen
  \bibfield  {author} {\bibinfo {author} {\bibfnamefont {G.}~\bibnamefont {Giudici}}, \bibinfo {author} {\bibfnamefont {J.~I.}\ \bibnamefont {Cirac}}, \ and\ \bibinfo {author} {\bibfnamefont {N.}~\bibnamefont {Schuch}},\ }\href {\doibase 10.1103/PhysRevB.106.035109} {\bibfield  {journal} {\bibinfo  {journal} {Phys. Rev. B}\ }\textbf {\bibinfo {volume} {106}},\ \bibinfo {pages} {035109} (\bibinfo {year} {2022})}\BibitemShut {NoStop}%
\bibitem [{\citenamefont {Verstraete}\ \emph {et~al.}(2006{\natexlab{b}})\citenamefont {Verstraete}, \citenamefont {Wolf}, \citenamefont {Perez-Garcia},\ and\ \citenamefont {Cirac}}]{computation_power_2006}%
  \BibitemOpen
  \bibfield  {author} {\bibinfo {author} {\bibfnamefont {F.}~\bibnamefont {Verstraete}}, \bibinfo {author} {\bibfnamefont {M.~M.}\ \bibnamefont {Wolf}}, \bibinfo {author} {\bibfnamefont {D.}~\bibnamefont {Perez-Garcia}}, \ and\ \bibinfo {author} {\bibfnamefont {J.~I.}\ \bibnamefont {Cirac}},\ }\href {\doibase 10.1103/PhysRevLett.96.220601} {\bibfield  {journal} {\bibinfo  {journal} {Phys. Rev. Lett.}\ }\textbf {\bibinfo {volume} {96}},\ \bibinfo {pages} {220601} (\bibinfo {year} {2006}{\natexlab{b}})}\BibitemShut {NoStop}%
\bibitem [{\citenamefont {Castelnovo}\ and\ \citenamefont {Chamon}(2008)}]{Deforemd_TC_2008}%
  \BibitemOpen
  \bibfield  {author} {\bibinfo {author} {\bibfnamefont {C.}~\bibnamefont {Castelnovo}}\ and\ \bibinfo {author} {\bibfnamefont {C.}~\bibnamefont {Chamon}},\ }\href {\doibase 10.1103/PhysRevB.77.054433} {\bibfield  {journal} {\bibinfo  {journal} {Phys. Rev. B}\ }\textbf {\bibinfo {volume} {77}},\ \bibinfo {pages} {054433} (\bibinfo {year} {2008})}\BibitemShut {NoStop}%
\bibitem [{\citenamefont {Castelnovo}\ \emph {et~al.}(2010)\citenamefont {Castelnovo}, \citenamefont {Trebst},\ and\ \citenamefont {Troyer}}]{Trebst_2009}%
  \BibitemOpen
  \bibfield  {author} {\bibinfo {author} {\bibfnamefont {C.}~\bibnamefont {Castelnovo}}, \bibinfo {author} {\bibfnamefont {S.}~\bibnamefont {Trebst}}, \ and\ \bibinfo {author} {\bibfnamefont {M.}~\bibnamefont {Troyer}},\ }\href {\doibase 10.1201/b10273-10} {\bibfield  {journal} {\bibinfo  {journal} {arXiv:0912.3272 [cond-mat]}\ }\textbf {\bibinfo {volume} {20103812}},\ \bibinfo {pages} {169} (\bibinfo {year} {2010})},\ \bibinfo {note} {arXiv: 0912.3272}\BibitemShut {NoStop}%
\bibitem [{\citenamefont {Kitaev}(2003)}]{kitaev_2002}%
  \BibitemOpen
  \bibfield  {author} {\bibinfo {author} {\bibfnamefont {A.}~\bibnamefont {Kitaev}},\ }\href {\doibase https://doi.org/10.1016/S0003-4916(02)00018-0} {\bibfield  {journal} {\bibinfo  {journal} {Annals of Physics}\ }\textbf {\bibinfo {volume} {303}},\ \bibinfo {pages} {2} (\bibinfo {year} {2003})}\BibitemShut {NoStop}%
\bibitem [{\citenamefont {Beach}\ and\ \citenamefont {Sandvik}(2006)}]{Sandvik_2006}%
  \BibitemOpen
  \bibfield  {author} {\bibinfo {author} {\bibfnamefont {K.}~\bibnamefont {Beach}}\ and\ \bibinfo {author} {\bibfnamefont {A.~W.}\ \bibnamefont {Sandvik}},\ }\href {\doibase https://doi.org/10.1016/j.nuclphysb.2006.05.032} {\bibfield  {journal} {\bibinfo  {journal} {Nuclear Physics B}\ }\textbf {\bibinfo {volume} {750}},\ \bibinfo {pages} {142} (\bibinfo {year} {2006})}\BibitemShut {NoStop}%
\bibitem [{\citenamefont {Chen}\ and\ \citenamefont {Poilblanc}(2018)}]{LR_RVB_Z2_2018}%
  \BibitemOpen
  \bibfield  {author} {\bibinfo {author} {\bibfnamefont {J.-Y.}\ \bibnamefont {Chen}}\ and\ \bibinfo {author} {\bibfnamefont {D.}~\bibnamefont {Poilblanc}},\ }\href {\doibase 10.1103/PhysRevB.97.161107} {\bibfield  {journal} {\bibinfo  {journal} {Phys. Rev. B}\ }\textbf {\bibinfo {volume} {97}},\ \bibinfo {pages} {161107} (\bibinfo {year} {2018})}\BibitemShut {NoStop}%
\bibitem [{\citenamefont {Ma}\ \emph {et~al.}(2019)\citenamefont {Ma}, \citenamefont {You},\ and\ \citenamefont {Meng}}]{Norther_DQCP_2019}%
  \BibitemOpen
  \bibfield  {author} {\bibinfo {author} {\bibfnamefont {N.}~\bibnamefont {Ma}}, \bibinfo {author} {\bibfnamefont {Y.-Z.}\ \bibnamefont {You}}, \ and\ \bibinfo {author} {\bibfnamefont {Z.~Y.}\ \bibnamefont {Meng}},\ }\href {\doibase 10.1103/PhysRevLett.122.175701} {\bibfield  {journal} {\bibinfo  {journal} {Phys. Rev. Lett.}\ }\textbf {\bibinfo {volume} {122}},\ \bibinfo {pages} {175701} (\bibinfo {year} {2019})}\BibitemShut {NoStop}%
\bibitem [{\citenamefont {Chepiga}\ and\ \citenamefont {Mila}(2017)}]{PhysRevB.96.054425}%
  \BibitemOpen
  \bibfield  {author} {\bibinfo {author} {\bibfnamefont {N.}~\bibnamefont {Chepiga}}\ and\ \bibinfo {author} {\bibfnamefont {F.}~\bibnamefont {Mila}},\ }\href {\doibase 10.1103/PhysRevB.96.054425} {\bibfield  {journal} {\bibinfo  {journal} {Phys. Rev. B}\ }\textbf {\bibinfo {volume} {96}},\ \bibinfo {pages} {054425} (\bibinfo {year} {2017})}\BibitemShut {NoStop}%
\bibitem [{\citenamefont {Cocchiarella}\ \emph {et~al.}(2025)\citenamefont {Cocchiarella}, \citenamefont {Yang}, \citenamefont {Zhang}, \citenamefont {Bañuls}, \citenamefont {Tu},\ and\ \citenamefont {Liu}}]{cocchiarella2025excitedstateslocaleffective}%
  \BibitemOpen
  \bibfield  {author} {\bibinfo {author} {\bibfnamefont {D.}~\bibnamefont {Cocchiarella}}, \bibinfo {author} {\bibfnamefont {M.}~\bibnamefont {Yang}}, \bibinfo {author} {\bibfnamefont {Y.}~\bibnamefont {Zhang}}, \bibinfo {author} {\bibfnamefont {M.~C.}\ \bibnamefont {Bañuls}}, \bibinfo {author} {\bibfnamefont {H.-H.}\ \bibnamefont {Tu}}, \ and\ \bibinfo {author} {\bibfnamefont {Y.}~\bibnamefont {Liu}},\ }\href {https://arxiv.org/abs/2511.16746} {\enquote {\bibinfo {title} {Excited states from local effective hamiltonians of matrix product states and their entanglement spectrum transition},}\ } (\bibinfo {year} {2025}),\ \Eprint {http://arxiv.org/abs/2511.16746} {arXiv:2511.16746 [cond-mat.str-el]} \BibitemShut {NoStop}%
\bibitem [{\citenamefont {Hu}\ \emph {et~al.}(2020)\citenamefont {Hu}, \citenamefont {Franco-Rubio},\ and\ \citenamefont {Vidal}}]{hu2020emergentuniversalitycriticalquantum}%
  \BibitemOpen
  \bibfield  {author} {\bibinfo {author} {\bibfnamefont {Q.}~\bibnamefont {Hu}}, \bibinfo {author} {\bibfnamefont {A.}~\bibnamefont {Franco-Rubio}}, \ and\ \bibinfo {author} {\bibfnamefont {G.}~\bibnamefont {Vidal}},\ }\href {https://arxiv.org/abs/2009.11383} {\enquote {\bibinfo {title} {Emergent universality in critical quantum spin chains: entanglement virasoro algebra},}\ } (\bibinfo {year} {2020}),\ \Eprint {http://arxiv.org/abs/2009.11383} {arXiv:2009.11383 [quant-ph]} \BibitemShut {NoStop}%
\bibitem [{\citenamefont {Zauner}\ \emph {et~al.}(2015)\citenamefont {Zauner}, \citenamefont {Draxler}, \citenamefont {Vanderstraeten}, \citenamefont {Degroote}, \citenamefont {Haegeman}, \citenamefont {Rams}, \citenamefont {Stojevic}, \citenamefont {Schuch},\ and\ \citenamefont {Verstraete}}]{Zauner_2015}%
  \BibitemOpen
  \bibfield  {author} {\bibinfo {author} {\bibfnamefont {V.}~\bibnamefont {Zauner}}, \bibinfo {author} {\bibfnamefont {D.}~\bibnamefont {Draxler}}, \bibinfo {author} {\bibfnamefont {L.}~\bibnamefont {Vanderstraeten}}, \bibinfo {author} {\bibfnamefont {M.}~\bibnamefont {Degroote}}, \bibinfo {author} {\bibfnamefont {J.}~\bibnamefont {Haegeman}}, \bibinfo {author} {\bibfnamefont {M.~M.}\ \bibnamefont {Rams}}, \bibinfo {author} {\bibfnamefont {V.}~\bibnamefont {Stojevic}}, \bibinfo {author} {\bibfnamefont {N.}~\bibnamefont {Schuch}}, \ and\ \bibinfo {author} {\bibfnamefont {F.}~\bibnamefont {Verstraete}},\ }\href {\doibase 10.1088/1367-2630/17/5/053002} {\bibfield  {journal} {\bibinfo  {journal} {New Journal of Physics}\ }\textbf {\bibinfo {volume} {17}},\ \bibinfo {pages} {053002} (\bibinfo {year} {2015})}\BibitemShut {NoStop}%
\bibitem [{\citenamefont {Hastings}(2006)}]{Hastings_2006}%
  \BibitemOpen
  \bibfield  {author} {\bibinfo {author} {\bibfnamefont {M.~B.}\ \bibnamefont {Hastings}},\ }\href {\doibase 10.1103/PhysRevB.73.085115} {\bibfield  {journal} {\bibinfo  {journal} {Phys. Rev. B}\ }\textbf {\bibinfo {volume} {73}},\ \bibinfo {pages} {085115} (\bibinfo {year} {2006})}\BibitemShut {NoStop}%
\bibitem [{\citenamefont {Wolf}\ \emph {et~al.}(2008)\citenamefont {Wolf}, \citenamefont {Verstraete}, \citenamefont {Hastings},\ and\ \citenamefont {Cirac}}]{PhysRevLett.100.070502}%
  \BibitemOpen
  \bibfield  {author} {\bibinfo {author} {\bibfnamefont {M.~M.}\ \bibnamefont {Wolf}}, \bibinfo {author} {\bibfnamefont {F.}~\bibnamefont {Verstraete}}, \bibinfo {author} {\bibfnamefont {M.~B.}\ \bibnamefont {Hastings}}, \ and\ \bibinfo {author} {\bibfnamefont {J.~I.}\ \bibnamefont {Cirac}},\ }\href {\doibase 10.1103/PhysRevLett.100.070502} {\bibfield  {journal} {\bibinfo  {journal} {Phys. Rev. Lett.}\ }\textbf {\bibinfo {volume} {100}},\ \bibinfo {pages} {070502} (\bibinfo {year} {2008})}\BibitemShut {NoStop}%
\bibitem [{\citenamefont {Kuwahara}\ \emph {et~al.}(2021)\citenamefont {Kuwahara}, \citenamefont {Alhambra},\ and\ \citenamefont {Anshu}}]{PhysRevX.11.011047}%
  \BibitemOpen
  \bibfield  {author} {\bibinfo {author} {\bibfnamefont {T.}~\bibnamefont {Kuwahara}}, \bibinfo {author} {\bibfnamefont {A.~M.}\ \bibnamefont {Alhambra}}, \ and\ \bibinfo {author} {\bibfnamefont {A.}~\bibnamefont {Anshu}},\ }\href {\doibase 10.1103/PhysRevX.11.011047} {\bibfield  {journal} {\bibinfo  {journal} {Phys. Rev. X}\ }\textbf {\bibinfo {volume} {11}},\ \bibinfo {pages} {011047} (\bibinfo {year} {2021})}\BibitemShut {NoStop}%
\bibitem [{\citenamefont {Cocchiarella}\ and\ \citenamefont {Ba\~nuls}(2025)}]{x2gx-wsp1}%
  \BibitemOpen
  \bibfield  {author} {\bibinfo {author} {\bibfnamefont {D.}~\bibnamefont {Cocchiarella}}\ and\ \bibinfo {author} {\bibfnamefont {M.~C.}\ \bibnamefont {Ba\~nuls}},\ }\href {\doibase 10.1103/x2gx-wsp1} {\bibfield  {journal} {\bibinfo  {journal} {Phys. Rev. Res.}\ }\textbf {\bibinfo {volume} {7}},\ \bibinfo {pages} {033184} (\bibinfo {year} {2025})}\BibitemShut {NoStop}%
\bibitem [{\citenamefont {Hasik}\ \emph {et~al.}(2020)\citenamefont {Hasik}, \citenamefont {Mbeng}, \citenamefont {Tu}, \citenamefont {Diop}, \citenamefont {Niu}, \citenamefont {Xi} \emph {et~al.}}]{peps-torch}%
  \BibitemOpen
  \bibfield  {author} {\bibinfo {author} {\bibfnamefont {J.}~\bibnamefont {Hasik}}, \bibinfo {author} {\bibfnamefont {G.~B.}\ \bibnamefont {Mbeng}}, \bibinfo {author} {\bibfnamefont {W.-L.}\ \bibnamefont {Tu}}, \bibinfo {author} {\bibfnamefont {S.-S.}\ \bibnamefont {Diop}}, \bibinfo {author} {\bibfnamefont {S.}~\bibnamefont {Niu}}, \bibinfo {author} {\bibfnamefont {Y.}~\bibnamefont {Xi}},  \emph {et~al.},\ }\href@noop {} {\enquote {\bibinfo {title} {{PEPS-torch}: a differentiable tensor network library for two-dimensional lattice models},}\ }\bibinfo {howpublished} {\url{https://github.com/jurajHasik/peps-torch}} (\bibinfo {year} {2020})\BibitemShut {NoStop}%
\bibitem [{\citenamefont {Xu}\ \emph {et~al.}(2025{\natexlab{a}})\citenamefont {Xu}, \citenamefont {Frías~Pérez},\ and\ \citenamefont {Yang}}]{Zenodo}%
  \BibitemOpen
  \bibfield  {author} {\bibinfo {author} {\bibfnamefont {W.-T.}\ \bibnamefont {Xu}}, \bibinfo {author} {\bibfnamefont {M.}~\bibnamefont {Frías~Pérez}}, \ and\ \bibinfo {author} {\bibfnamefont {M.}~\bibnamefont {Yang}},\ }\href {\doibase 10.5281/zenodo.17704973} {\enquote {\bibinfo {title} {Extracting conserved operators from a projected entangled pair state},}\ } (\bibinfo {year} {2025}{\natexlab{a}})\BibitemShut {NoStop}%
\bibitem [{\citenamefont {Francuz}\ \emph {et~al.}(2025)\citenamefont {Francuz}, \citenamefont {Schuch},\ and\ \citenamefont {Vanhecke}}]{Anna_2025}%
  \BibitemOpen
  \bibfield  {author} {\bibinfo {author} {\bibfnamefont {A.}~\bibnamefont {Francuz}}, \bibinfo {author} {\bibfnamefont {N.}~\bibnamefont {Schuch}}, \ and\ \bibinfo {author} {\bibfnamefont {B.}~\bibnamefont {Vanhecke}},\ }\href {\doibase 10.1103/PhysRevResearch.7.013237} {\bibfield  {journal} {\bibinfo  {journal} {Phys. Rev. Res.}\ }\textbf {\bibinfo {volume} {7}},\ \bibinfo {pages} {013237} (\bibinfo {year} {2025})}\BibitemShut {NoStop}%
\bibitem [{\citenamefont {Sandvik}(1997)}]{Sandvik_heisenberg_1997}%
  \BibitemOpen
  \bibfield  {author} {\bibinfo {author} {\bibfnamefont {A.~W.}\ \bibnamefont {Sandvik}},\ }\href {\doibase 10.1103/PhysRevB.56.11678} {\bibfield  {journal} {\bibinfo  {journal} {Phys. Rev. B}\ }\textbf {\bibinfo {volume} {56}},\ \bibinfo {pages} {11678} (\bibinfo {year} {1997})}\BibitemShut {NoStop}%
\bibitem [{\citenamefont {Abasto}\ \emph {et~al.}(2008)\citenamefont {Abasto}, \citenamefont {Hamma},\ and\ \citenamefont {Zanardi}}]{Fidelity_toric_code}%
  \BibitemOpen
  \bibfield  {author} {\bibinfo {author} {\bibfnamefont {D.~F.}\ \bibnamefont {Abasto}}, \bibinfo {author} {\bibfnamefont {A.}~\bibnamefont {Hamma}}, \ and\ \bibinfo {author} {\bibfnamefont {P.}~\bibnamefont {Zanardi}},\ }\href {\doibase 10.1103/PhysRevA.78.010301} {\bibfield  {journal} {\bibinfo  {journal} {Phys. Rev. A}\ }\textbf {\bibinfo {volume} {78}},\ \bibinfo {pages} {010301} (\bibinfo {year} {2008})}\BibitemShut {NoStop}%
\bibitem [{\citenamefont {Karle}\ \emph {et~al.}(2021)\citenamefont {Karle}, \citenamefont {Serbyn},\ and\ \citenamefont {Michailidis}}]{Scar_null_space_2021}%
  \BibitemOpen
  \bibfield  {author} {\bibinfo {author} {\bibfnamefont {V.}~\bibnamefont {Karle}}, \bibinfo {author} {\bibfnamefont {M.}~\bibnamefont {Serbyn}}, \ and\ \bibinfo {author} {\bibfnamefont {A.~A.}\ \bibnamefont {Michailidis}},\ }\href {\doibase 10.1103/PhysRevLett.127.060602} {\bibfield  {journal} {\bibinfo  {journal} {Phys. Rev. Lett.}\ }\textbf {\bibinfo {volume} {127}},\ \bibinfo {pages} {060602} (\bibinfo {year} {2021})}\BibitemShut {NoStop}%
\bibitem [{\citenamefont {Wegner}(1971)}]{Wegner_duality_1971}%
  \BibitemOpen
  \bibfield  {author} {\bibinfo {author} {\bibfnamefont {F.~J.}\ \bibnamefont {Wegner}},\ }\href {https://pubs.aip.org/aip/jmp/article-abstract/12/10/2259/465334/Duality-in-Generalized-Ising-Models-and-Phase?redirectedFrom=fulltext} {\bibfield  {journal} {\bibinfo  {journal} {Journal of Mathematical Physics}\ }\textbf {\bibinfo {volume} {12}},\ \bibinfo {pages} {2259} (\bibinfo {year} {1971})}\BibitemShut {NoStop}%
\bibitem [{\citenamefont {Trebst}\ \emph {et~al.}(2007)\citenamefont {Trebst}, \citenamefont {Werner}, \citenamefont {Troyer}, \citenamefont {Shtengel},\ and\ \citenamefont {Nayak}}]{Trebst_2007}%
  \BibitemOpen
  \bibfield  {author} {\bibinfo {author} {\bibfnamefont {S.}~\bibnamefont {Trebst}}, \bibinfo {author} {\bibfnamefont {P.}~\bibnamefont {Werner}}, \bibinfo {author} {\bibfnamefont {M.}~\bibnamefont {Troyer}}, \bibinfo {author} {\bibfnamefont {K.}~\bibnamefont {Shtengel}}, \ and\ \bibinfo {author} {\bibfnamefont {C.}~\bibnamefont {Nayak}},\ }\href {\doibase 10.1103/PhysRevLett.98.070602} {\bibfield  {journal} {\bibinfo  {journal} {Phys. Rev. Lett.}\ }\textbf {\bibinfo {volume} {98}},\ \bibinfo {pages} {070602} (\bibinfo {year} {2007})}\BibitemShut {NoStop}%
\bibitem [{\citenamefont {Wu}\ \emph {et~al.}(2023)\citenamefont {Wu}, \citenamefont {Khudorozhkov}, \citenamefont {Delfino}, \citenamefont {Green},\ and\ \citenamefont {Chamon}}]{U_1_SET_TC_2023}%
  \BibitemOpen
  \bibfield  {author} {\bibinfo {author} {\bibfnamefont {K.-H.}\ \bibnamefont {Wu}}, \bibinfo {author} {\bibfnamefont {A.}~\bibnamefont {Khudorozhkov}}, \bibinfo {author} {\bibfnamefont {G.}~\bibnamefont {Delfino}}, \bibinfo {author} {\bibfnamefont {D.}~\bibnamefont {Green}}, \ and\ \bibinfo {author} {\bibfnamefont {C.}~\bibnamefont {Chamon}},\ }\href {\doibase 10.1103/PhysRevB.108.115159} {\bibfield  {journal} {\bibinfo  {journal} {Phys. Rev. B}\ }\textbf {\bibinfo {volume} {108}},\ \bibinfo {pages} {115159} (\bibinfo {year} {2023})}\BibitemShut {NoStop}%
\bibitem [{\citenamefont {Cirac}\ \emph {et~al.}(2017)\citenamefont {Cirac}, \citenamefont {Pérez-García}, \citenamefont {Schuch},\ and\ \citenamefont {Verstraete}}]{CIRAC_MPDO_2017}%
  \BibitemOpen
  \bibfield  {author} {\bibinfo {author} {\bibfnamefont {J.}~\bibnamefont {Cirac}}, \bibinfo {author} {\bibfnamefont {D.}~\bibnamefont {Pérez-García}}, \bibinfo {author} {\bibfnamefont {N.}~\bibnamefont {Schuch}}, \ and\ \bibinfo {author} {\bibfnamefont {F.}~\bibnamefont {Verstraete}},\ }\href {\doibase https://doi.org/10.1016/j.aop.2016.12.030} {\bibfield  {journal} {\bibinfo  {journal} {Annals of Physics}\ }\textbf {\bibinfo {volume} {378}},\ \bibinfo {pages} {100} (\bibinfo {year} {2017})}\BibitemShut {NoStop}%
\bibitem [{\citenamefont {Haegeman}\ \emph {et~al.}(2015)\citenamefont {Haegeman}, \citenamefont {Zauner}, \citenamefont {Schuch},\ and\ \citenamefont {Verstraete}}]{haegeman2015shadows}%
  \BibitemOpen
  \bibfield  {author} {\bibinfo {author} {\bibfnamefont {J.}~\bibnamefont {Haegeman}}, \bibinfo {author} {\bibfnamefont {V.}~\bibnamefont {Zauner}}, \bibinfo {author} {\bibfnamefont {N.}~\bibnamefont {Schuch}}, \ and\ \bibinfo {author} {\bibfnamefont {F.}~\bibnamefont {Verstraete}},\ }\href {\doibase 10.1038/ncomms9284 | www.nature.com/naturecommunications} {\bibfield  {journal} {\bibinfo  {journal} {Nature communications}\ }\textbf {\bibinfo {volume} {6}},\ \bibinfo {pages} {8284} (\bibinfo {year} {2015})}\BibitemShut {NoStop}%
\bibitem [{\citenamefont {Crone}\ and\ \citenamefont {Corboz}(2020)}]{Corboz_TC_2020}%
  \BibitemOpen
  \bibfield  {author} {\bibinfo {author} {\bibfnamefont {S.~P.~G.}\ \bibnamefont {Crone}}\ and\ \bibinfo {author} {\bibfnamefont {P.}~\bibnamefont {Corboz}},\ }\href {\doibase 10.1103/PhysRevB.101.115143} {\bibfield  {journal} {\bibinfo  {journal} {Phys. Rev. B}\ }\textbf {\bibinfo {volume} {101}},\ \bibinfo {pages} {115143} (\bibinfo {year} {2020})}\BibitemShut {NoStop}%
\bibitem [{\citenamefont {Xu}\ \emph {et~al.}(2025{\natexlab{b}})\citenamefont {Xu}, \citenamefont {Pollmann},\ and\ \citenamefont {Knap}}]{Xu_2025}%
  \BibitemOpen
  \bibfield  {author} {\bibinfo {author} {\bibfnamefont {W.-T.}\ \bibnamefont {Xu}}, \bibinfo {author} {\bibfnamefont {F.}~\bibnamefont {Pollmann}}, \ and\ \bibinfo {author} {\bibfnamefont {M.}~\bibnamefont {Knap}},\ }\href {\doibase 10.1038/s41534-025-01030-z} {\bibfield  {journal} {\bibinfo  {journal} {npj Quantum Information}\ }\textbf {\bibinfo {volume} {11}},\ \bibinfo {pages} {74} (\bibinfo {year} {2025}{\natexlab{b}})}\BibitemShut {NoStop}%
\bibitem [{\citenamefont {Ok}\ \emph {et~al.}(2019)\citenamefont {Ok}, \citenamefont {Choo}, \citenamefont {Mudry}, \citenamefont {Castelnovo}, \citenamefont {Chamon},\ and\ \citenamefont {Neupert}}]{Topo_scar_2019}%
  \BibitemOpen
  \bibfield  {author} {\bibinfo {author} {\bibfnamefont {S.}~\bibnamefont {Ok}}, \bibinfo {author} {\bibfnamefont {K.}~\bibnamefont {Choo}}, \bibinfo {author} {\bibfnamefont {C.}~\bibnamefont {Mudry}}, \bibinfo {author} {\bibfnamefont {C.}~\bibnamefont {Castelnovo}}, \bibinfo {author} {\bibfnamefont {C.}~\bibnamefont {Chamon}}, \ and\ \bibinfo {author} {\bibfnamefont {T.}~\bibnamefont {Neupert}},\ }\href {\doibase 10.1103/PhysRevResearch.1.033144} {\bibfield  {journal} {\bibinfo  {journal} {Phys. Rev. Res.}\ }\textbf {\bibinfo {volume} {1}},\ \bibinfo {pages} {033144} (\bibinfo {year} {2019})}\BibitemShut {NoStop}%
\end{thebibliography}%

\onecolumngrid
\appendix

\section{Trivial solutions for the structure factor matrix $\mathcal{S}$}\label{app:Tri_sol}
In this section, we discuss the explicit form the trivial solutions of structure factor matrices. Knowing them is helpful because we can project out the trivial solutions and only focus on the non-trivial solutions.

For the 1-site structure factor matrix, there is only one trivial solution, which is the identity matrix. The trivial solutions of 2-site structure factor matrices of 2D quantum many-body states are the same as that of the 1D case~\cite{Yang_2023}, i.e., it is $X\otimes I-I\otimes X$ at momentum $\pmb{q}=(0,0)$, where $X$ can be any on-site Hermitian matrix. Similarly, the trivial solutions of 2-site structure factor matrices at momentum $\pmb{q}=(\pi,\pi)$ take the form $X\otimes I+I\otimes X$.

In the following, we give the form of the trivial solutions of 4-site (a $2\times 2$ plaquette) structure factor matrices. At momentum $\pmb{q}=(0,0)$, they are
\begin{align}
&\myarrayplaq{I}{I}{I}{I};\notag\\
&\myarrayplaq{A}{I}{I}{I}-\myarrayplaq{I}{I}{I}{A},\quad 
\mbox{and its all $C_4$ rotations;}\notag\\
&\myarrayplaq{A}{I}{I}{I}
-\myarrayplaq{I}{I}{A}{I},\quad \mbox{and} \quad 
\myarrayplaq{ I}{A}{ I}{ I}
-\myarrayplaq{I}{I}{I}{A};\notag\\
&\myarrayplaq{A}{I}{I}{B}-
\myarrayplaq{I}{A}{B}{I}, \quad \mbox{and} \quad 
\myarrayplaq{A}{B}{I}{I}
-\myarrayplaq{I}{I}{B}{A},\notag
\end{align} 
where $A$ and $B$ are arbitrary on-site Hermitian matrices.
At momentum $\pmb{q}=(\pi,\pi)$, the trivial solutions are 
\begin{align}
&\myarrayplaq{A}{I}{I}{I}+\myarrayplaq{I}{I}{I}{A},\quad 
\mbox{and its all $C_4$ rotations;}\notag\\
&\myarrayplaq{A}{I}{I}{I}
-\myarrayplaq{I}{I}{A}{I},\quad \mbox{and} \quad 
\myarrayplaq{ I}{A}{ I}{ I}
-\myarrayplaq{I}{I}{I}{A};\notag\\
&\myarrayplaq{A}{I}{I}{B}+
\myarrayplaq{I}{A}{B}{I}, \quad \mbox{and} \quad 
\myarrayplaq{A}{B}{I}{I}
+\myarrayplaq{I}{I}{B}{A}.\notag
\end{align}
At momentum $\pmb{q}=(\pi,0)$, the trivial solutions are
\begin{align}
 & \myarrayplaq{A}{I}{I}{I}
   + \myarrayplaq{I}{I}{I}{A},
   \quad \text{and}\quad
   \myarrayplaq{A}{I}{I}{I}
   - \myarrayplaq{I}{A}{I}{I}, \notag\\
 & \myarrayplaq{I}{A}{I}{I}
   + \myarrayplaq{I}{I}{A}{I},
   \quad \text{and}\quad
   \myarrayplaq{I}{I}{A}{I}
   - \myarrayplaq{I}{I}{I}{A}, \notag\\
 & \myarrayplaq{A}{I}{I}{I}
   + \myarrayplaq{I}{I}{A}{I},
   \quad \text{and}\quad
   \myarrayplaq{I}{A}{I}{I}
   + \myarrayplaq{I}{I}{I}{A}, \notag\\
 & \myarrayplaq{A}{I}{I}{B}
   - \myarrayplaq{I}{A}{B}{I},
   \quad \text{and}\quad
   \myarrayplaq{A}{B}{I}{I}
   + \myarrayplaq{I}{I}{B}{A}. \notag
\end{align}
The numbers of trivial solutions at different momenta are the same. At a momentum $\pmb{q}$, there are $28$ independent 4-site trivial solutions if the physical dimension of each site is $2$ and $153$ independent 4-site trivial solutions if the physical dimension of each site is $3$. 

Moreover, if one find a non-trivial solution satisfying $\hat{h}_{\pmb{x}}\ket{\Psi}\propto \ket{\Psi}$ and it commutes with the local terms of the trivial solutions: $[\hat{h}_{\pmb{x}},\hat{h}^{\text{tri}}_{\pmb{x}}]=0$, then $\sum_{\pmb{x}} \hat{h}_{\pmb{x}} \hat{h}^{\text{tri}}_{\pmb{x}}\ket{\Psi}\propto\sum_{\pmb{x}} \hat{h}^{\text{tri}}_{\pmb{x}}\ket{\Psi}=0$, so  $\sum_{\pmb{x}} \hat{h}_{\pmb{x}} \hat{h}^{\text{tri}}_{\pmb{x}}$ is also a solution. But the solutions $\sum_{\pmb{x}} \hat{h}_{\pmb{x}} \hat{h}^{\text{tri}}_{\pmb{x}}$ have the same physics as $\sum_{\pmb{x}} \hat{h}_{\pmb{x}}$ and we should exclude them.

\section{Deriving the relation between the static structure factor and the generating function}\label{App:derive_struct_factor}

We derive the following formula shown in the main text for evaluating the static structure factor without requiring the input iPEPS being normalized:
    \begin{align}
   S_{\alpha,\beta}(\pmb{q}) &=\sum_{\pmb{x}}e^{-i\pmb{q}\cdot\pmb{x}} \left[\frac{\bra{\Psi}\hat{o}_{\pmb{x}}^{\alpha}\hat{o}_{\pmb{0}}^{\beta}\ket{\Psi}}{\braket{\Psi}} - \frac{\bra{\Psi}e^{i\pmb{q}\cdot\pmb{0}}\hat{o}_{\pmb{x}}^{\alpha}\ket{\Psi}}{\braket{\Psi}}\frac{\bra{\Psi}e^{-i\pmb{q}\cdot\pmb{x}_q}\hat{o}_{\pmb{0}}^{\beta}\ket{\Psi}}{\braket{\Psi}}\right]  \notag \\
    &= \frac{\mathrm{d}}{\mathrm{d}\mu}\left[\frac{\bra{G^{\alpha}(\mu,\pmb{q})}\hat{o}_{\pmb{0}}^{\beta}\ket{\Psi}}{\braket{\Psi}} - \frac{\braket{G^{\alpha}(\mu,\pmb{q})}{\Psi}}{\braket{\Psi}}\frac{\bra{\Psi}\hat{o}_{\pmb{0}}^{\beta}\ket{\Psi}}{\braket{\Psi}}\right]  \bigg|_{\mu=0}  \notag \\
     &= \frac{\mathrm{d}}{\mathrm{d}\mu}\left[\frac{\bra{G^{\alpha}(\mu,\pmb{q})}\hat{o}_{\pmb{0}}^{\beta}\ket{\Psi}}{\braket{G^{\alpha}(\mu,\pmb{q})}{\Psi}}\frac{{\braket{G^{\alpha}(\mu,\pmb{q})}{\Psi}}}{\braket{\Psi}} - \frac{\braket{G^{\alpha}(\mu,\pmb{q})}{\Psi}}{\braket{\Psi}}\frac{\bra{\Psi}\hat{o}_{\pmb{0}}^{\beta}\ket{\Psi}}{\braket{\Psi}}\right]  \bigg|_{\mu=0}  \notag \\
     &= \frac{\mathrm{d}}{\mathrm{d}\mu}\frac{\bra{G^{\alpha}(\mu,\pmb{q})}\hat{o}_{\pmb{0}}^{\beta}\ket{\Psi}}{\braket{G^{\alpha}(\mu,\pmb{q})}{\Psi}}\bigg|_{\mu=0}.  \label{eq:derivative_derivation}
\end{align}
From the third line to the fourth line, we use the formula $\frac{\mathrm{d}}{\mathrm{d}x}[f(x)g(x)]=\frac{\mathrm{d}}{\mathrm{d}x}[f(x)]g(x)+f(x)\frac{\mathrm{d}}{\mathrm{d}x}[g(x)]$ to cancel the second term in the third line.
Moreover, when $\exp(\mathrm{i} \pmb{q}\cdot\pmb{x}_p)$ is real, i.e., $\pmb{q}=(0,0),(0,\pi),(\pi,0)$ or $(\pi,\pi)$, we can define a different generating function  $\bra{G^{\alpha}(\mu,\pmb{q})}=\bra{\Psi}\prod_{p}\exp(-\mathrm{i}\mu e^{-\mathrm{i}\pmb{q}\cdot\pmb{x}_p} O^{\alpha}_p)$, such that the iPEPO becomes it can be interpreted as a finite depth quantum circuit applying on $\ket{\Psi}$ and the generating function could be experimentally realized.

\begin{figure}
    \centering
    \includegraphics[width=\linewidth]{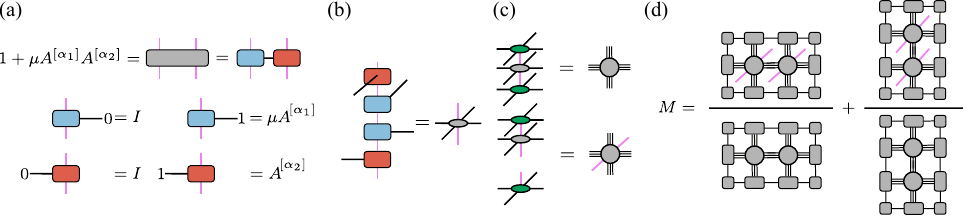}
    \caption{Schematics of the tensor network method for evaluating 2-site static structure factors. The black lines are the virtual indices and the purple lines are the physical indices. (a) Expressing the 2-site operator in the generating function in terms of a 2-site MPO. The non-zeros entries of the tensors are defined using the matrices.  (b) Obtaining iPEPO tensor from the MPO tensors. (c) Defining the triple-layer tensors with and without open physical indices. The upper and lower iPEPS tensors represented by the green ovals are from $\bra{\Psi}$ and $\ket{\Psi}$, respectively. (d) The matrix $M$ for evaluating one row of $S$, where the tensors represented by squares and rectangles are the corner and edge environment tensors obtained from CTMRG, respectively.}
    \label{fig:2_site_struct_factor}
\end{figure}
\begin{figure}
    \centering
    \includegraphics[width=\linewidth]{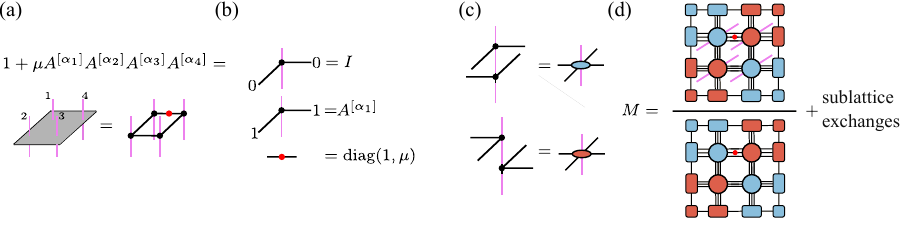}
    \caption{Schematics of the tensor network method for evaluating 4-site static structure factors. (a) Expressing the 4-site operator in the generating function in terms of a loop MPO.  (b) The definition of the non-zero entries of the MPO tensors at the left-top corner in terms of the matrices. The MPO tensors at other corners can be obtained similarity. (c) Obtaining the two iPEPO tensors from the MPO tensors.  (d) The matrix $M$ for evaluating one row of $S$. Notice that the tensor represented by the red dot can not be ignored when evaluating the CTMRG environment tensors.}
    \label{fig:4_site_struct_factor}
\end{figure}

\section{Relation between the static structure factor matrix and the fidelity suspectibility}

In this section, we derive the relation between the fidelity susceptibility and the static structure factor matrix. It is more convenient to consider a generating function that is slightly different from the one in the main text:
\begin{equation}
\bra{G(\pmb{\mu},\pmb{q})}=\bra{\Psi}\prod_{\pmb{x}}\exp(\mathrm{i}\sum_{\alpha}\mu_{\alpha} e^{\mathrm{i}\pmb{q}\cdot\pmb{x}} \hat{o}^{\alpha}_{\pmb{x}}),  
\end{equation}
where we generalize the scaler $\mu$ to a vector $\pmb{\mu}$ and we choose the iPEPO to be unitary such that the generating function is always normalized, $\braket{G(\pmb{\mu},\pmb{q})}{G(\pmb{\mu},\pmb{q})}=\braket{\Psi}{\Psi}=1$, as long as $\ket{\Psi}$ is normalized. 
Consider the Taylor expansion of the fidelity between the generating function and the original wavefunction: 
\begin{equation}
   \left|\braket{G(\pmb{\mu},\pmb{q})}{\Psi}\right|^2=1+\sum_{\alpha,\beta}\frac{\partial^2}{\partial \mu_\alpha\partial \mu_\beta}\left|\braket{G(\pmb{\mu},\pmb{q})}{\Psi}\right|^2\bigg|_{\pmb{\mu}=0} \frac{\mu_\alpha\mu_\beta}{2}+\cdots,
\end{equation}
where the first order term must be zero because of the relation~\cite{Fidelity_2020}:
\begin{equation}
    \frac{\partial}{\partial \mu_{\alpha}} \left|\braket{G(\pmb{\mu},\pmb{q})}{\Psi}\right|^2\bigg|_{\pmb{\mu}=0}= \frac{\partial}{\partial \mu_{\alpha}}\braket{G(\pmb{\mu},\pmb{q})}{G(\pmb{\mu},\pmb{q})}\bigg|_{\pmb{\mu}=0}=0.
\end{equation}
The coefficient of the second order term is called the \emph{quantum metric tensor}, and can be written as
\begin{align}
 &-\frac{\partial^2}{\partial \mu_\alpha\partial \mu_\beta}\left[\braket{G(\pmb{\mu},\pmb{q})}{\Psi}\braket{\Psi}{G(\pmb{\mu},\pmb{q})}\right]\bigg|_{\pmb{\mu}=0}\\
  =&\left[\braket{\frac{\partial}{\partial \mu_\alpha}G(\pmb{\mu},\pmb{q})}{\frac{\partial}{\partial \mu_\beta}G(\pmb{\mu},\pmb{q})}
  +\braket{\frac{\partial}{\partial \mu_\beta}G(\pmb{\mu},\pmb{q})}{\frac{\partial}{\partial \mu_\alpha}G(\pmb{\mu},\pmb{q})}
  -\braket{\frac{\partial}{\partial \mu_\alpha}G(\pmb{\mu},\pmb{q})}{\Psi}\braket{\Psi}{\frac{\partial}{\partial \mu_\beta}G(\pmb{\mu},\pmb{q})}
  -\braket{\frac{\partial}{\partial \mu_\beta}G(\pmb{\mu},\pmb{q})}{\Psi}\braket{\Psi}{\frac{\partial}{\partial \mu_\alpha}G(\pmb{\mu},\pmb{q})}\right]\bigg|_{\pmb{\mu}=0}\notag\\
  =&\sum_{i,j}\left[\bra{\Psi}e^{i\pmb{q}\cdot(\pmb{x}_i-\pmb{x}_j)}\hat{o}^{\alpha}_i\hat{o}^{\beta}_j\ket{\Psi}
  +\bra{\Psi}e^{i\pmb{q}\cdot(\pmb{x}_i-\pmb{x}_j)}\hat{o}^{\beta}_i\hat{o}^{\alpha}_j\ket{\Psi} -\bra{\Psi}e^{i\pmb{q}\cdot\pmb{x}_i}\hat{o}^{\alpha}_i\ket{\Psi}\bra{\Psi}e^{-i\pmb{q}\cdot\pmb{x}_j}\hat{o}^{\beta}_j\ket{\Psi}
  -\bra{\Psi}e^{i\pmb{q}\cdot\pmb{x}_i}\hat{o}^{\beta}_i\ket{\Psi}\bra{\Psi}e^{-i\pmb{q}\cdot\pmb{x}_j}\hat{o}^{\alpha}_j\ket{\Psi}\right]
    \notag\\
     =&N_{\text{sites}}\left[S_{\alpha,\beta}(\pmb{q})+S_{\beta,\alpha}(\pmb{q})\right]=2N_{\text{sites}}\mathcal{S}(\pmb{q}),
\end{align}
where $N_{\text{site}}$ is the number of sites and we have used the relation: 
\begin{align}
&\left[\braket{\frac{\partial^2}{\partial \mu_\alpha\partial \mu_\beta}G(\pmb{\mu},\pmb{q})}{G(\pmb{\mu},\pmb{q})}+\braket{G(\pmb{\mu},\pmb{q})}{\frac{\partial^2}{\partial \mu_\alpha\partial \mu_\beta}G(\pmb{\mu},\pmb{q})}+\braket{\frac{\partial}{\partial \mu_\alpha}G(\pmb{\mu},\pmb{q})}{\frac{\partial}{\partial \mu_\beta}G(\pmb{\mu},\pmb{q})}+\braket{\frac{\partial}{\partial \mu_\beta}G(\pmb{\mu},\pmb{q})}{\frac{\partial}{\partial \mu_\alpha}G(\pmb{\mu},\pmb{q})}\right]\bigg|_{\pmb{\mu}=0}=0.
\end{align}
Therefore the quantum metric tensor is proportional to the static structure factor matrix $\mathcal{S}$.
And the \emph{fidelity susceptibility} to tuning the parameter $\pmb{\mu}$ becomes~\cite{Fidelity_2020} 
\begin{equation}
\chi_F=N_{\text{sites}}\sum_{\alpha\beta}\frac{\mu_\alpha}{|\pmb{\mu}|}\frac{\mu_\beta}{|\pmb{\mu}|}\frac{S_{\alpha,\beta}(\pmb{q})+S_{\beta,\alpha}(\pmb{q})}{2}=N_{\text{sites}}\frac{\pmb{\mu}^\mathsf{T}\mathcal{S}\pmb{\mu}}{|\pmb{\mu}|^2},
\end{equation}
where $\frac{\pmb{\mu}}{|\pmb{\mu}|}$ is a unit vector. The \nameOs $\sum_{\pmb{x}}\sum_{\alpha}\mu_{\alpha} e^{\mathrm{i}\pmb{q}\cdot\pmb{x}} \hat{o}^{\alpha}_{\pmb{x}}$ are in the kernel of the static structure factor matrix $\mathcal{S}$, so we have $\chi_F=0$, which implies the fidelity is not sensitive to the deformation of the tensor network state $|\Psi\rangle$ generated by the \nameOs.

\section{Evaluating the static structure factor matrix $\mathcal{S}$ for multi-site operators from iPEPS}\label{App:iPEPS_gen_func}
In this section, we discuss the details on the calculation of the generating function in the last section by CTMRG and the differentiation of the generating function to obtain the structure factor matrix $\mathcal{S}$. In the following, we will choose $\pmb{q}=(0,0)$ as an illustration.

For the case of 1-site $\hat{o}_{\pmb{x}}^{\alpha}$, $\prod_{\pmb{x}}(\mathbbm{1}+\mu \hat{o}_{\pmb{x}}^{\alpha})$ is an on-site operator and it is straightforward to apply it to the iPEPS~\cite{Generating_iPEPS_2024}. For multi-site $\hat{o}_{\pmb{x}}^{\alpha}$, however, $\prod_{\pmb{x}}(\mathbbm{1}+\mu \hat{o}_{\pmb{x}}^{\alpha})$ is an iPEPO. To reduce the computational cost, it would be desirable to minimize the iPEPO bond dimension $D'$. Therefore, we choose $\hat{o}_{\pmb{x}}^{\alpha}$ to be a tensor product of on-site operators, i.e., Pauli (Gell-Mann) string when the physical bond dimension is $2$ ($3$). 

\subsection{Evaluating the generating function for the 2-site $\mathcal{S}$}
We choose the 2-site operator basis $\hat{o}^{\alpha}_{\pmb{x}}$ as $\hat{o}^{\alpha}_{\langle i,j\rangle}=A^{[\alpha_i]}_i\otimes A^{[\alpha_j]}_j$, with $\{A^{[\alpha_i]}_i\}$ being a complete orthonormal basis for Hermitian operators at site $i$. To express the operator $G^{\alpha}=\prod_{\langle i,j\rangle}(\mathbbm{1}+\mu A^{[\alpha_i]}_iA^{[\alpha_j]}_j)$ in term of an iPEPO, we first express the nearest-neighbor 2-site operator $\mathbbm{1}+\mu A^{[\alpha_i]}_iA^{[\alpha_j]}_j$ as a two-site matrix product operator (MPO) with bond dimension $D'=2$, as shown in Fig.~\ref{fig:2_site_struct_factor}a. Using the two-site MPO tensors, the iPEPO tensor for $G^{\alpha}$ with $D'=2$ can be constructed as in Fig.~\ref{fig:2_site_struct_factor}b. 

The denominator of the generating function for the static structure factor in Eq.~\eqref{eq:derivative_derivation} is a triple-layer tensor network defined by sandwiching the iPEPO tensor between the iPEPS $|\Psi\rangle$ and $\langle\Psi |$, and the numerator is the same tensor network except for the tensors at the origin have open physical legs, as shown in Fig.~\ref{fig:2_site_struct_factor}(c), which can be approximately contracted by using the environment tensors obtained from the single-site unit cell CTMRG. After the contraction, we obtain a $d^2\times d^2$ matrix $M^{\alpha}$, as shown in Fig.~\ref{fig:2_site_struct_factor}d. The static structure factors are obtained via $S_{\alpha,\beta}=\frac{\mathrm{d}}{\mathrm{d\mu}}\mathrm{Tr}\left[M^{\alpha}\left(A^{[\beta_1]}_1\otimes A^{[\beta_2]}_2\right)\right]\bigg|_{\mu=0}$.  

\subsection{Evaluating the generating function for the 4-site $\mathcal{S}$}
The idea of evaluating the generating function for the 4-site static structure factor is similar to that for the 2-site case, with only some details changed. We first express the operator $G^{\alpha}=\prod_{(i,j,k,l)\in p}\left(1+\mu \myarray{A^{[\alpha_i]}_i}{A^{[\alpha_j]}_j}{A^{[\alpha_k]}_k}{A^{[\alpha_l]}_l}\right)$ in terms of an iPEPO. We can express the 4-site operator in each plaquette $p$ in terms of a loop MPO with bond dimension $D'=2$, as shown in Fig.~\ref{fig:4_site_struct_factor}a, with the MPO tensors defined in Fig.~\ref{fig:4_site_struct_factor}b. Using the MPO tensors, two iPEPO tensors with bond dimension $D'=2$ can be constructed, as shown in Fig.~\ref{fig:4_site_struct_factor}c. A $d^4\times d^4$ matrix $M^{\alpha}$ can be obtained by contracting the tensor diagrams in Fig.~\ref{fig:4_site_struct_factor}d, where the triple-layer tensors are defined similarly to those in Fig.~\ref{fig:2_site_struct_factor}c and the environments of the triple-layer tensor network are obtained from CTMRG with a $2\times 2$ unit cell. The static structure factors are obtained via $S_{\alpha,\beta}=\frac{\mathrm{d}}{\mathrm{d\mu}}\mathrm{Tr}\left[M^{\alpha}\left(A^{[\beta_1]}_1\otimes A^{[\beta_2]}_2\otimes A^{[\beta_3]}_3\otimes A^{[\beta_4]}_4\right)\right]\bigg|_{\mu=0}$.

\subsection{Differentiation of the generating function}
\begin{figure}
    \centering
        \includegraphics[scale=0.5]{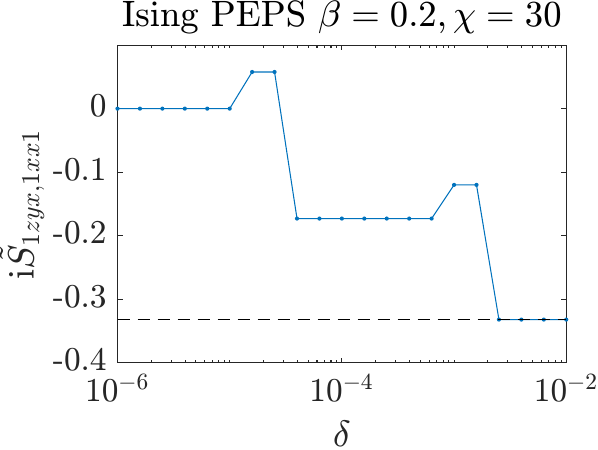}
    \caption{The static structure factor between two 4-site operators calculated from the Ising iPEPS with $\beta=0.2$, the CTMRG bond dimension is $\chi=30$. The dash line indicates the correct value of the structure factor.}
    \label{Fig:delta_influence}
\end{figure}

The derivative of the generating function with respect to $\mu$ can be in principle calculated using either the forward-mode or the backward-mode automatic differentiation (AD)~\cite{AD_2019}. However, we find that AD of Eq.~\eqref{eq:derivative_derivation} suffers from instability when applied to static structure factors of multi-site operators. The reason is that when $\mu=0$, the entries of the iPEPO tensors with the virtual index $1$ are exactly 0, as shown in Figs.~\ref{fig:2_site_struct_factor}(a,b) as well as Fig.~\ref{fig:4_site_struct_factor}(a,b). In the CTMRG iterations to contract the triple-layer tensor networks, we need to perform truncation in the singular value decomposition (SVD). At $\mu=0$, the matrix that will be SVDed has many rows and columns that are zeros, so many singular values will be exactly $0$. This is quite different from contracting the usual iPEPS where the sorted singular values from largest to smallest decay but not exactly zero, and we find that in our case this instability can not be fixed through approaches like Lorentz broadening~\cite{AD_2019} or the modified backward formula~\cite{Anna_2025}. Therefore, instead we use the central five-point formula for the derivative of a function $f(\mu)$:
\begin{equation}
    f'(\mu) = 
\frac{-f(\mu+2\delta) + 8f(\mu+\delta) - 8f(\mu-\delta) + f(\mu-2\delta)}{12\delta}+O(\delta^4),\notag
\end{equation}
where $\delta$ is the finite difference step. For the multi-site static structure factor, we should choose a suitable $\delta$ that is not too small, otherwise we can get wrong results, as shown in Fig.~\ref{Fig:delta_influence}, where we evaluate the structure factor of the two operators \myarrayplaq{I}{Z}{Y}{X} and \myarrayplaq{I}{X}{X}{I} from the Ising iPEPS at $\beta=0.2$ using CTMRG with $\chi=30$.

\subsection{Discussion of the cost and comparison with other contraction methods}
In the above method, the most expensive part is the CTMRG, and we need four different CTMRG contraction diagrams to calculate one $M$ matrix. Given an $M$ matrix we can easily obtain one row of the static structure factor matrix via taking the trace of the product between the $M$ matrix and the local basis operators at the origin. The number of calls for CTMRG contractions is therefore proportional to the size of the rows of the static structure factor matrix: $(d^2)^{k}$, where $d$ is the physical dimension and $k$ is the size of the support.

\begin{figure}[h]
    \centering
        \includegraphics[scale=0.6]{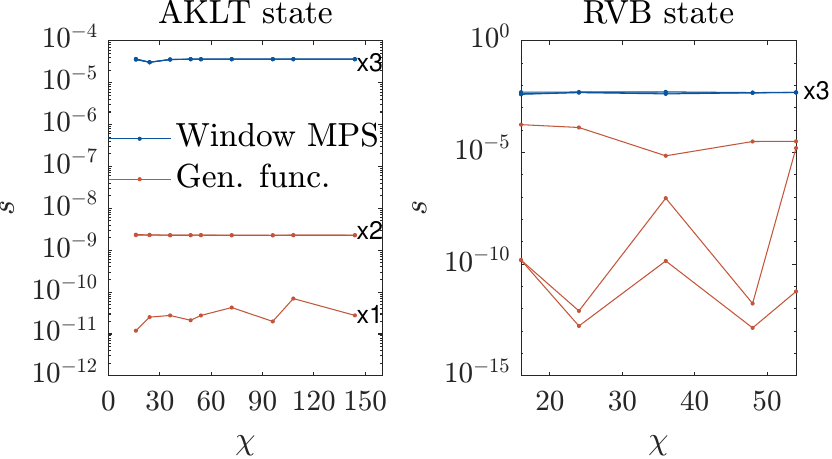}
    \caption{Compare the lowest three eigenvalues of the static structure factor matrix spectra of the AKLT (RVB) state calculated using the window MPS method and the generating function method, where $\chi$ is the environment bond dimension. The trivial solution (identity matrix) has been excluded. We mark the degeneracies on the right when levels are too close to each other.}
    \label{Fig:compare_win_MPS_gen_func}
\end{figure}
Additionally, we also compare the spectra of the 1-site static structure factor matrices evaluated using the generating function method and the window MPS method, as shown in Fig.~\ref{Fig:compare_win_MPS_gen_func}. The results indicate that the generating function method is more accurate than the window MPS method. However, for the critical RVB iPEPS, although the performance of the generating function method is better, both methods cannot find all three SU(2) symmetry generators to the same high precision.

\section{Parent Hamiltonians of the AKLT state on the square lattice}\label{app:AKLT_parent_Ham} 
In this section, we first give the form of the PEPS tensor of the 2D AKLT state and then discuss all of its 2-site frustration-free parent Hamiltonians.

Because the 2D spin-2 AKLT state on a square lattice can be viewed as putting many 1D spin-1 AKLT states along the square lattice both vertically and horizontally, and fusing two spin-1's of the 1D AKLT states at the same physical site to the spin-2 of the 2D AKLT state at the site, the 2D AKLT state can be exactly expressed as a PEPS whose tensor has the form
\begin{equation}
    T^i_{lurd}=\sum_{j,k}\braket{2,i}{1,j;1,k}\braket{1,j}{1/2,l;1/2,r}\braket{1,k}{1/2,u;1/2,d}
\end{equation}
on the vertices and spin-$1/2$ singlet on the edges, where $i$ is spin-2 physical index and $l,u,r,d$ are left, up, right, down spin-1/2 virtual indices. For the states $\ket{J_1,m_1}$ and $\ket{J_1,m_1;J_2,m_2}$, $J_i$ ($m_i$) denotes the quantum number for the spin ($S^z$) angular momentum, and the overlaps are simply the Clebsch–Gordan coefficients (the last two are the 1D AKLT MPS tensors). The non-zeros entries the tensor $T$ can be found in Tab.~\ref{tab:AKLT_tensor}.

\begin{table}[h]
\centering
\caption{The non-zero entries of the $S=2$ AKLT PEPS tensor on the square lattice.}

\setlength{\tabcolsep}{1.5em} 

\begin{tabular}{c c c c c}
\hline\hline
$S_z=+2$ & $S_z=+1$ & $S_z=0$ & $S_z=-1$ & $S_z=-2$ \\
\hline
$T_2(0,0,0,0)=1$
&
\(
\begin{aligned}
T_1(0,0,0,1)&=\tfrac12\\
T_1(0,0,1,0)&=\tfrac12\\
T_1(0,1,0,0)&=\tfrac12\\
T_1(1,0,0,0)&=\tfrac12
\end{aligned}
\)
&
\(
\begin{aligned}
T_0(0,0,1,1)&=\tfrac{1}{\sqrt6}\\
T_0(0,1,0,1)&=\tfrac{1}{\sqrt6}\\
T_0(0,1,1,0)&=\tfrac{1}{\sqrt6}\\
T_0(1,0,0,1)&=\tfrac{1}{\sqrt6}\\
T_0(1,0,1,0)&=\tfrac{1}{\sqrt6}\\
T_0(1,1,0,0)&=\tfrac{1}{\sqrt6}
\end{aligned}
\)
&
\(
\begin{aligned}
T_{-1}(0,1,1,1)&=\tfrac12\\
T_{-1}(1,0,1,1)&=\tfrac12\\
T_{-1}(1,1,0,1)&=\tfrac12\\
T_{-1}(1,1,1,0)&=\tfrac12
\end{aligned}
\)
&
$T_{-2}(1,1,1,1)=1$
\\[1em]
\hline
\end{tabular}\label{tab:AKLT_tensor}
\end{table}
The parent Hamiltonian of the AKLT state can be constructed by utilizing the following fusion rule 
\begin{equation}\label{eq:fusion_rule}
   2\otimes 2=\bigoplus_{S_{\text{tot}=0}}^{4}S_{\text{tot}}.
\end{equation}
Because two nearest neighboring sites contain 8 virtual spin-$1/2$'s, and the two virtual spin-$1/2$'s at the bond between the two sites fuse to spin-$0$ (singlet), the total spin of the two sites are determined by the rest 6 spin-1/2's which can fuse at most to spin-3, and hence the fusion channel $S_{\text{tot}}=4$ is forbidden.
Therefore, all 2-site frustration free parent Hamiltonians of the 2D spin-2 AKLT state can be expressed as:
\begin{align}\label{eq:2-site AKLT parent H}
    H(\alpha,M) &= \sum_{\langle i,j\rangle} U_{i,j} P(\alpha,M) U_{i,j}^{\dagger},\quad 
    U_{i,j} = \sum_{S_{\text{tot}}=0}^{4} \sum_{S^z_{\text{tot}}=-S_{\text{tot}}}^{S_{\text{tot}}}
    \sum_{S^z_i=-2}^{2} \sum_{S^z_j=-2}^{2}
    \ket{2,S^z_i;2,S^z_j} \langle2,S^z_i,2,S^z_j | S_{\text{tot}}, S^z_{\text{tot}}\rangle \bra{S_{\text{tot}}, S^z_{\text{tot}}}, \notag\\
    P(\alpha,M) &= \alpha \sum_{S_{\text{tot}}=0}^{3} \sum_{S_{\text{tot}}^z=-S_{\text{tot}}}^{S_{\text{tot}}} \ket{S_{\text{tot}}, S_{\text{tot}}^z} \bra{S_{\text{tot}}, S_{\text{tot}}^z}
    + \sum_{S_{\text{tot}}^z=-4}^{4} \sum_{S_{\text{tot}}^{z\prime}=-4}^{4} M_{S_{\text{tot}}^{z\prime},S_{\text{tot}}^z} \ket{4, S_{\text{tot}}^{z\prime}} \bra{4, S_{\text{tot}}^z},
\end{align}
where $M_{S_{\text{tot}}^{z\prime},S_{\text{tot}}^z}$ is an arbitrary Hermitian matrix. $\alpha$ can be any non-negative number, because it is the energy penalty for the fusion channels $S_{\text{tot}}=0,1,2,3$. $U_{i,j}$ is a unitary transformation which performs the basis transformation from two spin 2's at the sites $i$ and $j$ to the total spin $S_{\text{tot}}$, where $\langle2,S^z_i,2,S^z_j | S_{\text{tot}}, S^z_{\text{tot}}\rangle$ is nothing but the Clebsch–Gordan coefficient. Excluding the identity matrix, there are 80 linear indenpendent $M_{S_{\text{tot}}^{z\prime},S_{\text{tot}}^z}$. Taking $\alpha$ into account, there are $81$ non-trivial 2-site Hamiltonians for the AKLT state, which is consistent with the dimension of the kernel space of $\mathcal{S}$. Notice that if $P$ is semi-positive definite, the AKLT state is the ground state, otherwise it could be a quantum many-body excited scar state~\cite{Sanjay_2020}.   

\section{iPEPS study of the phase transition of the XX model in the staggered $Z$ field}

To approximate the ground state of the XX model in the staggered $Z$ field using 1-site unit cell iPEPS, one can perform a sublattice rotation $U=\prod_{i\in \text{odd sublattice}}X_i$ such that the Hamiltonian becomes $ H=-J\sum_{\langle i,j \rangle}(X_iX_j -Y_iY_j)-(1-J)\sum_{i} Z_i$. In the ground state optimization, we use $C_{4v}$ symmetric iPEPS tensors, i.e., it is invariant under the $C_4$ rotation (exchanging the left (up) and right (down) virtual legs and exchanging the left and down, right and up legs as well as exchanging the left and up, right and down legs). 

\begin{figure}
    \centering
    \includegraphics[width=0.8\linewidth]{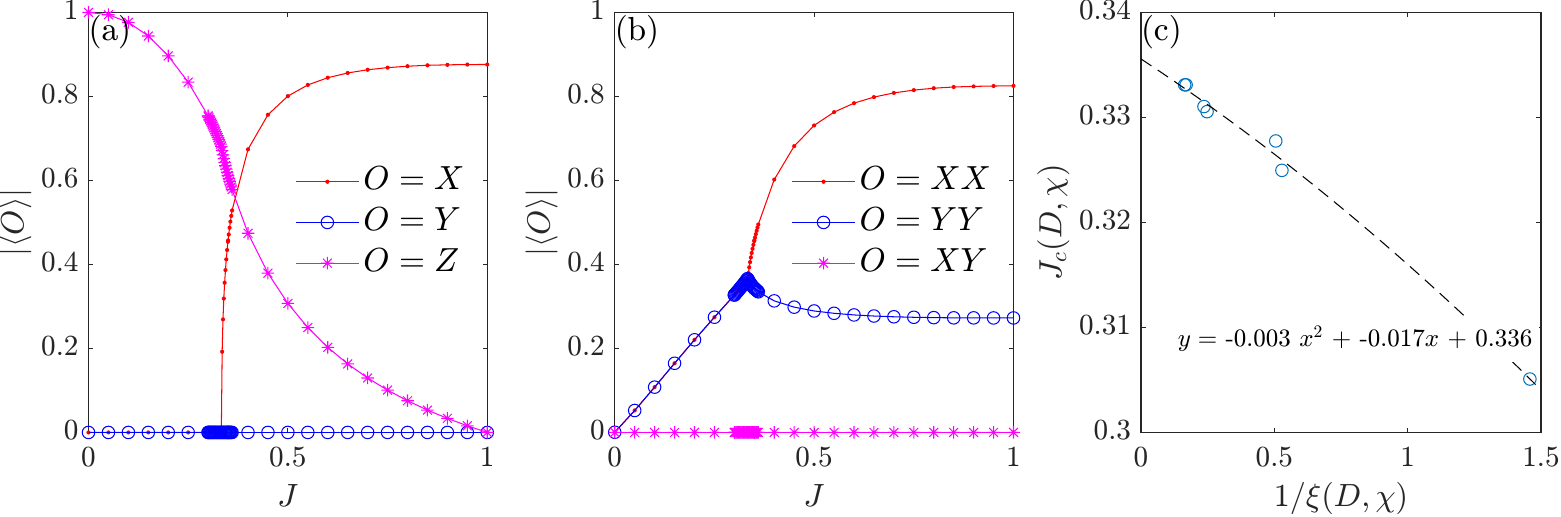}
    \caption{Results from the iPEPS optimization of the XX model in the staggered $Z$ field. (a) The expectation values of the on-site operators from the iPEPS with $(D,\chi)=(5,100)$. (b) The expectation values of two nearest neighboring operators $(D,\chi)=(5,100)$. (c) Extrapolate the position of the critical point, where $J_c(D,\chi)$ is the critical point determined by the order parameter evaluated from the iPEPS with bond dimensions $(D,\chi)$. $\xi(D,\chi)$ is the correlation length of the iPEPS with bond dimensions $D$ at $J_c(D,\chi)$ extracted using CTMRG with environment bond dimension $\chi$, where $(D,\chi)$ we have used are shown in the $x$ axis of Fig. 3b in the main text.}
    \label{fig:XX_app}
\end{figure}

The order parameter detecting the U(1) symmetry breaking is $\sqrt{\langle X\rangle^2+\langle Y\rangle^2}$. In the U(1) symmetry breaking phase, although spin can point towards any direction in the $x-y$ plane, we use real iPEPS tensor such that the spin can only spontaneously point to the $x$ direction, i.e., $\langle X\rangle\neq 0$ and $\langle Y\rangle=0$. Fig.~\ref{fig:XX_app}a shows the expectation values of the local operators, where $|\langle X\rangle|$ can be viewed as an order parameter which signals the spontaneous U(1) symmetry breaking. Fig.~\ref{fig:XX_app}b shows the expectation values of the geometrically 2-local operators, where $|\langle XX-YY\rangle|$ can be viewed as an alternative order parameter characterizing the spontaneous U(1) symmetry breaking. Fig.~\ref{fig:XX_app}c shows extrapolation of the critical point from results of iPEPS simulations with various bond dimensions, the value of the critical point we obtained, $J_c\approx 0.336$, is consistent with $J_c\approx 0.35$ obtained using density matrix renormalization group method in Ref.~\cite{XX_quantum_simulator_2024}.

\section{Supplementary results for the RVB state}
In this section, we give the form of the PEPS tensor for the short-range RVB state, and discuss how we reduce the size of the static structure factor matrix by imposing the SU(2) symmetry of the RVB state, as well as present the results from exactly contracting a finite periodic PEPS.
 
The RVB state can be exactly expressed as a PEPS with $d=2$ (spin-$1/2$: $\ket{\uparrow}$ and $\ket{\downarrow}$) and $D=3$ [spin $0\oplus 1/2$: $|0)$, $|\uparrow)$ and $|\downarrow)$]~\cite{RVB_PEPS_2012,systematic_SU2_PEPS_2016}, with the PEPS tensors at the vertices being
\begin{equation}
    \ket{\uparrow}\left[(\uparrow000|+(0\uparrow00|+(00\uparrow0|+(000\uparrow|\right]\\+\ket{\downarrow}\left[(\downarrow000|+(0\downarrow00|+(00\downarrow0|+(000\downarrow|\right]  
\end{equation}    
and the tensors at the edges being a singlet $|\uparrow\downarrow)-|\downarrow\uparrow)$~\cite{systematic_SU2_PEPS_2016}. Similar to the AKLT state, the SU(2) symmetry generators can be found from the 1-site solution of the structure factor matrix from both the periodic PEPS and the iPEPS. For the RVB state, all entries of the $4\times 4$ 1-site structure factor matrix $\mathcal{S}$ are zero, implying four solutions of \nameOsend: one is the identity operator and the other three are the three Pauli matrices.

It has been pointed out that the only possible SU(2) symmetric geometrically local operators consist of the Heisenberg terms and the 3-spin terms $\pmb{S}_i\cdot(\pmb{S}_j\times \pmb{S}_k)$, as well as their non-overlapping products~\cite{Locality_optim_2022}. Because the RVB state is real and the 3-spin interactions $\pmb{S}_i\cdot(\pmb{S}_j\times \pmb{S}_k)$ are purely imaginary, the 3-spin interactions $\pmb{S}_i\cdot(\pmb{S}_j\times \pmb{S}_k)$ cannot appear in the parent Hamiltonian of the RVB state. Notice that the only possible SU(2) symmetric 2-site solution is the Heisenberg model Hamiltonian, and there is no parameter to be tuned. It was also confirmed numerically that the ground state of the antiferromagnetic Heisenberg model on the square lattice is the Neel state~\cite{Sandvik_heisenberg_1997}, not the RVB state. So we only target for SU(2) symmetric 4-site-plaquette conserved operators, i.e.
 \begin{align}\label{eq:H_JJQQ_app}
    \hat{H}(J_1,J_2,Q_1,Q_2)=J_{1}\sum_{\langle ij\rangle}\pmb{S}_i\cdot \pmb{S}_j+J_{2}\sum_{\langle \langle ij\rangle\rangle}\pmb{S}_i\cdot \pmb{S}_j
    +Q_{1}\sum_p\sum_{\langle ij\rangle,\langle kl\rangle\in p }(\pmb{S}_i\cdot \pmb{S}_j)(\pmb{S}_k\cdot \pmb{S}_l)
    +Q_{2}\sum_p\sum_{\langle\langle ij\rangle\rangle,\langle\langle kl\rangle\rangle\in p}(\pmb{S}_i\cdot \pmb{S}_j)(\pmb{S}_k\cdot \pmb{S}_l).
\end{align}
We further decompose these terms into linear combinations of 39 operator string basis
\begin{align}\label{eq:SU_2_basis}
 \myarrayplaq{S^\alpha}{I}{I}{S^\alpha},\quad \myarrayplaq{S^\alpha}{I}{S^\alpha}{I},\quad 4\myarrayplaq{ S^\alpha}{ S^\alpha}{ S^\beta}{ S^\beta},\quad 4\myarrayplaq{ S^\alpha }{ S^\beta }{ S^\alpha }{ S^\beta }.
\end{align}
where $\alpha,\beta=x,y,z$ and all $C_{4v}$ rotations of above terms should be taking into consideration. From the above basis we get the best approximate solution whose corresponding eigenvector entries are $\tilde{J}_1,\tilde{J}_2,\tilde{Q}_1,\tilde{Q}_2$ for the four terms in Eq.~\eqref{eq:H_JJQQ_app}. When we use CTMRG to evaluate the static structure factors, the $C_{4v}$ symmetry might be slightly broken due to the approximation, so the coefficients should be understood as an average over 4-site terms transformed among each other under the $C_{4v}$ rotations. Therefore we can rewrite the solution as  
\begin{align}
    2\tilde{J}_1\sum_{\alpha,\beta}\myarrayplaq{S^\alpha}{I}{I}{S^\alpha}
    +\tilde{J}_2\sum_{\alpha,\beta}\myarrayplaq{S^\alpha}{I}{S^\alpha}{I}
    +4\tilde{Q}_1\sum_{\alpha,\beta}\myarrayplaq{S^\alpha}{S^\alpha}{S^\beta}{S^\beta}
    +4\tilde{Q}_2\sum_{\alpha,\beta}\myarrayplaq{S^\alpha}{S^\beta }{S^\alpha}{S^\beta}
    \,+ \,\text{all } C_{4v}\,\,\text{rotations}.
\end{align}
Because two neighboring plaquettes share a pair of nearest-neighbor sites, $\tilde{J}_1$ should be multiplied by an extra factor of $2$. Expressing the Hamiltonian in units of $J_1$ then gives  
$
J_2 = \frac{\tilde{J}_2}{2\tilde{J}_1}, \quad 
Q_1 = \frac{4\tilde{Q}_1}{2\tilde{J}_1}, \quad 
Q_2 = \frac{4\tilde{Q}_2}{2\tilde{J}_1}
$ in the Hamiltonian in Eq.~\eqref{eq:H_JJQQ_app}.
Under this convention of normalization, the quantum fluctuation per site of the \nameO in units of $J_1$ and the lowest eigenvalue $s_{\text{min}}$ of $\mathcal{S}$ are related via  $\left(\langle \hat{H}^2\rangle-\langle \hat{H}\rangle^2\right)/N_{\text{site}}=s_0/\!\big[(2\tilde{J}_1)^2\big]$, where $N_{\text{site}}$ is the number of the sites in the system.

In the main text, we have shown the results from approximately contracted iPEPS. Here we also show results from exactly contracted $4\times 4$ normalized periodic PEPS $\ket{\text{RVB}}$. Excluding trivial solutions, the lowest eigenvalue $s_{\text{min}}\approx 0.58\times 10^{-3}$ is close to zero, and the corresponding eigenvector gives a Hamiltonian $\hat{H}_0\equiv \hat{H}(J_1,J_2,Q_1,Q_2)$ in the form of Eq.~\eqref{eq:H_JJQQ_app} with $J_1=1$, $J_2\approx 0.3317$, $Q_1\approx -0.1698$ and $Q_2\approx 0.3562$. Since a small eigenvalue of the structure factor matrix implies small quantum fluctuation per site $(\Delta E_{\text{RVB}})^2=\left(\bra{\text{RVB}}\hat{H}_0^2\ket{\text{RVB}}-\bra{\text{RVB}}\hat{H}_0\ket{\text{RVB}}^2\right)/N_{\text{site}}= 2.154\times 10^{-3}$, $\hat{H}_0$ has $\ket{\text{RVB}}$ as an eigenstate approximately. Moreover, the ground state energy per site of $\hat{H}_0$ is $E_{\text{GS}}=-0.6261$, which is very close to the energy expectation value per site in the RVB state $E_{\text{RVB}}=\bra{\text{RVB}}\hat{H}_0\ket{\text{RVB}}/N_{\text{site}}=-0.6258$, so $\ket{\text{RVB}}$ is a good approximation of the ground state of $\hat{H}_0$, and meanwhile $\hat{H}_0$ serves as a good approximate parent Hamiltonian of $\ket{\text{RVB}}$. Nevertheless, since the parent Hamiltonian we found is approximate and non-frustration-free, it will depend on the system size and we need to extrapolate the coefficients $J_1,J_2,Q_1$ and $Q_2$ to the thermodynamic limit.

\section{Supplementary results for the deformed Ising wavefunctions}
In this section, we prove that $\ket{\Psi_{\text{Ising}}(\beta)}$ for all $\beta$ are in the null space of $\mathcal{S}$. We also show some evidence indicating that $\ket{\Psi(\beta)}$ for all $\beta$ might be quantum many-body scar states for the Hamiltonian we found. It is known that the family of the deformed Ising wavefunctions $\ket{\Psi(\beta)}$ is dual to the deformed toric code states, and we further discuss a dual model of the Hamiltonian we found.

A fact worth mentioning is that the norm  $\braket{\Psi_{\text{Ising}}}$ is the partition function of the 2D classical Ising model, such that the entries of $\mathcal{S}$ which correspond to the operators $Z$ and $Z\otimes Z$ are actually related to the magnetic susceptibility per site and the specific heat capacity per site of the classical Ising model~\cite{Fidelity_toric_code}, respectively. Since it is well known that these observables diverge at the critical point of classical Ising model, it indicates that some entries of $\mathcal{S}$ of critical iPEPS can diverge in the thermodynamic limit. Nevertheless, the \nameO we find is still well defined at the critical point, this is because in practice we always solve the eigenequation of $\mathcal{S}$ for different finite system sizes or finite bond dimensions, and then we can extrapolate the eigenvalues and eigenvectors to the thermodynamic limit.

\subsection{Proof on the exactly \nameOs of the deformed Ising wavefunctions}\label{app:Ising_PEPS}
In this section, we prove the numerically found \nameOs of the deformed Ising wavefunction are exact. We show the zero momentum case $\pmb{q}=(0,0)$ and other cases can be proved similarly. 

Since the deformed Ising wavefunction is defined as $\ket{\Psi_{\text{Ising}}}=\exp\left(\beta\sum_{\langle i,j\rangle}Z_iZ_j\right)\ket{++\cdots+}$, we only need to show $[\sum_{\pmb{x}}\hat{h}_{\pmb{x}},\sum_{\langle ij\rangle}Z_iZ_j]=0$ and $\sum_{\pmb{x}}\hat{h}_{\pmb{x}}\ket{\Psi}=0$. The solution found directly from the 4-site plaquette $\mathcal{S}$ is
\begin{align}\label{eq:momentun_0}
    \hat{h}'_{\pmb{x}}&=2\myarrayplaq{Z}{I}{Z}{I}-2\myarrayplaq{I}{Z}{I}{Z}-\myarrayplaq{Z}{X}{Z}{I}-\myarrayplaq{Z}{I}{Z}{X}+\myarrayplaq{X}{Z}{I}{Z}+\myarrayplaq{I}{Z}{X}{Z}\notag\\
    &=\myarrayplaq{Z}{P_{-}}{Z}{I}+\myarrayplaq{Z}{I}{Z}{P_{-}}-\myarrayplaq{P_{-}}{Z}{I}{Z}-\myarrayplaq{I}{Z}{P_{-}}{Z},
\end{align}
where $P_{-}=I-X$.

Let us first prove $[\sum_{\pmb{x}}\hat{h}'_{\pmb{x}},\sum_{\langle ij\rangle}Z_iZ_j]=0$. Look at the first line of the equation above. The first and the second terms of $\hat{h}'_{\pmb{x}}$ obviously commute with $\sum_{\langle ij\rangle}Z_iZ_j$. We group the third and the last terms as $\hat{h}'_{1,\pmb{x}}=-\myarrayplaq{Z}{X}{Z}{I}+\myarrayplaq{I}{Z}{X}{Z}$, and the forth and fifth terms as $\hat{h}'_{2,\pmb{x}}=-\myarrayplaq{Z}{I}{Z}{X}+\myarrayplaq{X}{Z}{I}{Z}$. One can check that  $[\sum_{\pmb{x}}\hat{h}'_{1,\pmb{x}},\sum_{\langle ij\rangle}Z_iZ_j]=0$ because of the following relations
\begin{align}\label{eq:commutator}
    \left[\myarraysix{I}{I}{I}{Z}{Z}{I},\myarraysix{I}{Z}{I}{I}{X}{Z}\right]&=2\times\myarraysix{I}{Z}{I}{Z}{ZX}{Z},\quad
     \left[\myarrayplaq{I}{Z}{Z}{I},\myarrayplaq{Z}{X}{Z}{I}\right]=2\times\myarrayplaq{Z}{ZX}{I}{I},\notag\\
      \left[\myarraysix{I}{I}{I}{I}{Z}{Z},\myarraysix{I}{Z}{I}{Z}{X}{I}\right]&=2\myarraysix{I}{Z}{I}{Z}{ZX}{Z},\quad
            \left[\myarrayplaq{I}{Z}{Z}{I},\myarrayplaq{I}{Z}{X}{Z}\right]=2\myarrayplaq{I}{I}{ZX}{Z},\notag\\
\end{align}
Similarly, one can show that $[\sum_{\pmb{x}}\hat{h}'_{2,\pmb{x}},\sum_{\langle ij\rangle}Z_iZ_j]=0$. Therefore, we have proved $[\sum_{\pmb{x}}\hat{h}'_{\pmb{x}},\sum_{\langle ij\rangle}Z_iZ_j]=0$. Then, it is easy to show that $\sum_{\pmb{x}}\hat{h}'_{\pmb{x}}\ket{++\cdots+}=0$ because $X$ is fixed to $1$. Hence, we have proved $\sum_{\pmb{x}}\hat{h}'_{\pmb{x}}\ket{\Psi_{\text{Ising}}}=0$ and the conserved operator is exact. 

We can actually regroup the terms in the Hamiltonian $\sum_{\pmb{x}}\hat{h}'_{\pmb{x}}$ to get new local term around a vertex
\begin{align}\label{eq:momentun_0}
    \hat{h}_{\pmb{x}}&=\myarraynineur{Z}{P_{-}}{Z}+\myarraynineld{Z}{P_{-}}{Z}-\myarrayninerd{P_{-}}{Z}{Z}-
   \myarraynineul{Z}{Z}{P_{-}}.
\end{align}
such that $\sum_{\pmb{x}}\hat{h}'_{\pmb{x}}=\sum_{\pmb{x}}\hat{h}_{\pmb{x}}$. From Eq.~\eqref{eq:commutator} it can be found that $[\hat{h}_{\pmb{x}},\sum_{\langle i,j\rangle}Z_iZ_j]=0, \forall \pmb{x}$. This also implying a sufficient condition for frustration-freeness~\cite{Sanjay_2020}, i.e., applying $\hat{h}_{\pmb{x}}$ on the blocked PEPS tensors annihilates it:
\begin{equation}
    \vcenter{\hbox{\includegraphics[width=4cm]{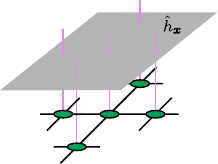}}}=0.
\end{equation}
However applying $\hat{h}'_{\pmb{x}}$ on a $2\times 2$-blocked PEPS tensor does not annihilates it.  Therefore $\hat{h}_{\pmb{x}}$ can be found using the conventional parent Hamiltonian constructions while $\hat{h}'_{\pmb{x}}$ can not.

In addition, since $\{\hat{h}'_{\pmb{x}},P\}=0$, where $P$ is the spatial inversion along the horizontal or vertical axis, the system has a ``particle-hole'' symmetry and the energy spectrum of $\sum_{\pmb{x}}\hat{h}_{\pmb{x}}$ and $-\sum_{\pmb{x}}\hat{h}_{\pmb{x}}$ must be the same, so the ground state energy of  $\sum_{\pmb{x}}\hat{h}_{\pmb{x}}$ must negative. But $P\ket{\Psi_{\text{Ising}}}=\ket{\Psi_{\text{Ising}}}$ implies  $\bra{\Psi_{\text{Ising}}}\hat{h}'_{\pmb{x}}\ket{\Psi_{\text{Ising}}}$ must be $0$. Hence $\ket{\Psi_{\text{Ising}}(\beta)}$ are not the ground states of the \nameO but it is an excited state lying in the middle of its energy spectrum. 
Indeed, the \nameO we found is not the usual parent Hamiltonian of the deformed Ising wavefunctions:
\begin{equation}
    H_{\text{parent}}(\beta)=-\sum_iX_i+\sum_i \exp\left(\beta\sum_{\{j|\langle i,j\rangle\}} Z_iZ_j\right),
\end{equation}
which involves 5-site terms. And it does not commute with $\sum_{\pmb{x}}\hat{h}_{\pmb{x}}$ we found because $[\sum_{\pmb{x}}\hat{h}_{\pmb{x}},\sum_iX_i]\neq0$.

Additionally, non-trivial solutions have the PEPS as excited state can be found from structure factor matrix at other momenta. At momentum $\pmb{q}=(\pi,\pi)$, we get 
\begin{align}
    \hat{h}_{\pmb{x}}''=2&\myarrayplaq{Z}{I}{Z}{I}+2\myarrayplaq{I}{Z}{I}{Z}-\myarrayplaq{Z}{X}{Z}{I}\notag-\myarrayplaq{Z}{I}{Z}{X}-\myarrayplaq{X}{Z}{I}{Z}-\myarrayplaq{I}{Z}{X}{Z}=\myarrayplaq{Z}{P_{-}}{Z}{I}+\myarrayplaq{Z}{I}{Z}{P_{-}}+\myarrayplaq{P_{-}}{Z}{I}{Z}+\myarrayplaq{I}{Z}{P_{-}}{Z}
\end{align}
At $\pmb{q}=(\pi,0)$, we get
\begin{align}
    \hat{h}_{\pmb{x}}'''=\myarrayplaq{Z}{X}{Z}{I}-\myarrayplaq{Z}{I}{Z}{X}-\myarrayplaq{X}{Z}{I}{Z}+\myarrayplaq{I}{Z}{X}{Z}=-\myarrayplaq{Z}{P_{-}}{Z}{I}+\myarrayplaq{Z}{I}{Z}{P_{-}}+\myarrayplaq{P_{-}}{Z}{I}{Z}-\myarrayplaq{I}{Z}{P_{-}}{Z}.
\end{align}
By regrouping the local terms in the Hamiltonian $\sum_{\pmb{x}}\exp(i\pmb{q}\cdot\pmb{x})\hat{h}''_{\pmb{x}}$ with $\pmb{q}=(\pi,\pi)$ and $\sum_{\pmb{x}}\exp(i\pmb{q}\cdot\pmb{x})\hat{h}'''_{\pmb{x}}$ with $\pmb{q}=(\pi,0)$, it can be checked that the local terms are the same as $\hat{h}_{\pmb{x}}$. So actually the solutions at different momenta are all resulted from $\hat{h}_{\pmb{x}}\ket{\Psi_{\text{Ising}}}=0$. 

\subsection{Spectrum of the non-trivial solution}
\begin{figure}
    \centering
    \includegraphics[width=0.5\linewidth]{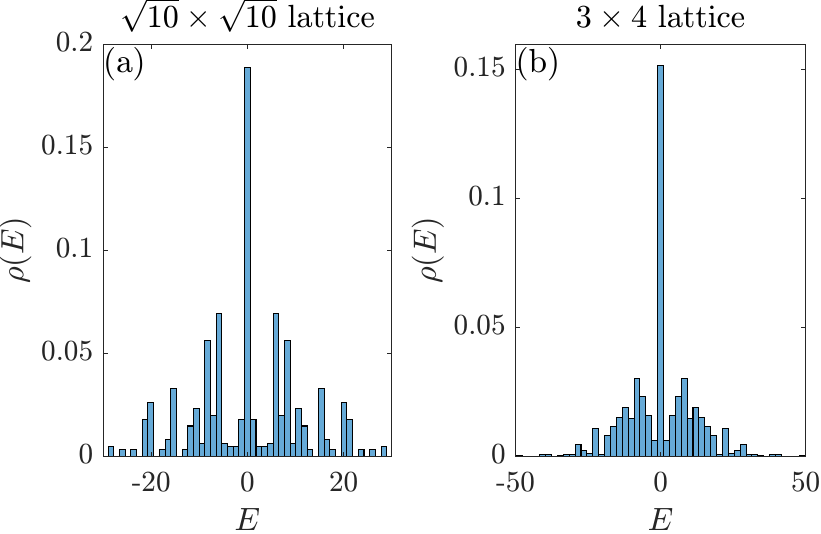}
    \caption{Density of states $\rho (E)$ obtained from the energy spectrum $E$ of $\sum_{\pmb{x}}\hat{h}_{\pmb{x}}$ on a torus, where the numbers of bins are $50$. (a) The $\sqrt{10}\times\sqrt{10}$ lattice which is bipartite. (b) The $3\times4$ lattice which is non-bipartite.}
    \label{fig_app:scar}
\end{figure}

To determine if the deformed Ising wavefunctions lying in the middle of the spectrum of $\sum_{\pmb{x}}\hat{h}_{\pmb{x}}$ are quantum many body scars, we should exclude the possibility that $\sum_{\pmb{x}}\hat{h}_{\pmb{x}}$ is an integrable model. It should be remarked that integrable models in 2D are rare, and the usual ones are free fermion systems or the stablizer codes. Nevertheless, decisive evidence of non-integrability should still be found from symmetry resolved level statistics of systems with more than 20 spins~\cite{scar_2018}, which is beyond the scope of this work and we leave it for future study. 

As an indication, in Fig.~\ref{fig_app:scar}, we show the density of states (DOS) obtained from the full energy spectrum of $\sum_{\pmb{x}}\hat{h}_{\pmb{x}}$ in a small system. It can be seen that the DOS is very large at $E=0$ (1160 zero energy eigenstates among all 4096 eigenstates for the $3\times 4$ lattice and 196 zero energy eigenstates among all 1024 eigenstates for the $\sqrt{10}\times \sqrt{10}$ lattice), which resembles the models proposed for quantum many-body scars~\cite{scar_2018,Scar_null_space_2021}.

\subsection{Dual model of the non-trivial solution}
For a system with a global internal $\mathbb{Z}_2$ symmetry, we can gauge it to obtain some $\mathbb{Z}_2$ lattice gauge theories~\cite{Wegner_duality_1971}. A well-known example is the Wegner duality between the transverse field Ising model and the toric code model in a $Z$ or $X$ field~\cite{Trebst_2007}. The Wegner duality maps $\mathbb{Z}_2$ spins on the vertices of the square lattice to $\mathbb{Z}_2$ spins on the edges of the lattice in the following way
\begin{equation}
    \myarrayninem{X}\mapsto\myvertex{X}{X}{X}{X},\quad  \myarraynineulnom{Z}{Z}\mapsto\myvertex{Z}{Z}{I}{I}.
\end{equation}
After the duality transformation, the Hilbert space has been enlarged but it is restricted by the gauge constraint $B_p=\myplaqgauge{Z}{Z}{Z}{Z}=1$. Since $\sum_{\pmb{x}}\hat{h}_{\pmb{x}}$ has a global $\mathbb{Z}_2$ symmetry $\prod_iX_i$, we can derive its dual model $\sum_{\pmb{x}}\hat{h}^{\text{dual}}_{\pmb{x}}$, where
\begin{align}\label{eq:dual_model}
    \hat{h}_{\pmb{x}}^{\text{dual}}&=2\myvertex{I}{Z}{Z}{I}-2\myvertex{I}{I}{Z}{Z}+\myvertex{X}{Y}{Y}{X}+\myvertex{Y}{X}{X}{Y}-\myvertex{X}{X}{Y}{Y}-\myvertex{Y}{Y}{X}{X}\notag\\
    &=2\myvertex{I}{Z}{Z}{I}-2\myvertex{I}{I}{Z}{Z}+4\left(\xwtvertex{\sigma^{-}}{\sigma^{-}}{\sigma^{+}}{\sigma^{+}}-\xwtvertex{\sigma^{-}}{\sigma^{+}}{\sigma^{+}}{\sigma^{-}}+\mathrm{h.c.}\right),
\end{align}
and $\sigma^{\pm}=(X\pm iY)/2$. The Pauli operators of the dual model is defined on the edges of the square lattice and $\hat{h}_{\pmb{x}}^{\text{dual}}$ lives on 4 edges neighboring to a vertex of the lattice. From the second line of the above equation it is clear the model has a U(1) symmetry generated by $\sum_e Z_e$, which is just the dual of $\sum_{\langle i,j\rangle}Z_iZ_j$. And the last four terms in the bracket in the second line resemble the model proposed in Ref.~\cite{U_1_SET_TC_2023}. 
Since on a torus the duality transformation partially preserves the energy spectrum, and the deformed Ising wavefunctions $\ket{\Psi_{\text{Ising}}(\beta)}$ are dual to the deformed toric code states $\ket{\Psi_{\text{TC}}(\beta)}=\exp(\frac{\beta}{2}\sum_e Z_e)\ket{\text{TC}}$, where $\ket{\text{TC}}$ is the ground state of the toric code model~\cite{kitaev_2002}, $\ket{\Psi_{\text{TC}}(\beta)}$ for all $\beta$ are also excited states lying in the middle of the spectrum of $\sum_{\pmb{x}}\hat{h}^{\text{dual}}_{\pmb{x}}$, and might be quantum many-body scars. Notice that in Ref.~\cite{Topo_scar_2019} a different Hamiltonian $\tilde{H}(\beta)$ which has $\ket{\Psi_{\text{TC}}(\beta)}$ as an excited eigenstate was proposed, but different from their $\tilde{H}(\beta)$, our $H_{\text{dual}}$ has no tuning parameter $\beta$.

\section{Evaluating the static structure factor matrix $\mathcal{S}$ for non-injective iPEPS}
The (blocked) iPEPS tensors define a linear mapping from the virtual space to the physical space, if this linear mapping is injective (non-injective), we say the iPEPS is injective (non-injective)~\cite{SCHUCH_2010}. Physically speaking, one class of non-injective iPEPS is cat-like states with long-range order, and another class of non-injective iPEPS is topologically ordered states. Let us discuss the two classes separately.

\subsection{iPEPS with long-range order}
One example of iPEPS with long-range order is the Ising PEPS when $\beta>\beta_c\equiv\log(1+\sqrt{2})/2$. In particular, when $\beta=+\infty$, it is the Greenberger–Horne–Zeilinger (GHZ) state. Such iPEPS can be viewed as a superposition $\ket{\Psi}=\ket{\Psi_{\uparrow}}+\ket{\Psi_{\downarrow}}$, where $\ket{\Psi_{\uparrow}}$ and  $\ket{\Psi_{\downarrow}}$ are the two $\mathbb{Z}_2$ ferromagnetic states. When we contract the Ising iPEPS, spontaneous symmetry breaking happens at the virtual level and the fixed point of the environment will converge to either $\ket{\Psi_{\uparrow}}$ or $\ket{\Psi_{\downarrow}}$. This will affect the static structure factor evaluation, i.e., for the GHZ state: $\sum_i\bra{\text{GHZ}}Z_iZ_0\ket{\text{GHZ}}-\bra{\text{GHZ}}Z_0\ket{\text{GHZ}}^2\neq\sum_i\bra{\uparrow\uparrow\cdots\uparrow}Z_iZ_0\ket{\uparrow\uparrow\cdots\uparrow}-(\bra{\uparrow\uparrow\cdots\uparrow}Z_0\ket{\uparrow\uparrow\cdots\uparrow})^2$. To faithfully evaluate the static structure factor matrix of such non-injective iPEPS, one need to reconstruct the broken symmetry. Denote the fixed point matrix product density operator (MPDO) of iPEPS as $\sigma$~\cite{CIRAC_MPDO_2017}, which is obtained from CTMRG or boundary MPS methods, and denote the virtual $\mathbb{Z}_2$ symmetry as $U_X$, one can evaluate the structure factors or correlation functions using $\rho+U_X\rho U_X^{\dagger}$ instead of $\rho$ which spontaneously breaks the $\mathbb{Z}_2$ symmetry. For other non-injective iPEPS with other discrete symmetries one can perform similar symmetrization.

Moreover, for an iPEPS which is the variational ground state of a Hamiltonian, e.g., the transverse field Ising model in the ferromagnetic phase, the iPEPS will converge to the symmetry breaking state $\ket{\Psi_{\uparrow}}$ or $\ket{\Psi_{\downarrow}}$, but not cat-like state, so variational iPEPS are usually injective and the CTMRG fixed point will be unique given the injective iPEPS. In practice, to avoid iPEPS at different Hamiltonian parameters spontaneously break to different states whose spins point towards different direction, one can initialize the iPEPS to be optimized by the converged iPEPS from the previous point, such that at every point in the symmetry breaking phase the iPEPS spontaneously symmetry breaks to the same direction.

\subsection{iPEPS with topological order}

One example of PEPS with topological order is the toric code PEPS. Similar to the Ising PEPS, the boundary MPDO $\rho$ of the toric code PEPS will also spontaneously break the weak virtual $\mathbb{Z}_2$ symmetry~\cite{haegeman2015shadows}, i.e., $U_X\rho\neq\rho, U_X\rho U_X^{\dagger}=\rho$. In this case, we claim that this will not affect the evaluation of the correlation functions. This can be proven by showing the correlation functions from the two fixed points $U_X\rho$ and $\rho$ are equal. 

To calculate the connected correlation function
\begin{equation}
C(\hat{o}_{\pmb{0}},\hat{o}_{\pmb{x}}')=\frac{\bra{\Psi}\hat{o}_{\pmb{0}}\hat{o}_{\pmb{x}}'\ket{\Psi}}{\braket{\Psi}}-\frac{\bra{\Psi}\hat{o}_{\pmb{0}}\ket{\Psi}\bra{\Psi}\hat{o}_{\pmb{x}}'\ket{\Psi}}{\braket{\Psi}^2}
\end{equation}
we denote the transfer operator of the double-layer toric code PEPS as $\mathbb{T}$ and the transfer operator containing the local operator $\hat{o}$ as $\mathbb{T}_o$. From the pulling through condition for topological iPEPS tensor~\cite{SCHUCH_2010} we have $(U_X\otimes I)\mathbb{T}=\mathbb{T}(U_X\otimes I)$ and  $(U_X\otimes I)\mathbb{T}_o=\mathbb{T}_o(U_X\otimes I)$. Notice that the latter relation is not valid for the Ising PEPS.  Therefore, we can prove that the correlation functions from the two fixed points are equal, i.e.
\begin{align}
    &\frac{(\rho_L|(U_X\otimes I)\mathbb{T}_o\mathbb{T}^{L}\mathbb{T}_{o'}(U_X\otimes I)|\rho_R)}{(\rho_L|(U_X\otimes I)\mathbb{T}^{L+2}(U_X\otimes I)|\rho_R)}-\frac{(\rho_L|(U_X\otimes I)\mathbb{T}_o(U_X\otimes I)|\rho_R)(\rho_L|(U_X\otimes I)\mathbb{T}_{o'}(U_X\otimes I)|\rho_R)}{(\rho_L|(U_X\otimes I)\mathbb{T}(U_X\otimes I)|\rho_R)^2}\notag\\
    =& \frac{(\rho_L|\mathbb{T}_o\mathbb{T}^{L}\mathbb{T}_{o'}|\rho_R)}{(\rho_L|\mathbb{T}^{L+2}|\rho_R)}-\frac{(\rho_L|\mathbb{T}_o|\rho_R)(\rho_L|\mathbb{T}_{o'}|\rho_R)}{(\rho_L|\mathbb{T}|\rho_R)^2},
\end{align}
where $(\rho_L|$ and $|\rho_R)$ are the left and right vectorized fixed point MPDO, respectively, and $L$ is the distance between $\hat{o}_{\pmb{0}}$ and $\hat{o}'_{\pmb{x}}$ in the horizontal direction. Notice that here we illustrate using the boundary MPS formalism for convenience and it is the same for the CTMRG case. This is consistent with the fact the topologically degenerate ground states can not be distinguished using local observables and the situations for other topological orders can be considered similarly. We conclude that the spontaneous symmetry breaking at the virtual level of topological iPEPS will not affect the evaluation of the static structure factors, which is different from the cat state case. Notice that, when topological order and the conventional spontaneous symmetry breaking at the virtual level coexist, one still needs to consider the symmetrization of the fixed point MPDO's. 

For the toric code model and its generalizations, i.e., the quantum double models~\cite{kitaev_2002}, the degrees of freedom are at the edges of the lattice. We can alternatively consider the medial lattice so that the the degrees of freedom are now at the vertices of the medial lattice. However, the iPEPS on the medial lattice has a sublattice structure and does not have the 1-site translation symmetry~\cite{Corboz_TC_2020,Xu_2025}:
\begin{equation}
    \includegraphics[width=4cm]{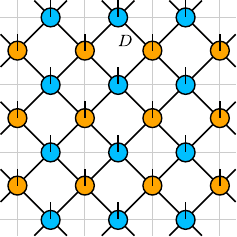},
\end{equation}
where the gray lines are the primal lattice, and the yellow and blues circles are iPEPS tensors on the medial lattice.

In this case, if we want to find solutions which have the same form as the parent Hamiltonian of the toric code model, i.e., summation of some operators on the vertices and the plaquettes of the primal lattice, we can design the ansatz for the conserved operators on the medial lattice as
\begin{equation}\label{eq:ansatz_AB_sublat}
    \hat{H}=\sum_{\pmb{x}\in A}e^{i\pmb{q}\cdot\pmb{x}}\left( \sum_{\alpha}h_{\alpha}\myarraysix{\sigma^{\alpha_1}}{\sigma^{\alpha_4}}{I}{\sigma^{\alpha_2}}{\sigma^{\alpha_3}}{I}+\sum_{\alpha'}h_{\alpha'}\myarraysix{I}{\sigma^{\alpha'_1}}{\sigma^{\alpha'_4}}{I}{\sigma^{\alpha'_2}}{\sigma^{\alpha'_3}}\right)_{\pmb{x}},
\end{equation}
where $\sigma^{\alpha_i}$ ($\alpha_i=x,y,z$) are Pauli matrices, and $A$ is the set of sites for one of the two sublattices, and $\pmb{q}=(\frac{2\pi n}{\sqrt{2}L_x}, \frac{2\pi m}{\sqrt{2}L_y})$. The two terms in Eq.~\eqref{eq:ansatz_AB_sublat} correspond to the vertices and plaquettes terms in the primal lattice. Instead of using all operator basis of the $2\times 3$ rectangle, Eq.~\eqref{eq:ansatz_AB_sublat} uses only a subset of the basis
\begin{equation}
    \mathcal{O}=\left\{\hat{o}^\alpha~\Bigg\vert~\myarraysix{\sigma^{\alpha_1}}{\sigma^{\alpha_4}}{I}{\sigma^{\alpha_2}}{\sigma^{\alpha_3}}{I} \quad \mathrm{or} \quad \myarraysix{I}{\sigma^{\alpha_5}}{\sigma^{\alpha_8}}{I}{\sigma^{\alpha_7}}{\sigma^{\alpha_6}}  \right\}
\end{equation} 
The static structure factors to be used to extract such ansatz become
\begin{align}\label{eq:def_struct_factor_AB}
S_{\alpha,\beta}(\pmb{q})
=& \sum_{\pmb{x}\in A} e^{-i\pmb{q}\cdot\pmb{x}}
C(\hat{o}^{\alpha}_{\pmb{x}}, \hat{o}^{\beta}_{\pmb{0}}),
\end{align}
where $\pmb{0}$ is a chosen site in $A$ and  $\hat{o}^{\alpha}, \hat{o}^{\beta}\in \mathcal{O}$. The generating function then becomes
\begin{equation}
\bra{G^{\alpha}(\mu,\pmb{q})}=\bra{\Psi}\prod_{\pmb{x}\in A}(\mathbbm{1}+\mu e^{-i\pmb{q}\cdot\pmb{x}} \hat{o}^{\alpha}_{\pmb{x}}),
\end{equation}
and the static structure factor is still calculated through
    \begin{align}\label{Eq:iPEPS_struct_factor_AB}
   S_{\alpha,\beta}(\pmb{q}) &=\frac{\mathrm{d}}{\mathrm{d}\mu}\frac{\bra{G^{\alpha}(\mu,\pmb{q})}\hat{o}^{\beta}_{\pmb{0}}\ket{\Psi}}{\braket{G^{\alpha}(\mu,\pmb{q})}{\Psi}}\Bigg|_{\mu=0}.
\end{align} 
So our approach can be also applied to the topologically ordered iPEPS. And the solution shown in Eq.~\eqref{eq:dual_model}, which only includes the vertex terms, can be found by setting $h_{\alpha'}=0$ in Eq.~\eqref{eq:ansatz_AB_sublat}.

\end{document}